\documentclass[11pt,a4paper]{article}
\pdfoutput=1
\usepackage{jheppub}
\newcommand{\non}{\nonumber \\}

\newcommand{\alp}{\alpha}     \newcommand{\bet}{\beta}
     
\newcommand{\eps}{\epsilon}   
      \renewcommand{\th}{\theta}

   \newcommand{\ome}{\omega}

\newcommand{\Gam}{\Gamma}     
     
\newcommand{\Sig}{\Sigma}

\newcommand{\cA}{{\cal A}}    \newcommand{\cB}{{\cal B}}

\newcommand{\cG}{{\cal G}}

\newcommand{\cO}{{\cal O}}

\newcommand{\RR}{\mathbb{R}}

\newcommand{\pa}{\partial}

\newcommand{\rar}{\rightarrow}

\newcommand{\gsim}{ \lower .75ex \hbox{$\sim$} \llap{\raise .27ex \hbox{$>$}} }
\newcommand{\lsim}{ \lower .75ex \hbox{$\sim$} \llap{\raise .27ex \hbox{$<$}} }

\def\be{\begin{equation}}
\def\ee{\end{equation}}
\def\bea{\begin{eqnarray}}
\def\eea{\end{eqnarray}}

\begin{document}

%\preprint{{\bf ***~\jobname.tex~***}}

\title{Lattice potentials and fermions in holographic  non Fermi-liquids: hybridizing local quantum criticality}

\author{Yan Liu,}
\author{Koenraad Schalm,}
\author{Ya-Wen Sun,}
\author{and Jan Zaanen}
\affiliation{
~\\
\mbox{Institute Lorentz for Theoretical Physics, Leiden University}\\
\mbox{P.O. Box 9506, Leiden 2300RA, The Netherlands }}
%\affiliation{}

\emailAdd{liu, kschalm, sun, jan@lorentz.leidenuniv.nl.}

%\date{~\\[.5in]}

\abstract{
We study lattice effects in strongly coupled  systems of fermions at a finite density described by a holographic dual consisting of fermions in Anti-de-Sitter space in the 
presence of a Reissner-Nordstr\"om black hole. The lattice effect is encoded by a periodic modulation of the chemical potential with a wavelength of order of the 
intrinsic length scales of the system. This corresponds with a highly complicated ``band structure" problem in AdS, which we only manage to solve in the weak potential
limit. The ``domain wall" fermions in AdS encoding for the Fermi surfaces in the boundary field theory diffract as usually against the periodic lattice, giving rise to band gaps.
However, the deep infrared of the field theory as encoded by the near horizon AdS${}_2$  geometry in the bulk reacts in a surprising way to the weak potential. The hybridization
of the fermions bulk dualizes into a linear combination of CFT$_1$ ``local quantum critical" propagators in the bulk, characterized by momentum dependent exponents
displaced by lattice Umklapp vectors. This has the consequence that the metals showing quasi-Fermi surfaces cannot be localized in band insulators.    
In the AdS${}_2$ metal regime, where the conformal dimension of the fermionic operator is large and no Fermi surfaces are present at low $T/\mu$, the lattice gives rise to 
a characteristic dependence of the energy scaling as a function of momentum. We predict crossovers from a high energy standard momentum AdS${}_2$ scaling to a low 
energy regime where exponents found associated with momenta ``backscattered" to a lower Brillioun zone in the extended zone scheme. We comment on how these findings
can be used as a unique fingerprint for the detection of AdS${}_2$ like ``pseudogap metals" in the laboratory.}

\maketitle{}

%\begin{multicols}{2}

\renewcommand{\cA}{A}
\renewcommand{\cB}{B}
%%%%%%%%%%%%%%%%%%%%%%%%%%%%%%%%%%%%%%
%%%%%%%%%%%%%%%%%%%%%%%%%%%%%%%%%%%%%%
\section{Introduction}
%%%%%%%%%%%%%%%%%%%%%%%%%%%%%%%%%%%%%%
%%%%%%%%%%%%%%%%%%%%%%%%%%%%%%%%%%%%%%

The explanation of the difference between metals and insulators is one of the early successes of quantum mechanics.
This is a very familiar affair: the periodic potential of the ions acts like a diffraction grating for the single electron wave functions,  giving rise to gaps in the dispersion relations at the Brillioun zone boundary. Upon filling up the bands according to the Pauli principle,
the Fermi energy  can either end up in the middle of the bands or in the band gap, and the result is a Fermi-liquid metal or a band  
insulator, respectively.  Recently, more than 75 years later, 
a metallic state of matter has been found which appears to be completely unrelated 
to this free-fermion physics. Using a minimal bottom up AdS/CFT construction  \cite{Lee:2008xf, arXiv:0903.2477, arXiv:0904.1993},  it was discovered 
that a strongly interacting  system of conformal fields when forced to finite (fermion) density turns into a metallic state that is controlled by an emergent, local quantum 
criticality \cite{Faulkner:2009wj}.  The question arises how such a metal reacts to a static and periodic background potential. The low energy excitations
are no longer behaving like quantum mechanical waves diffracting against the background potential. Instead, these are highly collective critical modes controlled 
by the emergent, purely temporal scale invariance. 

In the AdS/CFT construction the strongly coupled conformal physics is described in terms of a gravitational dual consisting of Dirac fermions with special holographic boundary conditions, propagating in the Anti-de-Sitter Reissner-Nordstr\"om 
black hole (AdS-RN) background. Adding a spatially 
modulated potential turns out to be a highly complicated band structure-like problem, that we found intractable except in the limit of a weak potential. 
Exploiting perturbation theory using the strength of the potential as a small parameter we manage to solve the problem, and the outcomes are very surprising. 
The ``domain wall" fermions of the AdS bulk dual to the propagating ``background fermions" encoding  for quasi-fermi surfaces in the boundary theory diffract 
in the usual way against the periodic potential giving rise to
standard band gaps splitting the Fermi surfaces (Fig. \ref{cartoon1},
Fig. \ref{cartoon2}). However, the deep infrared associated with
the AdS$_2$ near horizon geometry of the RN black hole which encodes for the fermion self-energy in the bulk is a completely different story. This is most clearly 
exposed in the ``AdS$_2$ metal" regime. This the regime where the fermionic operators have such high scaling dimension that they never form Fermi surfaces in the deep IR despite the presence of a chemical potential. The low-frequency single fermion spectral function in the strongly coupled boundary CFT is in that case just proportional to the local quantum critical
CFT$_1$ propagators associated with the near horizon AdS$_2$ geometry of the extremal AdS black hole coding for a non-zero chemical potential. They are therefore characterized by a scaling form $A(\ome,\vec{k}) \sim \ome^{2\nu_{\vec{k}}}$ with the particular holographic feature that the exponent is momentum dependent. The presence of a periodic potential causes these propagators to ``hybridize'' into a linear combination of  CFT$_1$s. The consequence is that the effects of breaking translational symmetry are found in the {\em energy scaling} behavior of this ``algebraic pseudo-gap" spectrum. At ``high" energy
where the weak potential does not exert influence one finds the usual $\omega^{2\nu_{\vec{k}}}$ scaling of the fermion spectral functions. The novel momentum dependence of the exponents persists and also the Umklapp momenta $\vec{K}$ end 
up in a similar exponentiated form. The effect is that upon descending in energy a cross-over follows to a regime governed by the smaller (less irrelevant) exponent $\nu_{\vec{k}-\vec{K}}$.% ; 

The question arises whether the ``pseudogap metals" found in cuprate
superconductors \cite{cupr1,cupr2} and the less well known manganites
have anything to do with the AdS$_2$ metals \cite{mang1,mang2,mang3}.

To separate the technical details in the computation from the interesting physics that the holographic single fermion spectral function in the presence of a lattice potential describes
% The 
% remainder of the paper is organized as follows. 
we will first summarize % the outline of the approach and
the physics of our results in the next section, which is hopefully 
also comprehensible for the non-experts. In Section \ref{sec2}-\ref{sec4} we will preform the detailed computation. Section \ref{sec2} formulates the basic set-up for the holographic gravity system which encodes the lattice effect  in terms of a modulated chemical potential and how a Dirac fermion probes this geometry to study the corresponding spectral function.  In Section \ref{sec3},  we will recall how to obtain the $\omega \ll \mu_0$ retarded Green's function by the near-far matching method \cite{Faulkner:2009wj} and collect all the important ingredients. In section \ref{sec4}  we will compute the concrete semi-analytical full retarded fermion Green's function in the presence of a lattice.  
Some details % of Section \ref{sec3}
are relegated to the appendix.
     
\section{Summary of results and Outlook}  \label{sec2sum}
   
The AdS/CFT dictionary strongly suggests that % the RN black hole
local quantum criticality
is a robust feature of fermionic critical systems. The holographic encoding of critical finite density systems in terms of the gravitational bulk is natural and ubiquitous in that the finite chemical potential in the boundary simply dualizes to an electrical charge carried by a black hole in the AdS bulk. This AdS Reissner-Nordstr\"om black hole is extremal and it is distinctly characterized
by an emergent near-horizon AdS$_2 \times  \RR^d$ geometry ($d$ is the
number of space dimensions). This near-horizon region in turn governs
the deep infrared of the field theory. What the ``AdS$_2$'' factor
shows is that the deep infrared preserves criticality but only in
purely temporal regards: through the duality it encodes a 0+1
dimensional CFT, a theory with dynamical critical exponent $z=\infty$ \cite{Faulkner:2009wj}. This ``locally" quantum  critical metallic behavior is quite suggestive 
with regard to the strange metals representing the central mystery in the condensed matter laboratories: both in the quantum critical heavy fermion
systems and the normal state of optimally doped high T$_\text{c}$ superconductors,  experimental indications for such a behavior have been around for a long
time. 

The fermionic responses of this state can be studied by computing the single fermion propagators of the field theory, which dualize in infalling solutions of 
the Dirac equation in the bulk \cite{Lee:2008xf,arXiv:0904.1993,arXiv:0903.2477}. One of the main novelties of holographic dual description of finite density systems is that the (anomalous) scaling dimensions become arbitrary free parameters, rather than small deviations from near free field values. In the dual holographic description it is simply the mass parameter of the field in the AdS bulk. Pending the charge to mass ratio of the Dirac field, one then finds quite different behaviors. When the charge is much smaller than the Dirac mass in the bulk, corresponding to highly irrelevant
scaling dimensions $\Delta$ of the fermion fields at zero density, one finds that the fermion spectral densities (Im$G (\omega,k)$) at low $\omega/\mu$
are directly proportional to the infrared CFT$_1$ densities (Im${\cal G} (\omega,k)$) associated with the  AdS$_2$ near horizon geometry \cite{Faulkner:2009wj}:
\begin{equation}
{\rm Im} G(\omega,k) \simeq {\rm Im} {\cal G} (\omega,k) = {\rm Im} c_ke^{i \phi_k} \omega^{2 \nu_k}~~{\rm with} ~~%\nu_k=\sqrt{\frac{k^2}{\mu^2}+ \Delta^2 - q^2}
\nu_k=\sqrt{\frac{2k^2}{\mu^2}+ \frac{m^2}{6} -\frac{q^2}{3}}
\label{AdS2metal}
\end{equation}
where the ultraviolet ($\mu=0$) scaling dimension of the fermionic operator $\Delta$ is related to the mass in units of the AdS radius as $\Delta = mL+\frac{d}{2}$ and $q$ is the fermion charge.
The oddity of this ``AdS$_2$ metal" is that the momentum dependence enters exclusively through the momentum dependence of the {\em infrared} scaling dimensions $\nu_k$. In the small charge-to-mass regime $\nu_k$ is always positive, and one therefore finds at all momenta an ``algebraic pseudo gap" at the chemical potential, where just the algebraic rise of density of states is momentum dependent.  Although
this pure ``AdS$_2$ metal" gets less attention --- presumably because it looks quite unfamilar --- it is a serious part of the anti-de-Sitter%holography
/condensed matter (AdS/CMT) agenda. For instance, the AdS$_2$ metal appears to be generic outcome of fermions in top-down implementations of AdS/CFT where the weakly coupled system known in detail \cite{Gauntlett:2011mf, Belliard:2011qq, Gauntlett:2011wm}.

In the other scenario, when the charge-to-mass ratio is large one
enters a different regime\footnote{There is also an intermediate oscillatory region where the spectral function is a periodic funciton of $\ln \omega$ and it has a non-vanishing weight at $\omega=0$ \cite{Faulkner:2009wj}. We will not consider this regime here.} 
characterized by ``quasi Fermi-surfaces" defined by the appearance of poles in the retarded Green's function $G(\omega,k)$. For small $\omega/\mu$ one can show that in this regime the holographically computed Green's function behaves as \cite{arXiv:0903.2477,arXiv:0904.1993,Faulkner:2009wj}
\begin{eqnarray}
G (\omega,k) & \simeq & \frac{Z}{v_F (k - k_F)  - \omega - \Sigma (\omega,k)} +\ldots, \non
\Sigma (\omega,k) & = &  c_k e^{i \phi_k} \omega^{2 \nu_k} +\ldots.
\label{quasiFEmetal}
\end{eqnarray}
Here the $AdS_2$ deep infrared enters as a non-(Landau-)Fermi-liquid self energy, while the Fermi-surface information is encoded as if there is a free Fermi gas
decaying into this critical infrared \cite{arXiv:1001.5049}.
In the bulk the origin of the emergent nearly free Fermi gas is clear. 
There the emergence is ``geometrized'' into the existence of Dirac fermions living at zero energy 
at the geometrical domain wall which interpolates between the 3+1 dimensional critical theory in the UV and the 0+1 dimensional local critical theory in the IR.   The appearance of these quasi-Fermi surfaces simultaneously signals the proximity
to ``fermion hair" instabilities in the bulk (the electron star \cite{Hartnoll:2009ns,Hartnoll:2010gu}, Dirac hair \cite{Cubrovic:2010bf,Cubrovic:2011xm} {or quantum fermion models \cite{Sachdev:2011ze,Allais:2012ye}}) 
dual to the instability of the AdS$_2$ metal in the boundary towards a stable fermionic state, which appears to be a real Fermi-liquid. 

We now wish to ask the question: how do both these non-Fermi-liquid metals react to the presence of a static periodic potential characterized by a wave-vector of order of 
the Fermi-momentum? This is complementary to recent similar
studies of lattice effects in holographic duals of strongly coupled
theories: Hartnoll and Hofman consider the lattice
effect on momentum relaxation and show that it equals $\Gam \sim
T^{2\Delta_{k_L}}$ with $\Delta_{k_L}$ the scaling dimension of the
charge density operator $J(\omega,k)$ in the locally critical theory
at the lattice momentum $k_L$  \cite{Hartnoll:2012rj}. From the
momentum relaxation one can directly extract the DC conductivity. The
broadening of $\omega=0$ delta-function into a Drude peak in the AC conductivity due to lattice
effects was numerically studied by Horowitz, Santos and Tong with the
remarkable observation that the holographic computation has a lot in
common with experimental results in cuprates
\cite{Horowitz:2012ky}. Similar to the latter study, we shall use a simple and natural
encoding of a spatially varying static ``lattice potential''.  One just adds a spatial variation of the electric field on the black hole horizon, dualizing
into a spatial variation of the chemical potential in the boundary
field theory. This approach was  first introduced by Aperis {\em et al.}
\cite{Aperis:2010cd} and  Flauger {\em et al.} \cite{Flauger:2010tv}. 
who considered the responses of scalar fields in the regime where the periodicity of the potential is large compared to the intrinsic length scales of the system.
Here we will focus in on the fermionic responses, where the wavevector of the periodic modulation is of order of the chemical potential, i.e. of order of the the 
Fermi momentum when a Fermi surface is formed. Spatially modulated chemical potentials are also considered in different holographic contexts in e.g. \cite{
Keranen:2009ss,
  arXiv:1101.3326,Maeda:2011pk,Hutasoit:2011rd,GarciaGarcia:2012zd,Ganguli:2012up}.\footnote{Different ways to encode lattices in holographic set-ups are with intersecting branes \cite{Kachru:2009xf} and holographic monopoles \cite{Bolognesi:2010nb}.}
We will assume that this is a simple unidirectional potential with only a one-dimensional periodicity,
\begin{equation}
\mu (x,y)  =  \mu_0 + 2\epsilon \cos (\frac{x}{a}) ,
\label{perpotential}
\end{equation}
corresponding with an Umklapp wavevector (lattice momentum) in momentum space $K = 1/a$.

This corresponds to a bulk problem which is conceptually simple, but
technically quite hazardous (section \ref{sec2}-\ref{sec4}). Firstly,
this spatial modulation of the electric charge would correspond to a
non-spherically symmetric black-hole. We shall ignore this
backreaction, which ought to be a good approximation as long as $\eps
\ll \mu_0$.\footnote{The numerical results by Horowitz, Santos and
  Tong \cite{Horowitz:2012ky} do include the full backreaction but
  they needed a spatially varying scalar field to induce a modulation
  in the chemical potential. We have been informed that they also have
numerical results for the full backreacted geometry in the set-up we
use here. It would be interesting to compare our analytic predictions
with their numerical work.} 
In the present approximation, the free Dirac fermions in the bulk experience this electrical field 
modulation just as a diffracting potential in a spatial direction. On the other hand, the curved geometry along the radial direction and the boundary conditions
associated with the dictionary translate this to a very complicated ``band structure'' problem in the bulk that we did not manage to tackle in general. However, as in the case
of elementary band structure, it is expected that the weak potential scattering limit is representative for its most salient effects. Using $\epsilon/\mu_0$ again as the small parameter, 
we performed a weak potential scattering perturbation theory for the fermions in the bulk. As will be explained in section \ref{sec3}, this second approximation still corresponds
with a surprisingly tedious problem that we nevertheless managed to solve qualitatively.

\begin{figure}[t!]
\begin{center}
\begin{tabular}{cc}
\includegraphics[width=0.4\textwidth]{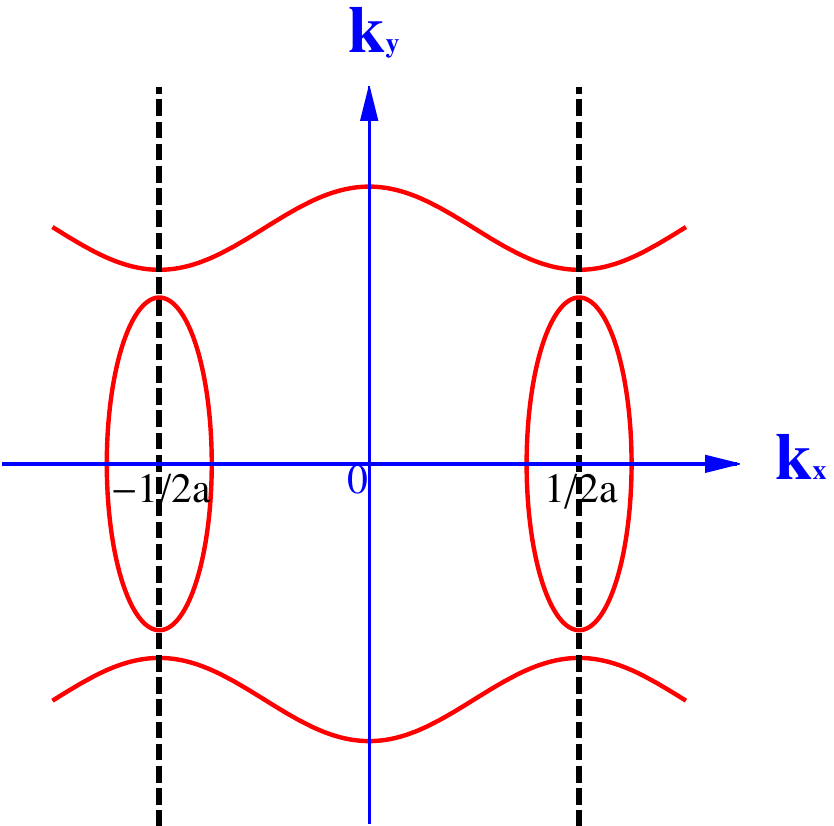}
\includegraphics[width=0.4\textwidth]{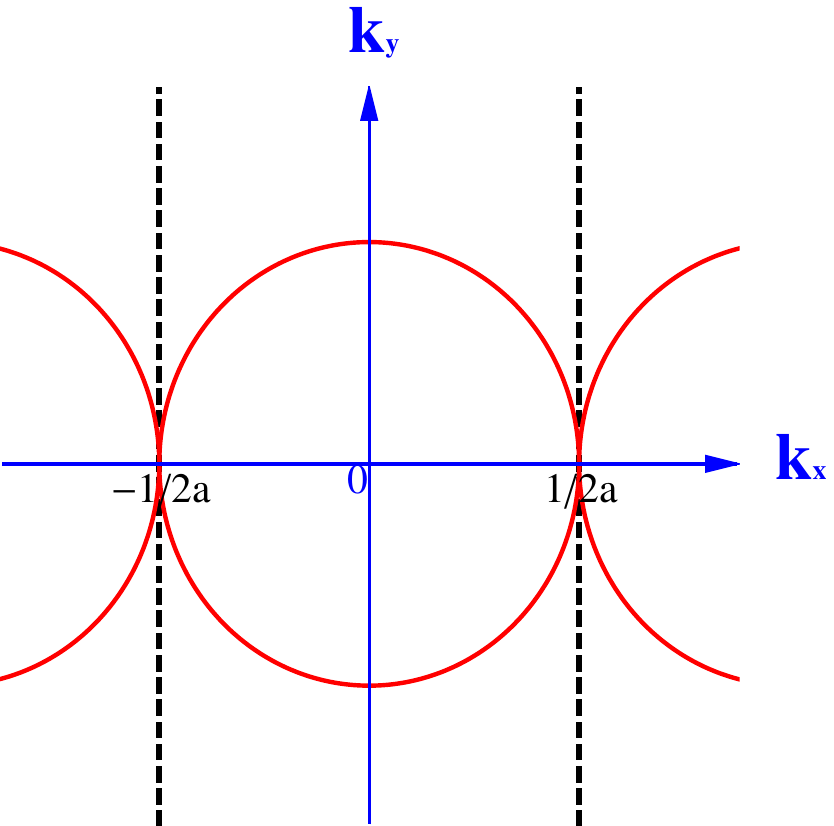}
\end{tabular}
\caption{A cartoon of our results of the band structure for different $k_F$. The system under consideration has a lattice structure only in $x$ direction.  The red line curve is the Fermi surface ($\omega=0$) and the black dashed line is the first BZ boundary. The left plot is for $k_F>\frac{K}{2}$ and the right plot is for $k_F=\frac{K}{2}$. We have a band gap at the first Brillouin Zone boundary $k_x=\pm\frac{K}{2}$ (the black dashed line) for generic $k_F>\frac{K}{2}$ (i.e. $k_y\neq 0$) which will close when $k_F=\frac{K}{2}$ 
(i.e. $k_y= 0$)
.
  At generic $k_x$, the self-energy receives a second order correction related to the lattice effect, i.e. $\Sigma=\alpha_{\vec{k}}\, \omega^{2 \nu_{\vec{k}}}  + \beta^{(-)}_{\vec{k}}\omega^{2 \nu_{\vec{k}-\vec{K}}}+\beta^{(0)}_{\vec{k}}\omega^{2 \nu_{\vec{k}}}\ln\ome+\beta^{(+)}_{\vec{k}}\omega^{2 \nu_{\vec{k}
 +\vec{K}}} +\ldots$.  (See Eq.~\eqref{CFT1generalSelf}).
Note that this picture is a result in the %extended 
periodic Brillouin zone scheme.}
\label{cartoon1}
\end{center}
\end{figure}

The outcomes are quite unexpected and to emphasize the physics over the technical steps we summarize them here.  
The predictable part is associated with the domain wall fermions responsible for the background free-fermion
dispersion in Eq. (\ref{quasiFEmetal}). These react to the presence of the periodic potential just as free fermions. For a unidirectional potential along the $x$-direction we find that a bandgap opens in the ``background'' dispersion  right at the scattering wavevector $(K, k_y)$. For small $\epsilon$ the size of the gap behaves as (see (\ref{gapp})),
\begin{equation}
\Delta (K, k_y) \simeq \epsilon \sqrt{1 - \frac{1}{\sqrt{ 1 + (2\frac{k_y}{K})^2} }}.
\label{gap}
\end{equation}
The band gap is vanishing for this unidirectional potential when the transversal momentum $k_y = 0$. This might look unfamiliar but holographic fermions have a chiral property that causes them to react to potentials in the same way as the helical surface states of three dimensional topological insulators: at $k_y=0$ the gap disappears since such 
fermions do not scatter in a backward direction.    
The new quasi-Fermi-surface in the presence of the weak potential can now be constructed as usual (Fig. \ref{cartoon1}). The Umklapp surfaces are straight lines in the $k_y$
direction, centered at the Umklapp momenta. The gaps are centered at these Umklapp surfaces and when these intersect the Fermi-surface, 
the latter split in Fermi-surface ``pockets'', see Fig. \ref{cartoon1}.  

%%%%%%%%%%%%%%%%%%%%%%%%%%%%%%%
%%%%%%%%%%%%%%%%%%%%%%%%%%%%%%%%%%

The deep infrared AdS$_2$ sector is the interesting part. Although in a less direct way, one can discern that the physics in the bulk is still about the 
``hybridization'' of free Dirac waves at momenta $\vec{k}$ and $\vec{k} \pm \vec{K}$, just as in the weak potential scattering problem of the solid state textbooks. The meaning of this hybridization  
in the dual field theory is however quite surprising. 
The AdS/CFT dictionary associates the free problem in the bulk with the physics of strongly interacting conformal fields in the boundary. In the deep infrared AdS$_2\times \RR^2$ in the absence of a lattice, one associates an {\em independent} CFT$_1$ to every point in momentum space 
(cf. Eq. (\ref{AdS2metal})).\footnote{The CFT$_1$'s are not completely independent. From the Green's function of the system, one reads of that they are spatially correlated over a length scale $\xi_{space}=\frac{\sqrt{2}}{\mu\nu_{k=0}}$ \cite{Iqbal:2011in}.} Upon breaking the continuous translational invariance, these CFT$_1$'s start to interact and this has interesting consequences. As it corresponds in essence  with quantum mechanical level mixing or hybridization in the bulk, we suggest to call the physics in the boundary ``hybridization" as well, but the novelty is the way that this now refers to the reorganization of the locally critical CFT$_1$'s. 

The holographic solution of this problem of ``weak potential'' hybridization of CFT$_1$'s 
is as follows (section \ref{sec4}). The CFT$_1$ propagator, either associated with the AdS$_2$ metal spectral density
or the quasi Fermi-surface regime self energy, ``hybridizes'' due to the Umklapp potential into a ``linear combination",
\begin{eqnarray}
{\cal G} (\omega, \vec{k}) &=& \alpha_{\vec{k}}\cG_0(\omega,\vec{k}) + \cG_1(\ome,\vec{k}) \nonumber\\
 &=& \alpha_{\vec{k}}\, \omega^{2 \nu_{\vec{k}}}  + \beta^{(-)}_{\vec{k}}\omega^{2 \nu_{\vec{k}-\vec{K}}}+\beta^{(0)}_{\vec{k}}\omega^{2 \nu_{\vec{k}}}\ln\ome+\beta^{(+)}_{\vec{k}}\omega^{2 \nu_{\vec{k}
 +\vec{K}}} +\ldots
\label{CFT1general}
\end{eqnarray}
% \begin{eqnarray}
% \Sig (\omega, \vec{k}) &=& \alpha_{\vec{k}}\Sig_0(\omega,\vec{k}) + \Sig_1(\ome,\vec{k}) \nonumber\\
%  &=& \alpha_{\vec{k}}\, \omega^{2 \nu_{\vec{k}}}  + \beta^{(-)}_{\vec{k}}\omega^{2 \nu_{\vec{k}-\vec{K}}}+\beta^{(0)}_{\vec{k}}\omega^{2 \nu_{\vec{k}}}\ln\ome+\beta^{(+)}_{\vec{k}}\omega^{2 \nu_{\vec{k}
%  +\vec{K}}} +\ldots
% \label{CFT1general}
% \end{eqnarray}
The leading term with coefficient $\alpha_{\vec{k}}$ is the zero'th order result as before, and the amplitudes $\alpha_{\vec{k}}, \beta_{\vec{k}}^{(\pm,0)}$ are depending
on UV data with $\beta \ll \alpha$ in general for $\eps\ll \mu_0$. (The complete answers are given in section~\ref{sec-self}.)

\begin{figure}[t!]
\begin{center}
\begin{tabular}{cc}
\includegraphics[width=0.6\textwidth]{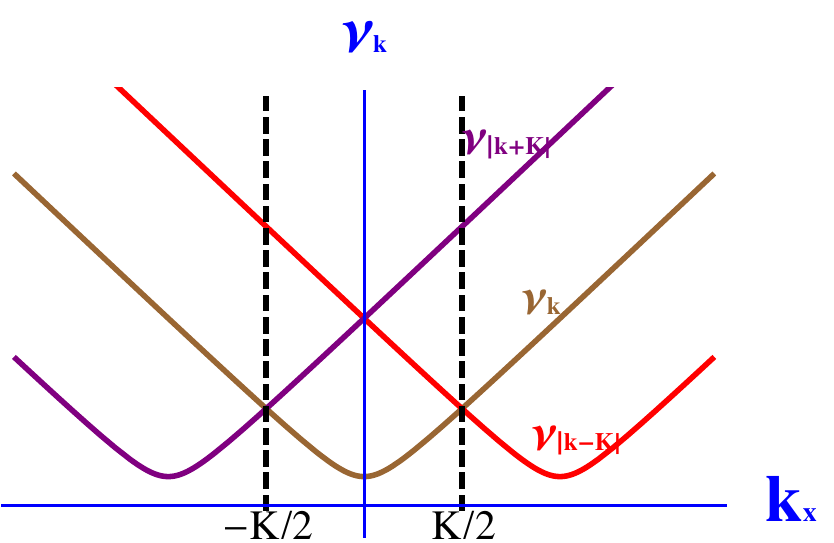}
\end{tabular}
\caption{The behavior of the different powers of $\omega^{2\nu_{k-K}},\omega^{2\nu_k},\omega^{2\nu_{k+K}}$ in the lattice AdS${}_2$-metal spectral function as a function of $k$. The IR of the Green's function is controlled by the lowest branch: $\omega^{2\nu_{k+K}}$ in the $\ell=-1$ Brillioun zone, $\omega^{2\nu_{k}}$ in the $\ell=0$ Brillioun zone, and $\omega^{2\nu_{k-K}}$ in the $\ell=1$ Brillioun zone.}
\label{cartoonnu}
\end{center}
\end{figure}

The physical meaning of Eq. (\ref{CFT1general}) is as follows. Starting in the AdS${}_2$ metal at the ``high'' energies where the potential is not exerting influence yet, one finds the 
usual scaling of fermion spectral function in the ``algebraic pseudogap'' fashion $\sim \omega^{2\nu_{\vec{k}}}$. However, at some scale associated with the ratio of the prefactors
$\beta_{\vec{k}}^{(...)}$ a cross over follows to the ``less irrelevant''   $\sim \omega^{2\nu_{\vec{k}-\vec{K}}}$; since $\nu_{\vec{k}} \sim k$ the ``hybridization of the CFT$_1$'s'' implies that the CFT$_1$'s
associated with the Brillioun zone with an index lower by one $K$ in the extended zone scheme takes over at low energy. In this way, the periodic potential gets completely
encoded in the energy scaling behavior of the electron propagators! Stepping back in the extended zone all the way to the first Brillioun zone one encounters an oddity:
one cannot subtract a further $K$ and therefore in the first zone the $\omega^{2\nu_{\vec{k}-\vec{K}}}$ contribution is not present. The ramification is that instead the next order in the
scaling hierarchy takes over: in the first zone one finds instead a crossover to the $\omega^{2\nu_k}+\omega^{2\nu_k} | \ln (\omega) |$ scaling (Fig. \ref{cartoonnu}). This logarithmic correction 
{ is easily understood as the leading term in the expansion }
\begin{equation}
\ome^{2\nu_k(\mu)}= \ome^{2\nu_k(\mu_0)}(1-4k^2\delta\mu(x,y)\ln\omega+\ldots);
\end{equation}
it is present in the higher Brillioun corrections as well, but for $\ell>0$ these corrections are subleading compared to the Umklapp correction.
The logarithmic correction has an interesting collusion right at the Brillioun zone boundary. Due to the ``resonant'' condition in the bulk, at the boundary yet another scaling arises, associated with a factor
$\omega^{2\nu_k} | \ln (\omega) |^2$. This takes over in the deep IR at the first Brillioun zone boundary, in a regime
$\delta k_x \ll \nu_k/|\ln(\omega)|$, where $\delta k_x $ denotes the deviation of the momentum from the boundary of the first Brillioun zone.
This understanding allows us to immediately guess what the answer will be at higher order in perturbation theory. At every next order in perturbation theory one can Umklapp to one further Brillioun zone:
\begin{eqnarray}
{\cal G}_{\text{full}} (\omega, \vec{k}) % &=& \alpha_{\vec{k}}\cG_0(\omega,\vec{k}) + \cG_1(\ome,\vec{k}) \nonumber\\
 &\sim   \omega^{2 \nu_{\vec{k}}}  &+ \eps^2 (\omega^{2 \nu_{\vec{k}-\vec{K}}}+\omega^{2 \nu_{\vec{k}
 +\vec{K}}})+\eps^4 (\omega^{2 \nu_{\vec{k}-2\vec{K}}}+2\omega^{ \nu_{\vec{k}
 +2\vec{K}}}) \non
&&+\ldots +\eps^{2n} (\omega^{2 \nu_{\vec{k}-n\vec{K}}}+\omega^{2 \nu_{\vec{k}
 +n\vec{K}}})+\ldots.
\label{CFT1generalfull}
\end{eqnarray}

%%%%%%%%%%%%%%%%%%%%%%%%%%%%%%%
%\begin{minipage}[t!]{\textwidth}
\begin{figure}[t!]
\begin{center}
\begin{tabular}{ccc}
\includegraphics[width=0.3\textwidth]{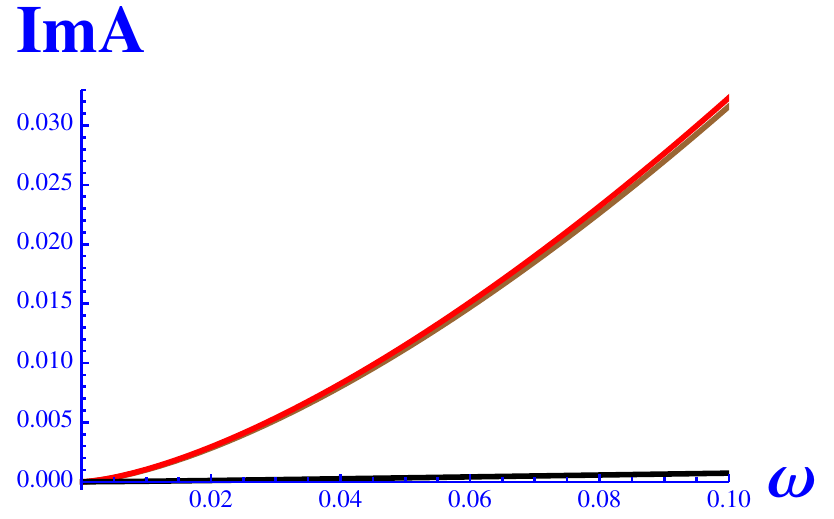}
\includegraphics[width=0.3\textwidth]{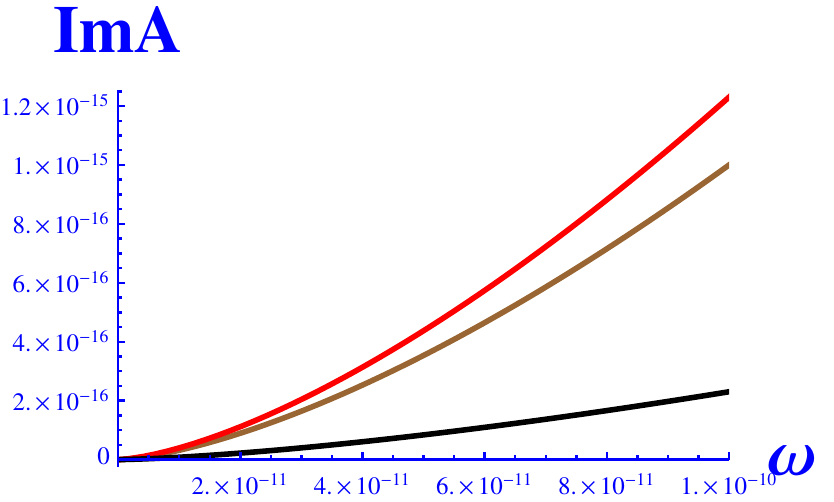}
\includegraphics[width=0.3\textwidth]{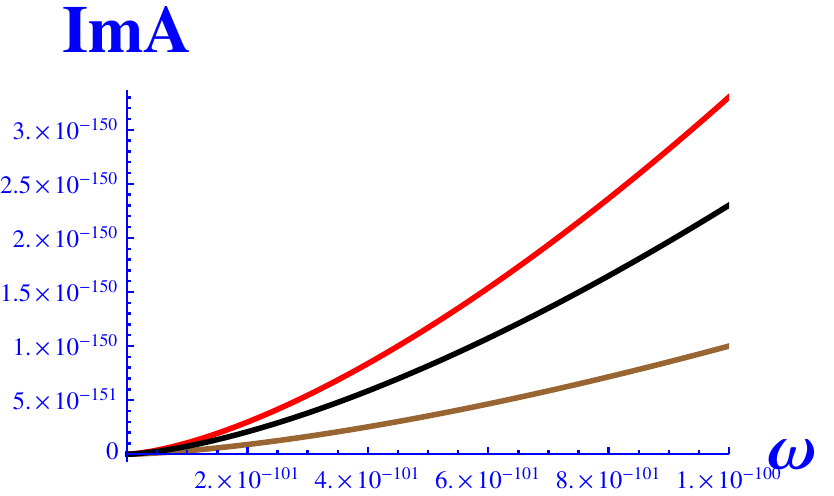}
\end{tabular}
\caption{This sequence of AdS$_2$ metal spectral functions for a fixed generic $k$ shows how the Umklapp contribution takes over at low frequencies. 
For $\ell=0$, inside the 1st BZ, we show the full corrected spectral function $A_{\text{full}}(\ome,\vec{k})\sim \text{Im} G$ (red), the original ``bare'' holographic spectral function $A_{\text{pure AdS}}(\ome,\vec{k})\sim \text{Im} {G}_0$ (brown), and the Umklapp correction due to the periodic chemical potential modulation $\delta A_{\text{lattice}}(\ome,\vec{k})\sim \text{Im} \delta G$ (black).}
\label{cartoon5}
\end{center}
%\end{figure}
%%%%%%%%%%%%%%%%%%%%%%%%%%%%%%%%%%%%

%%%%%%%%%%%%%%%%%%%%%%%%%%%%%%%%%%%%
%\begin{figure}[h]
\begin{center}
\begin{tabular}{c}
 \includegraphics[width=0.8\textwidth]{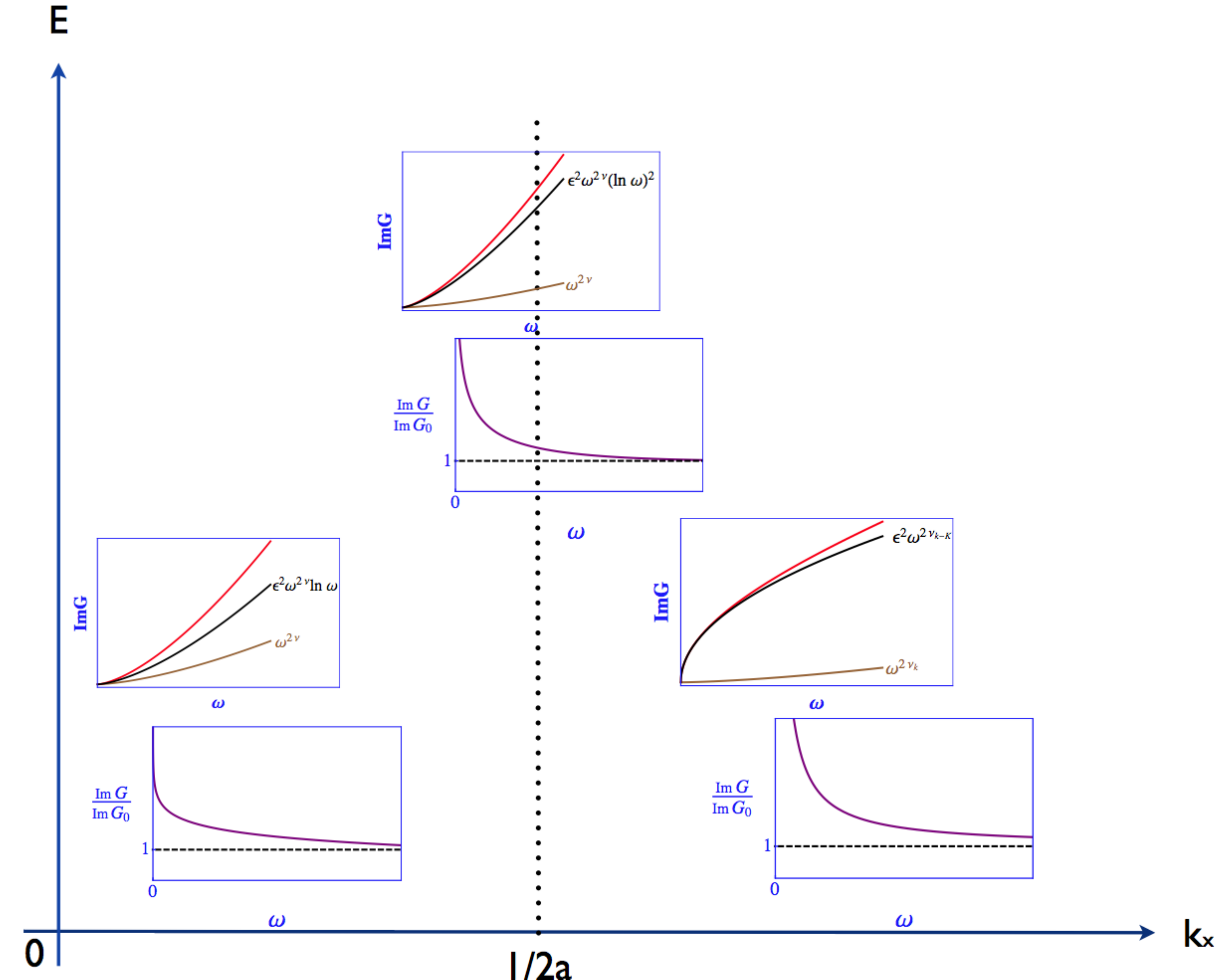}
\end{tabular}
\caption{The behavior of the AdS$_2$ metal. The spectral function in three distinct regimes: the first Brillion zone, the Brillioun edge and higher Brillioun zones. In each case the full spectral function $\text{Im}{G}(\ome, \vec{k}) =\text{Im} {G}_0+\text{Im} \delta{G} $ is plotted in red, the ``bare'' component $G_0$ in brown and the Umklapp contribution, $\text{Im} \delta{G}$ in black. The frequency scale is chosen such that the contributions are comparable by zooming as in Fig. \ref{cartoon5}. Below the ratio of the full spectral function in units of the ``bare'' spectral function is plotted for the %same 
full range of frequencies. This shows the excess states at low frequency that appear due to the lattice.}
\label{cartoon3}
\end{center}
\end{figure}
%\end{minipage}
%%%%%%%%%%%%%%%%%%%%%%%%%%%%%%%%%%%%%%

This phenomenon of Umklapp imprinting on the scaling behavior of the fermion propagators is most easily discerned in the AdS${}_2$ metal. 
As we already emphasized, the AdS$_2$-metal is the ultimate ``algebraic pseudo-gap'' state, where the fermion spectra are characterized by pseudogaps at all 
momenta, but where the algebraic rise of the spectral function is characterized by the momentum dependence of the exponents. This result therefore predicts that upon adding a 
periodic potential, the power law responses acquire generically subdominant corrections. However, in the deep IR it is these subdominant corrections  which control the density of states (Fig.\ref{cartoon3}). Consistency argues that this can only be an enhancement of the number of states at low energy (otherwise there would be a zero in the spectral weight at small but finite $\omega$). 
In the first Brillioun zone ($\ell=0$) the logarithmic correction which becomes dominant when $|\omega| \lesssim \epsilon |\ln |\ome||$ just accounts for the correction $\delta\mu(x,y)$. But in the higher Brillioun zones it is a true Umklapp correction which becomes dominant when $\omega \lesssim \eps$.

In the other regime, when $q \gg m$ and the system has quasi-Fermi surfaces to start, these effects are somewhat masked but they are still at work. Now the CFT$_1$ propagators modified by the potential determine the self-energy \eqref{quasiFEmetal}, i.e.
\begin{eqnarray}
\Sig (\omega, \vec{k}) &=& \alpha_{\vec{k}}\Sig_0(\omega,\vec{k}) + \Sig_1(\ome,\vec{k}) \nonumber\\
 &=& \alpha_{\vec{k}}\, \omega^{2 \nu_{\vec{k}}}  + \beta^{(-)}_{\vec{k}}\omega^{2 \nu_{\vec{k}-\vec{K}}}+\beta^{(0)}_{\vec{k}}\omega^{2 \nu_{\vec{k}}}\ln\ome+\beta^{(+)}_{\vec{k}}\omega^{2 \nu_{\vec{k}
 +\vec{K}}} +\ldots.
\label{CFT1generalSelf}
\end{eqnarray}
The lattice therefore alters the dispersive behavior of the ``domain wall'' free fermions. As we already argued, the latter react as usual to the potential by forming band gaps, but in addition one has the 
AdS$_2$ ``heat bath'' that reacts independently: it does not gap; the lattice only modifies its energy scaling. Since the modification entails extra states at low energies, the lattice enhances the dispersion of ``domain wall'' fermions and they become less well-defined. 

%%%%%%%%%%%%%%%%%%%%%%%%%%%%%%%%%%%%%%
\begin{figure}[ht!]
\begin{center}
\begin{tabular}{cc}
\includegraphics[width=0.6\textwidth]{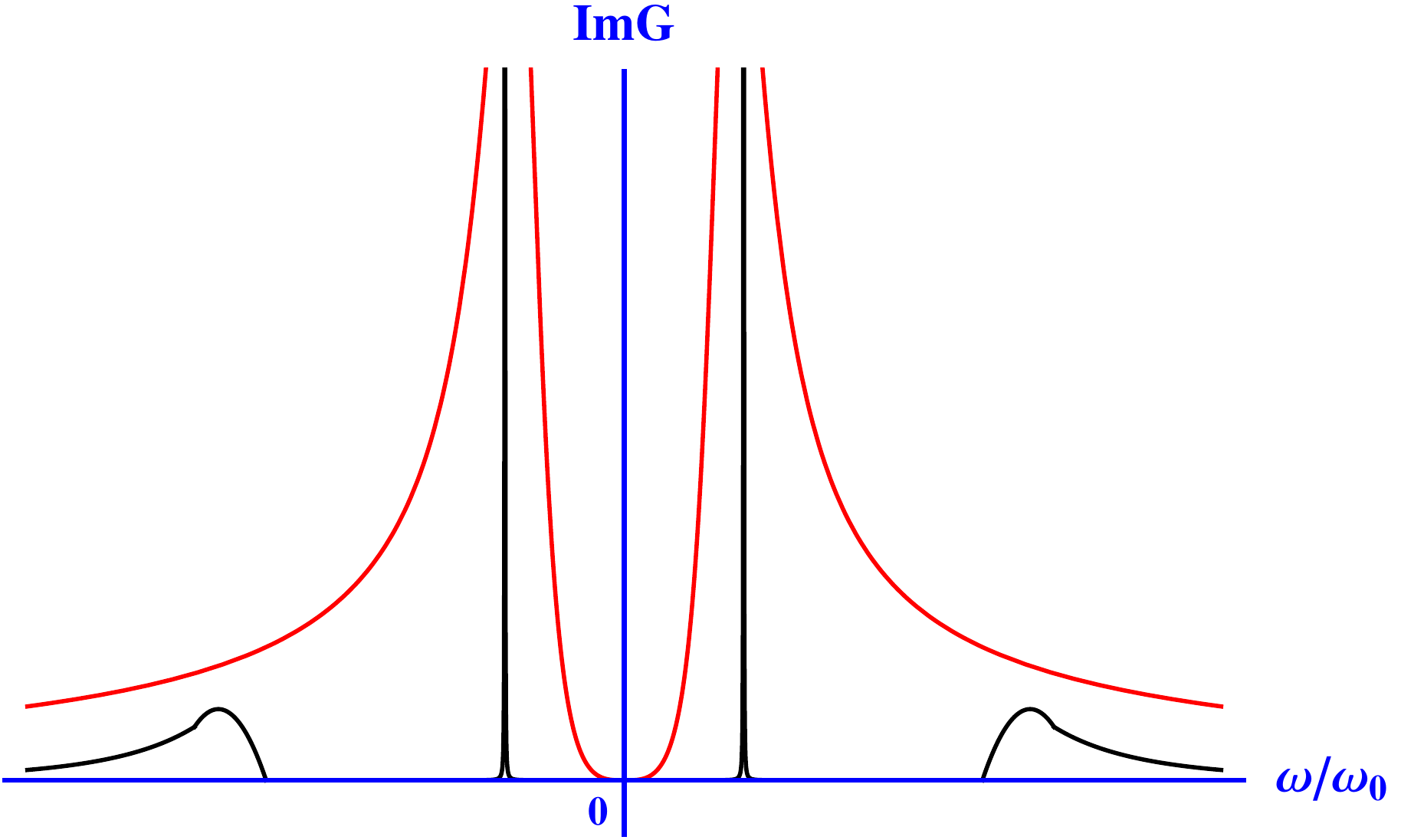}
\end{tabular}
\caption{A cartoon of the dispersion driven pseudogap: standard band
  gap (black)vs a pseudogap driven by self-energy corrections that
  persist to the lowest frequencies (red), i.e. the spectral function
  is only singular at one point right at the chemical potential.}
\label{cartoon2}
\end{center}
\end{figure}
%%%%%%%%%%%%%%%%%%%%%%%%%%%%%%%%%%%%%%

This implies that also in this regime the hard gaps of free fermions turn into soft pseudogaps with a small but non-zero spectral weight extending all the way to lowest energies (Fig.~\ref{cartoon2}).
Due to the proximity of the free fermion poles 
the information regarding the modification  of the AdS$_2$ sector as it resides in the self-energy appears in a somewhat scrambled  form in the full fermion spectral functions
but in principle it can be retrieved by careful analysis of the latter. An interesting general consequence of these results is that different from a real Fermi-liquid this quasi 
Fermi-surface ``strange metal" {\em is not localizable by a periodic potential in a band insulator}. The AdS$_2$ deep infrared is just not describing diffraction of quantum 
mechanical waves, but instead it is about highly collective critical functions that loose their purely temporal scale invariance when translational symmetry is broken, acquiring
the energy scaling highlighted in Eq. (\ref{CFT1general}) and  Fig. \ref{cartoon3}. This is yet another way of sharply distinguishing the ``quasi Fermi surface'' strange metals from true Fermi-liquids. 
At present that the ``hard wall'' confining electron star as constructed by Sachdev \cite{Sachdev:2011ze} is the only obvious 
holographic encoding of a straightforward Fermi-liquid, 
and one infers immediately that in the boundary field theory the quasiparticles will react like the domain wall fermions, forming conventional hard gapped band insulators in principle. 
The physical argument put forward here would argue that in the true groundstate of the quasi-Fermi surface regime -- the Lifshitz-type infrareds associated with the ``deconfining'' electron stars \cite{Hartnoll:2010gu} -- a similar soft pseudo-gap enhancement will happen in the presence of a periodic potential. These Lifshitz-type infrareds presumably encode for 
 fractionalized Fermi surfaces which arise from Fermions interacting with a gaplless ``heat-bath''. Qualitatively this is similar to the AdS${}_2$ bath in the AdS RN-BH set-up explored here. It remains an interesting open question to verify this.

\subsection*{Outlook}

Perhaps the most interesting ramification of our findings is that they can be used as a {\em diagnostic} for the realization of holographic metals in laboratory systems.
The main challenge  of the AdS/CMT development is to find out whether the holographic strange metals have any relevancy towards the strongly interacting 
electron systems as realized in the condensed matter laboratories. It appears that the workings of periodic potentials, nesting and so forth, might be used as 
a potentially powerful diagnostic in this regard. Although we have only control over the weak potential limit, it appears that the most salient feature we have identified
might be quite general: static potentials are encoded in the energy scaling, while the periodic potential  information is encoded in the way 
that this scaling varies in the extended zone scheme. At the least it suggests a quite unconventional to analyze angular resolved photoemission data (ARPES); in a 
future publication we will address how these matters act out in the real space fermionic response as measured by scanning tunneling spectroscopy (STS).

ARPES gives in its standard mode access to the extended Brillioun zone. It is usually taken for granted that the electron spectral functions are (modulo the spectral weight)
just copies of each other in different zones.  This is actually a flawed intuition based on non-interacting fermions. For free fermions it is of course a fact that the periodicity 
of the potential in real space dualizes in a perfect periodicity in momentum space in the form of the extended Brillioun zone. Actually, this is also still the case for the 
Fermi-liquid in the scaling limit. Because of the adiabatic continuity, the emergent quasiparticles have to behave in this regard like the free fermions: although one sees
many copies of the Fermi surface in the extended zone scheme, all Fermi-surfaces including their IR fixed point physics (e.g. the $\omega^2$ self energies) are identical 
in all the zones. There is ``only one Fermi surface'' in the Fermi liquid which can be measured in any zone, and the only variations that are allowed in the extended zone
are associated with the spectral weights in the photoemission (the ``shadow band'' motive). However, in a true non-Fermi liquid where the adiabatic continuation is not 
at work, single particle momentum is no longer a meaningful quantum number and there is no fundamental reason for the system to be precisely periodic in the extended single 
particle momentum Brillioun zone. This is what one sees so clearly at work in the above holographic non Fermi-liquids!
 Quite generally, the non periodicity of the fermion spectral functions in the extended zone
is therefore a direct experimental probe of the non-Fermi liquid nature of electron matter. 

To find out whether the phenomenological marginal Fermi liquid behavior found in optimally doped cuprates has any relation with the AdS${}_2$ metals \cite{Faulkner:2010zz},
it would be quite interesting to study whether the self energies are actually varying going to higher Brillioun zones -- to the best of our knowledge this has never been attempted
systematically.   Another empirical issue coming into view is whether the pseudogap features seen by e.g. ARPES and STS in underdoped cuprates have any dealing with
an AdS${}_2$  deep infrared while experiencing translational symmetry breaking potentials. A case has been evolving on the course of time that these pseudogaps behave in quite 
mysterious ways. A famous example is the ``nodal-antinodal dichotomy'', referring  to the rather sharp change from relatively coherent, marginal Fermi-liquid behavior in a region
in momentum space close to the ``nodal points" where the massless
Bogoliubov excitations are found in the superconductor, to a highly
incoherent antinodal momentum space regime \cite{cupr2}. 
This sudden change seems to happen at the location of the Umklapp surface associated e.g. with the two sublattice 
antiferromagnetism of the insulating cuprates \cite{cupr2}. Could this be some strong potential version of the weak potential effects we have identified in this paper?
 Similarly, underdoped cuprates
do have a reputation  for ``refusing to localize'', as examplified by
very slowly (logarithmically) diverging resistivities in the
underdoped regime \cite{cupr3}, the hardship to destroy 
superconductivity by well developed stripe order \cite{cupr4}, etc. Much more work is needed to settle these issues.

Finally, our findings might also be instrumental to interrogate other systems that appear to be characterized by pseudogap single electron behavior. A problem 
for the experimentalists is that pseudogap-like behavior can easily arise as an experimental artefact, associated with bad surfaces and so forth. A point in case are the 
bilayer manganites. Famous for their colossal magneto-resistance effects, these are very bad metals at low temperature that have been around for a while as 
non-superconducting bad metals \cite{mang4}. It first appeared that in
the best samples one does find signs of very heavy but coherent
Fermi-liquid quasiparticles \cite{mang1}, but recently 
it was demonstrated that these quasiparticles are actually artefacts
of stacking faults \cite{mang3}. The best manganite surfaces in fact show a quite neat pseudogap behavior 
throughout the Brillioun zone. Could these be realizations of the AdS${}_2$ metal? We look forward to a very careful study of the scaling behavior of the pseudogap: the 
AdS${}_2$ metal should signal itself through an algebraic scaling behavior as function of energy which should show non-trivial variations in momentum space.

%%%%%%%%%%%%%%%%%%%%%%%%%%%%%%%%%%%%%%
%%%%%%%%%%%%%%%%%%%%%%%%%%%%%%%%%%%%%%
\section{Lattice holographic metals through chemical potential modulation}\label{sec2}
%%%%%%%%%%%%%%%%%%%%%%%%%%%%%%%%%%%%%%
%%%%%%%%%%%%%%%%%%%%%%%%%%%%%%%%%%%%%%

We consider the most elementary holographic dual to a fermion system at a finite density in 2+1 dimensions. This is an extremal Reissner-Nordstr\"{o}m black hole solution of the AdS Einstein-Maxwell-Dirac theory:
\bea
ds^2&=&-r^2 f(r)dt^2+\frac{dr^2}{r^2 f(r)}+r^2(dx^2+dy^2),\\
A_t&=&\mu_0\big(1-\frac{1}{r}\big),\\
f(r)&=&1-\frac{1+Q^2}{r^3}+\frac{Q^2}{r^4},\\
\mu_0&=&2Q,
\eea
where $\mu_0$ denotes the chemical potential of the system. We have set both the cosmological radius and the horizon to unity. The Hawking temperature of this background is
\be\label{tem}
T=\frac{1}{4\pi}(3-Q^2).
\ee

We now wish to take lattice effects into consideration. We do so by simply modifying the chemical potential {\em alone}:
\bea
ds^2&=&-r^2 f(r)dt^2+\frac{dr^2}{r^2 f(r)}+r^2(dx^2+dy^2),\\
A_t&=&\big[\mu_0+\mu_1(x)\big]\big(1-\frac{1}{r}\big),\\
f(r)&=&1-\frac{1+Q^2}{r^3}+\frac{Q^2}{r^4},\\
\mu_0&=&2Q.
\eea
Here $\mu_1(x)$ is a small periodic potential which denotes the effect of the lattice. In the % weak potential
limit $\mu_1(x)\ll \mu_0$, we can ignore the backreaction of this weak nonuniform potential $\mu_1(x)$ to the background metric. Strictly speaking this is not a solution to the equations of motion of the system, even in a scaling limit. Nevertheless, if the energy density due to $\mu_1$ is small, combined with the intuition that the lattice modulation should decrease into the interior and the knowledge that it is precisely the (qualitative) IR physics in the interior that we are interested in, ignoring the backreaction should be a reasonable approximation to the true physics
\footnote{This is similar to the case of  \cite{Hartnoll:2012rj}  where the authors also assumed that 
the lattice effect is irrelevant and the IR physics is still governed by the
AdS$_2$ geometry. In a qualitatively similar setup where the lattice
periodicity is imposed through a periodic modulation of a scalar order
parameter which backreacts on both the gauge field and the geometry
\cite{Horowitz:2012ky}  the numerical solution shows explicitly that in the deep IR is
indeed governed by the AdS$_2$ geometry. 
The authors of \cite{Horowitz:2012ky}, who have also numerically solved the
chemical potential modulation set-up we employ here, have privately
communicated to us that this indeed also happens here as surmised.

Alternatively, In our case, one could imagine adding a neutral 
sector whose energy density is periodic with precisely
the opposite phase. 
Even in cases where the backreaction of a periodic potential may
change the geometry qualitatively, the feature of 
hybridyzation will not change as this is only related to the fact that we are 
considering the nearest neighbor interaction in the momentum
space. What will change, if the backreaction changes qualitatively is
all results that depend on the near-horizon AdS$_2$ geometry.}.
%\footnote{This is similar to the case of \cite{Hartnoll:2012rj} where the authors also assumed that the lattice effect is irrelavant and the IR physics is still governed by the
%AdS$_2$ geometry. In our case, alternatively one could imagine adding a neutral 
%sector whose energy density is periodic with precisely
%the opposite phase. In fact, even when the backreaction of the periodic potential is
%strictly considered and the geometry changes qualitatively, the main feature of 
%hybridyzation will not change as this is only related to the fact that we are 
%considering the nearest neighbor interaction in the momentum space.
%Alternatively one could imagine adding a neutral sector whose energy density is periodic with precisely the opposite phase.}   
In the following, we will study the system at zero temperature, i.e. $Q=\sqrt{3}$ and the result we obtain should be a good approximation to the finite but low temperature case. For simplicity, we shall only consider a one-dimensional lattice with the other direction continuous.

Such a periodic potential $\mu_1(x) = \mu_1(x+2\pi a)$ ought to capture the leading effect of the weak potential generated by periodically spaced sources (ion cores).  For the sake of simplicity and to grasp the main feature of the system we can approximate $\mu_1(x)$ by a simple $\cos(\frac{x}{a})$ function. This  lowest frequency contribution in the Fourier series of $\mu_1(x)$ is usually also the largest contribution. This kind of approximation also corresponds to the nearest neighbor approximation in momentum space.  In fact, we know that in the weak potential limit of a condensed matter lattice system, the boundary of the $n$-th Brilliouin zone will open a gap whenever the $n$-th Fourier coefficient in the Fourier series of the potential is not zero. To analyze the effect of the periodic potential to the holographic system, the lowest order term in the Fourier series of $\mu_1(x)$ can already capture the interesting physics that we want to show. Thus we choose \footnote{Note that the gauge field's equation of motion is also satisfied as long as $\mu_1(x)$ is a harmonic function and here we are choosing the largest Fourier contribution in $\mu_1(x)$ to the harmonic function as a proper approximation.}
\be\label{mu1} \mu_1(x)=2\epsilon \cos\frac{x}{a}. 
\ee
with $\eps \ll \mu_0$ the small parameter that controls the modulation.

We probe this zero-temperature periodically modulated background with a Dirac fermion with mass $m$ and charge $q$:
\be (\Gamma^a e_a^{~\mu}\mathcal{D}_\mu-m)\Psi=0 \ee
where $\mathcal{D}_\mu=\partial_\mu+\frac{1}{4}\omega_{ab\mu}\Gamma^{ab}-iqA_\mu$ and the indices $\mu$ and $a$ denote the bulk spacetime and tangent space index respectively.  The Dirac field corresponds to a fermionic operator of dimension $\Delta=\frac{3}{2}+m$ in the field theory. Solving the Dirac equation with infalling boundary condition at the horizon for Dirac field, we can extract the field theory two-point retarded Green's function for this fermionic operator by the AdS/CFT dictionary.  

To solve the Dirac equation we choose a basis of gamma matrices 
\be
\label{gammat}
\Gamma^{\underline{t}}=\begin{pmatrix}
i\sigma^1    & 0\\
0 & i\sigma^1
\end{pmatrix}, ~~~
\Gamma^{\underline{r}}=\begin{pmatrix}
-\sigma^3    & 0\\
0 &-\sigma^3
\end{pmatrix}, ~~~\Gamma^{\underline{x}}=\begin{pmatrix}
-\sigma^2   & 0\\
0 & \sigma^2
\end{pmatrix},  ~~~\Gamma^{\underline{y}}=\begin{pmatrix}
0 & -i\sigma^2  \\
i\sigma^2   & 0
\end{pmatrix}.
\ee
A redefinition of $\Psi$ 
% Along the $x$ direction we can choose the following boundary condition for $\Psi$,
\be% \Psi(r,t, x+2\pi N_1a,y)=
\Psi(t,r,x,y)=(-gg^{rr})^{-\frac{1}{4}}\begin{pmatrix}
\Phi_1 \\
\Phi_2
\end{pmatrix}, \ee
with each $\Phi_\alpha,~\alp=1,2$ a two-component spinor
effectively removes the spin connection from the equation of motion  \cite{arXiv:0903.2477}.
Due the explicit $x$-dependence in $\mu_1(x)$ the different momentum
modes are no longer independent, but will mix those which differ by an
inverse lattice vector $K=1/a$%\comment{CHECK}
. At the same time, the periodicity in the $x$-direction due to $\mu_1(x)$ ought to reflect itself in the solutions $\Phi_\alpha$. These features can be accounted for by a Bloch expansion 
\be
\Phi_\alpha =\int\frac{d\omega dk_x dk_y}{2\pi}\sum_{\ell\in \text{Z}}\Phi_{\alpha}^{(\ell)}(r,\omega,k_x,k_y)e^{-i\omega t+i \big[(k_x+\ell K)x+k_y y\big]},~~\alpha=1,2,
\ee
where $k_x\in[-\frac{K}{2},\frac{K}{2}]$ and $\ell$ characterizes the momentum level or Brillouin zone.% . Note that $k_x=-\frac{K}{2}$ and $k_x=\frac{K}{2}$ are just the boundaries of the first Brillouin zone. After constraining $k_x$ to reside in the first Brillouin zone, we can label a state with $k_x$ and the level $\ell$ where $k_x+\ell K$ is the real momentum along the $x$ direction, i.e. we will work in the extended Brillouin zone scheme.

In Bloch waves, the Dirac equation of motion reduces to
\bea\label{eom1}
&&(\partial_r+m\sqrt{g_{rr}}\sigma^3)\Phi_\alpha^{(\ell)}-i\sqrt{\frac{g_{rr}}{g_{tt}}}\bigg(\omega+q\mu_0(1-\frac{1}{r})\bigg)\sigma^2\Phi_\alpha^{(\ell)}
-(-1)^\alpha\sqrt{\frac{g_{rr}}{g_{xx}}}(\frac{\ell}{a}+k_x)\sigma^1\Phi_\alpha^{(\ell)} \nonumber\\
&&-i (-1)^\alpha\sqrt{\frac{g_{rr}}{g_{xx}}} k_y\sigma^1\Phi_\beta^{(\ell)}-iq\epsilon \sqrt{\frac{g_{rr}}{g_{tt}}}\sigma^2(1-\frac{1}{r})(\Phi_\alpha^{(\ell-1)}+\Phi_\alpha^{(\ell+1)})=0,
\eea
where $\alpha=1,2; \beta=3-\alpha$ and $\epsilon$ is the Fourier coefficient in $\mu_1(x)=2\epsilon \cos\frac{x}{a}$. We explicitly see how the periodicity mixes modes whose momentum differ by $K$. As a result, the momentum is no longer conserved and, as is well known, only $k_x$ is the conserved quantity.

%%%%%%%%%%%%%%%%%%%%%%%%%%%%%%%%%%%%%%
%%%%%%%%%%%%%%%%%%%%%
\section{Retarded Green's function}\label{sec3}
%%%%%%%%%%%%%%%%%
%%%%%%%%%%%%%%%%%%%%%%%%%%%%%%%%%%%%%%
%%%%%%%%%%%%%%%%%%%%%%%%%%%%%%%%%%%%%%

We now solve the EOM (\ref{eom1}) with the infalling boundary condition for each $\Phi_\alpha^{(\ell)}$ to extract the corresponding retarded Green's function in the dual strongly interacting boundary field theory. From (\ref{eom1}), we can see that generally the different spinor components of the Dirac field mix with each other for nonzero $k_y$. To simplify the equations, we make the following rotation in $\alpha,~\beta$ space
(recall that each $\Phi_\alpha$ is still a two-component spinor and it is in that spinor space that the Pauli matrices act. This transformation rotates  the two separate two-component spinors into each other.)
%\commently{YL: made it unitary. Matrix M corrected.  The corresponding code `diagnew.nb' is in the dropbox.KS: Shouldn't %there be an additional $k_y$ in the lower right hand entry? (i.e. the matrix should be the identity when $k_y=0$.}
%\be
%\begin{pmatrix}
%-i(k_x+\frac{\ell}{a}+k_\ell) ~&& k_y\\
%\frac{i(-k_x-\frac{\ell}{a}+k_\ell)}{k_y} ~&& 1
%\end{pmatrix}
%\begin{pmatrix}
%\Phi_1^{(\ell)} \\
%\Phi_2^{(\ell)}
%\end{pmatrix}\to
%\begin{pmatrix}
%\Phi_1^{(\ell)} \\
%\Phi_2^{(\ell)}
%\end{pmatrix},
%\ee
\be\label{mmatrix}
\begin{pmatrix}
\frac{(k_x+\frac{\ell}{a}+k_\ell)}{\sqrt{k_y^2+(k_x+\frac{\ell}{a}+k_\ell)^2}} ~&& \frac{ik_y}{\sqrt{k_y^2+(k_x+\frac{\ell}{a}+k_\ell)^2}} \\
\frac{i(-k_x-\frac{\ell}{a}+k_\ell)}{k_y\sqrt{1+(k_x+\frac{\ell}{a}-k_\ell)^2/k_y^2}} ~&& \frac{1}{\sqrt{1+(k_x+\frac{\ell}{a}-k_\ell)^2/k_y^2}} 
\end{pmatrix}
\begin{pmatrix}
\Phi_1^{(\ell)} \\
\Phi_2^{(\ell)}
\end{pmatrix}\to
\begin{pmatrix}
\Phi_1^{(\ell)} \\
\Phi_2^{(\ell)}
\end{pmatrix},
\ee
where $k_\ell=\sqrt{(k_x+\frac{\ell}{a})^2+k_y^2}$.  When $k_y\to 0$, this transformation matrix is $\begin{pmatrix}
1\ 0 \\
0\ 1
\end{pmatrix}$ for $k_x>0$, while $\begin{pmatrix}
0\ i \\
i\ 0
\end{pmatrix}$ for $k_x<0$.

After this rotation, the equation of motion (\ref{eom1}) simplifies to
\bea\label{eom2}
&&(\partial_r+m\sqrt{g_{rr}}\sigma^3)\Phi_\alpha^{(\ell)}-i\sqrt{\frac{g_{rr}}{g_{tt}}}(\omega+q\mu_0(1-\frac{1}{r}))\sigma^2\Phi_\alpha^{(\ell)}-(-1)^\alpha\sqrt{\frac{g_{rr}}{g_{xx}}}k_\ell\sigma^1\Phi_\alpha^{(\ell)} \nonumber\\
&&-iq\epsilon\sqrt{\frac{g_{rr}}{g_{tt}}}\sigma^2(1-\frac{1}{r})\sum_{\beta; \ell'=\ell\pm1}M_{\alpha\beta,\ell\ell'}\Phi_{\beta}^{(\ell')}=0,
\eea
with $M_{\alpha\beta,\ell\ell'}=(M_{\alpha1,\ell\ell'}, M_{\alpha2,\ell\ell'})$ and
%\be\label{mn} M_{\alpha1, \ell\ell'}=k_y^{1-\alpha}\bigg(\frac{1}{2}+
%\frac{\frac{\ell-\ell'}{a}-(-1)^\alpha k_{\ell}}{2k_{\ell'}}\bigg),~~~
%M_{\alpha2,\ell\ell'}=k_y^{2-\alpha}\bigg(\frac{1}{2}-\frac{\frac{\ell-\ell'}{a}-(-1)^\alpha k_{\ell}}{2k_{\ell'}}
%\bigg). \ee
\bea\label{mn} M_{\alpha1, \ell\ell'}&=&-i^{3-\alpha}\text{sign}(k_y^{3-\alpha})
\frac{\bigg(\frac{\ell-\ell'}{a}+k_{\ell}-(-1)^{\alpha}k_{\ell'}\bigg)\sqrt{k_{\ell'}^2-(-1)^\alpha k_{\ell'}(k_x+\frac{\ell'}{a})}}{2k_{\ell'}\sqrt{k_\ell^2+k_\ell(k_x+\frac{\ell}{a})}},\nonumber\\
M_{\alpha2,\ell\ell'}&=&(-i)^{\alpha}\text{sign}(k_y^{\alpha})
\frac{\bigg(\frac{\ell-\ell'}{a}-k_{\ell}-(-1)^{\alpha}k_{\ell'}\bigg)\sqrt{k_{\ell'}^2-(-1)^\alpha k_{\ell'}(k_x+\frac{\ell'}{a})}}{2k_{\ell'}\sqrt{k_\ell^2-k_\ell(k_x+\frac{\ell}{a})}}.\eea
Note that $M_{\alpha\beta,\ell\ell'}$ have smooth limits as $k_y\rar 0$. %despite the $k_y^{\beta-\alpha}$ prefactor.
This rotation makes explicit that mixing of the different levels is controlled by $k_\ell$. In terms of symmetries, the coupling between different momentum levels is related to the continuous translational symmetry breaking in the $x$ direction while the mixing between different spinor indices is related to the broken $SO(2)$ symmetry in the $x,y$ plane. 

Equation (\ref{eom2}) describes a set of infinite coupled equations with infinite many fields to be solved, corresponding to the full range of $\ell=0,\pm 1,\pm 2,...,\pm\infty$. In principle we need to diagonalize these equations to solve them. To do so explicitly is non-trivial because of the $r$-radial coordinate dependence of the coefficients in the equation. Instead we shall solve them in perturbation theory in the limit $\epsilon\ll \mu_0$.
For $\epsilon$ is a small parameter, we can solve the EOM (\ref{eom2}) order by order in $\epsilon$ in the whole spacetime.  

Expanding each level $\Phi_\alpha^{(\ell)}$ in $\epsilon$ as
\begin{equation}
\Phi_{\alpha}^{(\ell)}=\Phi_{\alpha}^{(0,\ell)}+\epsilon\Phi_{\alpha}^{(1,\ell)}+\epsilon^2\Phi_{\alpha}^{(2,\ell)}+...
\end{equation} 
the zeroth order equation is just the free Dirac equation without the interaction term, and each level $\ell$ decouples at this order. 
\be
(\partial_r+m\sqrt{g_{rr}}\sigma^3)\Phi_\alpha^{(0,\ell)}-i\sqrt{\frac{g_{rr}}{g_{tt}}}(\omega+q\mu_0(1-\frac{1}{r}))\sigma^2\Phi_\alpha^{(0,\ell)}
-(-1)^\alpha\sqrt{\frac{g_{rr}}{g_{xx}}}k_\ell\sigma^1\Phi_\alpha^{(0,\ell)} =0,
\ee
Substituting the lowest order solutions into the first order equation to obtain the first order equations, we obtain the hierarchy of equations
\bea
&&(\partial_r+m\sqrt{g_{rr}}\sigma^3)\Phi_\alpha^{(s,\ell)}-i\sqrt{\frac{g_{rr}}{g_{tt}}}(\omega+q\mu_0(1-\frac{1}{r}))\sigma^2\Phi_\alpha^{(s,\ell)}-(-1)^\alpha\sqrt{\frac{g_{rr}}{g_{xx}}}k_\ell\sigma^1\Phi_\alpha^{(s,\ell)} \nonumber\\
&&~~~~~~~~~-iq\sqrt{\frac{g_{rr}}{g_{tt}}}\sigma^2(1-\frac{1}{r})\sum_{\beta;\ell'=\ell\pm1}M_{\alpha\beta,\ell\ell'}\Phi_\beta^{(s-1,\ell')}=0
\eea
with $s=1,2,\ldots$
%\bea
%&&(\partial_r+m\sqrt{g_{rr}}\sigma^3)\Phi_\alpha^{(2,\ell)}-i\sqrt{\frac{g_{rr}}{g_{tt}}}(\omega+q\mu_0(1-\frac{1}{r}))
%\sigma^2\Phi_\alpha^{(2,\ell)}-(-1)^\alpha\sqrt{\frac{g_{rr}}{g_{xx}}}k_\ell\sigma^1\Phi_\alpha^{(2,\ell)} \nonumber\\
%&&~~~~~~~~~-iq\sqrt{\frac{g_{rr}}{g_{tt}}}\sigma^2(1-\frac{1}{r})\sum_{\beta;\ell'=\ell\pm1}M_{\alpha\beta,\ell\ell'}\Phi_%\beta^{(1,\ell')}=0.
%\eea

Previous studies have shown that the relative size of $q$ to $m$ determines the characteristics of the system \cite{arXiv:0904.1993,Faulkner:2009wj}. We shall assume that this will remain the case with the lattice modulation and use this liberty to choose $m=0$ from here on.

%%%%%%%%%%%%%%%%%%%%%%%%%%%%%%%%%%%%%%
\subsection{Boundary conditions}\label{subsec31}
%%%%%%%%%%%%%%%%%%%%%%%%%%%%%%%%%%%%%%

To compute the retarded field theory Green's function we need to
impose infalling boundary conditions at the extremal horizon $r\equiv
r_h=1$.\footnote{We have chosen $\mu_0=2\sqrt{3}$. } The number of linearly independent infalling boundary conditions are equal to the number of fields in the system, which is formally infinite. Because the fields in the system can be labeled by the level $\ell$ and the spinor component $\alpha$, we can also label the boundary conditions using $\alpha\ell$. The following is the definition of the $\alpha\ell$-th boundary condition: we pick a single infalling spinor component at zeroth order: $\Phi^{(0,\ell)}_\alpha(r_h) \neq 0$ for a fixed component $\alpha$ and fixed level $\ell$ with the other component and all other levels vanishing. % the $\alpha\ell$-th boundary condition corresponds to imposing ``1" infalling wave for the $\alpha\ell$-th field at the zeroth order while 0 infalling waves for other levels or spinor components. The explicit meaning of ``1" infalling wave will be manifested in the next subsection.
% Under the $\alpha\ell$-th boundary condition, at the zeroth order, only $\Phi^{(0,\ell)}_{\alpha}$ is nonzero, and
\newcommand{\fix}{\mathrm{fix}}
At lowest order in $\eps$, all other modes decouple from these boundary conditions and only the field $\Phi^{(0,\ell^\fix)}_{\alpha^\fix}(r)$ will be non-zero.
At the first order in $\eps$,  we can see from the equations (\ref{eom1}) and (\ref{eom2}) that now also the fields $\Phi^{(1,\ell^\fix\pm 1)}_{\alpha^\fix}(r)$ are nonzero when $k_y=0$ and when $k_y\neq0$ both $\Phi^{(1,\ell^\fix\pm 1)}_{\alpha^\fix}(r_h)$ and $\Phi^{(1,\ell\pm 1)}_{\beta^\fix\neq\alpha^\fix}(r_h)$ are nonzero. At the second order, we also excite $\Phi^{(2,\ell^\fix\pm1\pm1)}_{\alpha}(r_h)$ and $\Phi^{(2,\ell^\fix\pm1\pm1)}_{\beta}(r_h)$. When we excite these higher order modes, we should also impose infalling boundary conditions for these modes. As we shall see, it will suffice for us to compute the solution up to second order only.

Letting $\ell^\fix$ and $\alpha^\fix$ range over all possible values, we obtain an infinite set of linearly independent boundary conditions. Thus one can get the most general solution to the Dirac equation with infalling boundary conditions.
The full solution is therefore an $\infty\times\infty$ matrix $\Phi_{\alpha\ell(\beta\ell')}^{(s)}(r)$, whose $\alpha\ell\beta\ell'$ element gives the solution $\Phi_{\alpha}^{(s,\ell)}(r)$ under the $\beta\ell'$-th infalling horizon boundary condition.

At AdS infinity --- the conformal boundary at $r\rar \infty$ corresponding to the UV in the field theory --- the equations of all levels and all orders $\Phi^{(s,\ell)}_{\alpha}$ decouple and are of the same form as the free Dirac equation, i.e. for any  $\beta\ell'$-th horizon boundary condition all $\Phi^{(s,\ell)}_{\alpha}$ behave on the conformal boundary as
\be
\Phi^{(s)}_{\alpha\ell(\beta\ell')}= \cA_{\alpha\ell,\beta\ell'}^{(s)} r^m  \begin{pmatrix}
0 \\
1
\end{pmatrix}+\cB_{\alpha\ell,\beta\ell'}^{(s)} r^{-m}
\begin{pmatrix}
1\\
0
\end{pmatrix} +\ldots
\ee
Note that the superscript $(s)$ means the order of $\epsilon$, and not the Bloch wave indexed by $\ell$.
The retarded Green's function is then computed from these two matrices
$\cA_{\alpha\ell,\beta\ell'}\equiv \sum_s \eps^s \cA_{\alpha\ell,\beta\ell'}^{(s)}$ and $\cB_{\alp\ell,\bet\ell'}\equiv \sum_s \eps^s \cB_{\alpha\ell,\beta\ell'}^{(s)}$ as
\be G_R={\cB}{\cA}^{-1}.\ee 
The spectral weight can be easily obtained from the Green's function matrix. In the following we focus on the diagonal momentum Green's function $G_R(\omega;\vec{k},\vec{k})$ and therefore the diagonal momentum spectral weight \footnote{Since the lattice breaks translation invariance,
  momentum is no longer a good quantum number. The Green's function
  will therefore also have non-zero value for operators are momenta
  that differ by a lattice vector. Formally we should diagonalize the
  system into a new ``non-momentum'' basis of modes. 
  Since one in ARPES experiments the photoelectron propagates in the galilean 
  continuum it has  a definite
  momentum, one keeps the
  Green's function in the momentum-basis. The dominant response is
  still in the diagonal momentum-channel, though this now contains a
  mixing with other momentum-modes.
}.
In both  ${\cA}$ and ${\cB}$ matrices, the leading order contribution to the off-diagonal elements are of order $\epsilon$ whereas the leading order contributions in the diagonal elements are of order $\epsilon^0$. For consistency in determining the effect of $\mu_1(x)$ in the spectral weight up to the second order, we therefore have to keep both the zeroth and second order solutions in the diagonal elements of the solution matrix as well as the first order solution in the off diagonal elements. This way we see that the modulation $\mu_1(x)$ affects the diagonal momentum Green's function at the second order of $\epsilon$. This is the case for a generic momentum $k_x$. At the boundary of the first Brillouin zone, we will need more care as we will see later.

Thus generally the  ${\cA}$ and ${\cB}$ matrices look like
\begin{eqnarray}
\label{matrices}
%\mathcal{A}_{\alpha\ell,\alpha'\ell'}= \begin{pmatrix}
%... ~~~&~&~& ~&~&~& \\
%&\cO(\eps)&&&&&\\
% &(1+\eps^2)\delta_{\alp\alp'}&\eps P_{\alp\alp'}&&&& \\
% &\eps P_{\alp\alp'}&(1+\eps^2)\delta_{\alp\alp'}&\eps P_{\alp\alp'}&&&\\
% &&\eps P_{\alp\alp'}&(1+\eps^2)\delta_{\alp\alp'}&&&&\\
% &&&\eps P_{\alp\alp'}&&&\\
% &&&&&&\ldots\\
%\end{pmatrix},\non
{\cA}_{\alpha\ell,\alpha'\ell'}= \begin{pmatrix}
... ~~~&~&~& ~&~&~& \\
&\eps F^{\cA}_{\alp\alp'}&&&&&\\
 &(Z^{\cA}+\eps^2S^{\cA})\delta_{\alp\alp'}&\eps F^{\cA}_{\alp\alp'}&&&& \\
 &\eps F^{\cA}_{\alp\alp'}&(Z^{\cA}+\eps^2S^{\cA})\delta_{\alp\alp'}&\eps F^{\cA}_{\alp\alp'}&&&\\
 &&\eps F^{\cA}_{\alp\alp'}&(Z^{\cA}+\eps^2S^{\cA})\delta_{\alp\alp'}&&&&\\
 &&&\eps F^{\cA}_{\alp\alp'}&&&\\
 &&&&&&\ldots\\
\end{pmatrix},\non
{\cB}_{\alpha\ell,\alpha'\ell'}= \begin{pmatrix}
... ~~~&~&~& ~&~&~& \\
&\eps F^{\cB}_{\alp\alp'}&&&&&\\
 &(Z^{\cB}+\eps^2S^{\cB})\delta_{\alp\alp'}&\eps F^{\cB}_{\alp\alp'}&&&& \\
 &\eps F^{\cB}_{\alp\alp'}&(Z^{\cB}+\eps^2S^{\cB})\delta_{\alp\alp'}&\eps F^{\cB}_{\alp\alp'}&&&\\
 &&\eps F^{\cB}_{\alp\alp'}&(Z^{\cB}+\eps^2S^{\cB})\delta_{\alp\alp'}&&&&\\
 &&&\eps F^{\cB}_{\alp\alp'}&&&\\
 &&&&&&\ldots\\
\end{pmatrix},
\end{eqnarray}
where the zeroth, first and second order $\cA$ and $\cB$ coefficients $Z^{\cA,\cB}$, $F^{\cA,\cB}$ and $S^{\cA,\cB}$ do depend on $\ell$ and $\ell'$ in contrast to this sketch of their form.
 
In the following sections we will compute these two matrices using the near-far matching method.

%%%%%%%%%%%%%%%%%%%%%%%%%%%%%%%%%%%%%%
\subsection{Matching}
%%%%%%%%%%%%%%%%%%%%%%%%%%%%%%%%%%%%%%

The macroscopic properties of this system are determined by the low
energy modes.  We shall therefore focus on the small $\omega/\mu_0$
region. This enables us to use the near-far matching method to
determine the spectral function following \cite{Faulkner:2009wj}. We
divide the spacetime into two regions: the near region and the far
region. The near region is the near horizon region, i.e. $ r-1 \ll 1$. Here the geometry
becomes that of AdS$_2\times \RR^2$.\footnote{Recall that we are
  ignoring gravitational backreaction. However, because the breaking
  of translational invariance due to the lattice should be a UV
  effect, even with gravitational backreaction the strict near horizon
  IR region should be approximately AdS$_2\times \RR^2$. Physically we
are interested in the qualitative effects due to the periodic
potential and intuitively these should already be there with the
periodically modulated electrostatic potential. Including the induced modulation
in the spacetime curvature due to the varying energy-density in the
electric field will presumably yield only quantitative and not
qualitative changes.} The far region is defined as the $\frac{\omega/\mu_0}{r-1}\ll 1$ region. We can solve the equations in the near and far region separately. For $\ome\ll \mu_0$ the near and far region have a non-zero overlap $\ome/\mu_0 \ll r-1 \ll 1$ and we can match the separate approximations within this region. 

We will first study the near horizon region in detail.

%%%%%%%%%%%%%%%%%%%%%%%%%%%%%%%%%%%%%%
\subsubsection{Near region solutions for $\Phi_{\alpha}^{(0,\ell)}$}
%%%%%%%%%%%%%%%%%%%%%%%%%%%%%%%%%%%%%%

The zeroth order solution (no lattice modulation) is the simplest case and the results were already obtained in \cite{Faulkner:2011tm,Faulkner:2009wj}. We collect their solutions here.

Near the horizon, we have $g_{tt}=g^{rr}\simeq 6 (r-1)^2$ and $g_{xx}\simeq 1.$  and the equation \eqref{eom2} for $\Phi_{\alpha}^{(0,\ell)}$ is by construction identical to extremal RN AdS black hole case with $k_\ell=\sqrt{(k_x+\frac{\ell}{a})^2+k_y^2}$ \cite{Faulkner:2009wj}:
\be
\partial_r\Phi_\alpha^{(0,\ell)}-\frac{i}{6(r-1)^2}(\omega+q\mu_0(r-1))\sigma^2\Phi_\alpha^{(0,\ell)}
-(-1)^\alpha\frac{1}{\sqrt{6}(r-1)}k_\ell\sigma^1\Phi_{\alpha}^{(0,\ell)}=0.
\ee
After performing a coordinate transformation, $x=\frac{\omega}{r-1}$, this first order spinor equation is equivalent to a second order one for each of the separate components:
\be\label{eomnear0}
\partial_x^2\Phi_{\alpha j}^{(0,\ell)}+(\frac{2}{x}-\frac{\partial_x h_{\alpha j}}{h_{\alpha j}})\partial_x \Phi_{\alpha j}^{(0,\ell)}
+\frac{h_{\alpha 1}h_{\alpha 2}}{x^4}\Phi_{\alpha j}^{(0,\ell)}=0~.
\ee
Here \be\label{hnew} h_{\alpha j}=-x(\frac{x}{6}+\frac{q\mu_0}{6}+(-1)^{\alpha+j+1}\frac{k_\ell}{\sqrt{6}})\ee 
and the equation is solvable \cite{Faulkner:2011tm} in terms of Whittaker functions.%  Note that (\ref{eomnear0}) does not depend on $\omega$ as the explicit term cancels with the $1/\ome$ dependence in $h_{\alp j}$. The definition for $h_{\alp j}$ including the explicit $1/\omega$ factor as given will be useful below, however.

For the two independent components of the spinor, we have the infalling solutions 
\bea\label{infalling}
\Phi_\alpha^{(0,\ell)} &=&
\begin{pmatrix}
\Phi_{\alpha 1}^{(0,\ell)} \\
\Phi_{\alpha 2}^{(0,\ell)}
\end{pmatrix} \non
&=&\begin{pmatrix}
c_{(\alpha,\text{in})} x^{-1/2}\bigg[-W_{\frac{1}{2}+i\frac{q\mu_0}{6},\nu_{k_\ell}}(-2i\frac{x}{6})+\frac{i(-1)^\alpha k_\ell}{\sqrt{6}}
W_{-\frac{1}{2}+i\frac{q\mu_0}{6},\nu_{k_\ell}}(-2i\frac{x}{6})\bigg]\\
c_{(\alpha,\text{in})} i x^{-1/2}\bigg[W_{\frac{1}{2}+i\frac{q\mu_0}{6},\nu_{k_\ell}}(-2i\frac{x}{6})+\frac{i(-1)^\alpha k_\ell}{\sqrt{6}}
W_{-\frac{1}{2}+i\frac{q\mu_0}{6},\nu_{k_\ell}}(-2i\frac{x}{6})\bigg]
\end{pmatrix}
\eea
and outgoing solutions:
\bea\label{outgoing}
\begin{pmatrix}
\Phi_{\alpha 1}^{(0,\ell)} \\
\Phi_{\alpha 2}^{(0,\ell)}
\end{pmatrix}
=\begin{pmatrix}
c_{(\alpha,\text{out})} x^{-1/2}\bigg[-W_{\frac{1}{2}-i\frac{q\mu_0}{6},\nu_{k_\ell}}(2i\frac{x}{6})-\frac{i (-1)^\alpha k_\ell}{\sqrt{6}}
W_{-\frac{1}{2}-i\frac{q\mu_0}{6},\nu_{k_\ell}}(2i\frac{x}{6})\bigg]\\
c_{(\alpha,\text{out})} i x^{-1/2}\bigg[-W_{\frac{1}{2}-i\frac{q\mu_0}{6},\nu_{k_\ell}}(2i\frac{x}{6})+\frac{i (-1)^\alpha k_\ell}{\sqrt{6}}
W_{-\frac{1}{2}-i\frac{q\mu_0}{6},\nu_{k_\ell}}(2i\frac{x}{6})\bigg]
\end{pmatrix}
\eea
where $\nu_{k_\ell}=\sqrt{\frac{k_\ell^2}{6}-q^2(\frac{\mu_0}{6})^2}$ and $c_{(\alpha,\text{in})},c_{(\alpha,\text{out})}$ are arbitrary constants.

In the coordinate $x$ the horizon is located at $x\to \infty$ and the matching region is at $x\to 0.$
 Imposing the infalling boundary condition for $\Phi_{\alpha}^{(0,\ell)}$ near the horizon, in the matching region we find the following analytic solution for
 $\Phi_{\alpha}^{(0,\ell)}$:
\be\label{matchingphi0} \Phi_{\alpha}^{(0,\ell)}=v_{-\alpha}^{(0,\ell)}\big(\frac{1}{r-1}\big)^{-\nu_{k_\ell}}+v_{+\alpha}^{(0,\ell)} \mathcal{G}_{\alpha\text{IR}}^{(0, \ell)}(\omega)\big(\frac{1}{r-1}\big)^{\nu_{k_\ell}}+\ldots\ee
where we have chosen the normalizalition of the leading coefficient to equal\be v_{\pm\alpha}^{(0,\ell)}=\tilde{c}_0\begin{pmatrix}
\pm \nu_{k_\ell}\\
\frac{q\mu_0}{6}-(-1)^\alpha \frac{k_\ell}{\sqrt{6}}~
\end{pmatrix}.\ee
This determines the ``pure AdS IR Green's function'' to be \be {\mathcal{G}}_{\alpha \text{IR}}^{(0, \ell)}(\omega)=
e^{-i\pi \nu_{k_\ell}}\frac{\Gamma(-2\nu_{k_\ell})\Gamma(1+\nu_{k_\ell}-\frac{i\mu_0 q}{6})((-1)^\alpha i\sqrt{6}k_\ell-6\nu_{k_\ell}-i\mu_0 q)}{\Gamma(2\nu_{k_\ell})\Gamma(1-\nu_{k_\ell}-\frac{i\mu_0 q}{6})((-1)^\alpha i\sqrt{6}k_\ell+6\nu_{k_\ell}-i\mu_0 q)}\big(\frac{\omega}{3}\big)^{2\nu_{k_\ell}}.\ee
It will be useful to introduce a shorthand for the prefactor ${\mathcal{G}}_{\alpha\text{IR}}^{(0,\ell)}(\ome)=\tilde {\mathcal{G}}_{\alpha\text{IR}}^{\ell}\omega^{2\nu_{k_\ell}}$.

%%%%%%%%%%%%%%%%%%%%%%%%%%%%%%%%%%%%%%
\subsubsection{Near region solutions for $\Phi_{\alpha}^{(1,\ell)}$}
%%%%%%%%%%%%%%%%%%%%%%%%%%%%%%%%%%%%%%

We now consider the near region solutions for $\Phi_{\alpha}^{(1,\ell)}$.  The equation of motion for $\Phi_{\alpha}^{(1,\ell)}$ in the near region reduces to
\bea
&&\partial_r\Phi_\alpha^{(1,\ell)}-\frac{i}{6(r-1)^2}(\omega+q\mu_0(r-1))\sigma^2\Phi_\alpha^{(1,\ell)}
-(-1)^\alpha\frac{1}{\sqrt{6}(r-1)}k_\ell\sigma^1\Phi_{\alpha}^{(1,\ell)}\nonumber\\
&&~~~~~~~~~-iq\sigma^2\frac{1}{6(r-1)}\sum_{\beta;\ell'=\ell\pm 1}M_{\alpha\beta,\ell\ell'}\Phi_\beta^{(0,\ell')}=0.
\eea
% The components of the spinors are
% \be
% \Phi_\alpha^{(0,\ell)}=
% \begin{pmatrix}
% \Phi_{\alpha 1}^{(0,\ell)} \\
% \Phi_{\alpha 2}^{(0,\ell)}
% \end{pmatrix},~~~\Phi_\alpha^{(1,\ell)}=
% \begin{pmatrix}
% \Phi_{\alpha 1}^{(1,\ell)} \\
% \Phi_{\alpha 2}^{(1,\ell)}
% \end{pmatrix} \ee
%Eliminating one of the spinor components $\Phi_\alpha^{(1,\ell)}=
% \begin{pmatrix}
 %\Phi_{\alpha 1}^{(1,\ell)} \\
 %\Phi_{\alpha 2}^{(1,\ell)} \end{pmatrix}$
%we can get two independent second order equations for each of the components separately
%\be\label{peteom1}
%\pa_r^2(\Phi_{\alpha j}^{(1,\ell)})-\frac{h'_{\alpha j}}{h_{\alpha j}}\pa_r(\Phi_{\alpha j}^{(1,\ell)})+\frac{h_{\alpha 1}h_{\alpha 2}}{\ome^2}\Phi_{\alpha j}^{(1,\ell)}+\frac{X_{\alpha j}^%{\ell}}{\ome^2}=0, ~~~(j=1,2)\ee
%where
%\bea
%&&h_{\alpha j}=-\omega\bigg(\frac{\omega}{6(r-1)^2}+\frac{q\mu_0}{6(r-1)}+\frac{(-1)^{\alpha+j+1} k_\ell}{\sqrt{6}(r-1)}\bigg),\nonumber\\
%&& X_{\alpha j}^{\ell}=
%-\frac{q\ome h_{\alpha j}}{6(r-1)}\sum_{\beta;\ell'=\ell\pm 1}M_{\alpha\beta,\ell\ell'}\Phi_{\beta j}^{(0,\ell')} + (-1)^j \sum_{\beta;\ell'=\ell\pm 1}\omega h_{\alpha j}\partial_r\bigg[\frac%{q}{6(r-1)h_{\alpha 1}}M_{\alpha\beta,\ell\ell'}\Phi_{\beta l}^{(0,\ell')}
%\bigg]  \nonumber\\
%\eea
%with $l=3-j$. This time we have inhomogeneous second order equations, in which the zeroth order terms appear as source terms.

Similar to the $\Phi^{(0,\ell)}_{\alp}$ case, we perform a coordinate transformation $x=\frac{\omega}{r-1}$ and obtain 
%\footnote{When considering the $\omega$ corrections to the near horizon region solutions, we have the similar equation. But %those corrections are not as important as $\epsilon$ corrections, even not as the zeroth order solution
%discussion in this paper.}}
\be\label{peteom2}
\partial_x^2\Phi_{\alpha j}^{(1,\ell)}+(\frac{2}{x}-\frac{\partial_x h_{\alpha j}}{h_{\alpha j}})\partial_x \Phi_{\alpha j}^{(1,\ell)}
+\frac{h_{\alpha 1}h_{\alpha 2}}{x^4}\Phi_{\alpha j}^{(1,\ell)}+ \frac{1}{x^4}X_{\alpha j}^{\ell}=0,
\ee
with $h_{\alpha j}$ as before (see eq. \eqref{hnew})
% \be h_{\alpha j}=-\frac{x}{\omega}(\frac{x}{6}+\frac{q\mu_0}{6}+(-1)^{\alpha+j+1}\frac{k_\ell}{\sqrt{6}})\ee
%and
%\bea \label{xsource} X_{\alpha j}^{\ell}&=& \sum_{\beta; \ell'=\ell\pm 1} X_{(\alpha \ell,\beta\ell')j} \nonumber \\
%X_{(\alpha \ell,\beta\ell')j} &\equiv&
%-\frac{qh_{\alpha j}x}{6}M_{\alpha\beta,\ell\ell'}\Phi_{\beta j}^{(0,\ell')}
 %-(-1)^jx^2 h_{\alpha j}\partial_x\bigg[\frac{qx}{6 h_{\alpha j}}M_{\alpha\beta, \ell\ell'}\Phi_{\beta l}^{(0,\ell')}\bigg]
%\eea
%with $\Phi_{\alpha j}^{(0,\ell)}$ being the zeroth order solution.
and
\bea \label{xsource} X_{\alpha j}^{\ell}&=& \sum_{\beta; \ell'=\ell\pm 1} {X_{(\alpha j,\beta \ell')}^{\ell}}, \nonumber \\
X_{(\alpha j,\beta \ell')}^{\ell} &\equiv&
-\frac{qh_{\alpha j}x}{6}M_{\alpha\beta,\ell\ell'}\Phi_{\beta j}^{(0,\ell')}
 -(-1)^jx^2 h_{\alpha j}\partial_x\bigg[\frac{qx}{6 h_{\alpha j}}M_{\alpha\beta, \ell\ell'}\Phi_{\beta l}^{(0,\ell')}\bigg]
\eea
with $\Phi_{\alpha j}^{(0,\ell)}$ being the zeroth order solution and $l=3-j$.%\commently{convention changed here to match the following calculations.}

We know that the solutions of inhomogeneous equations can be derived
from the solutions of the corresponding homogeneous equations. Let us
denote the two linear independent solutions of the $\alpha\ell,j$-th
homogeneous equation as ${\eta_1}_{\alpha j}^{\ell}$ and
${\eta_2}_{\alpha j}^{\ell}$. Specifically we shall choose
${\eta_1}_{\alpha j}^{\ell} = \Phi^{(0,\ell), \text{in}}_{\alpha}$ to be the
infalling solution and ${\eta_2}_{\alpha j}^{\ell}= \Phi^{(0,\ell),
 \text{out}}_{\alpha}$ to be the outgoing solution given in eqns
\eqref{infalling} and \eqref{outgoing}. Then the special solution
${\eta_s}_{\alpha j}^\ell$ to the inhomogeneous equations
(\ref{peteom2}) is obtained using the standard Green's function
\be\label{gfinhom}
G^{\ell}_{\alp j}(x,x')= \frac{{\eta_1}_{\alpha j}^{\ell}(x){\eta_2}_{\alp
    j}^\ell(x')\th(x-x') + {\eta_1}_{\alpha j}^{\ell}(x'){\eta_2}_{\alp
    j}^\ell(x)\th(x'-x) }{W(x)}
\ee
where $W(x)$ is the Wronskian of the system $W(x)={\eta_1}_{\alpha
  j}^{\ell}\partial_x{\eta_2}_{\alpha j}^{\ell}-{\eta_2}_{\alpha
  j}^{\ell}\partial_x{\eta_1}_{\alpha j}^{\ell}$; it is proportional to $W(x)=
g_0 \frac{ h_{\alpha j}}{x^2}$ with $g_0$ a constant independent of $x$.
We find (recall that $x \in [0,\infty)$)
\bea
\label{specialsolution-can}
{\eta_{s,\mathrm{canonical}}}_{(\alpha j,\beta \ell')}^{(1,\ell)}(x)&=& {\eta_1}_{\alpha j}^{\ell}(x)\int_{0}^xdx'\frac{ {\eta_2}_{\alpha j}^{\ell}
X_{(\alpha j,\beta \ell')}^{\ell}}{{x'}^4 W(x')}+{\eta_2}_{\alpha j}^{\ell}(x)\int^{\infty}_xdx'\frac{
{\eta_1}_{\alpha j}^{\ell} X_{(\alpha j,\beta \ell')}^{\ell}}{{x'}^4
W(x')}. % \nonumber\\
% &=& {\eta_1}_{\alpha j}^{\ell}(x)\int_{0}^xdx'\frac{ {\eta_2}_{\alpha j}^{\ell}
% X_{(\alpha j,\beta \ell')}^{\ell}}{{x'}^4 W(x')}-{\eta_2}_{\alpha
% j}^{\ell}(x)\int^{x}_0 dx'\frac{
% {\eta_1}_{\alpha j}^{\ell} X_{(\alpha j,\beta \ell')}^{\ell}}{{x'}^4
% W(x')} + c_1 {\eta_1}_{\alp j}^{\ell}(x)
\eea
%\commently{Note in (4.19-4.20), the source term is $X=\sum...$, while in (4.22), we only pick one term in the sum to be the sourse term. This is just because of the boundary condition we choose: when we pick $\alpha\ell$-th b.c., there is only one non-trivial 0th order field.  Don't we need to point it out here?}
%\comment{Of course. In the future, just do so straight away. You know the details far better than I. The ``golden rule'' in collaborative writing is always to suggest specific text. ``open-ended'' comments that someone else than has to translate into text is very inefficient.}
 %[KS I DO NOT UNDERSTAND. THE ABOVE IS THE CANONICAL SOLUTION. THIS IS THE ANSWER BY DEFINITION. THEN YOU CAN ADD HOMOGENEOUS TERMS ETC]}
Note that we have restricted the solution to only one of the source terms in (\ref{xsource}). This is a consequence of the boundary conditions we choose: for a specific $\alpha\ell$-th b.c., only one zeroth order field is non-vanishing.

The special solution for a second order inhomogeneous equation is not
unique, however. Different special solutions can differ by additional
homogeneous contributions $c_1 \eta_1 +c_2 \eta_2$, where $c_1$ and $c_2$ are
arbitrary constants. The coefficients in front of the outgoing homogeneous solutions must be fixed by the boundary
conditions, while the coefficients in front of the ingoing homogeneous solutions will not affect the final result. Note that this degree of freedom can be absorded by
changing the fixed limits of the integrations in
(\ref{specialsolution-can}): special solutions obtained with different
limits of integrations %are 
differ precisely by homogenous terms.

 Here we are instructed to choose the infalling boundary condition % for this special solution
 at the horizon, $x=\infty$. Since ${\eta_1}_{\alp j}^\ell$ is the homogeneous
 infalling solution, it is easily seen that this fixes the upper
 limit of the second integration in \eqref{specialsolution-can} term
 to be $\infty$. 
For convenience we choose the {\em undetermined} limit of the first integration
term to also be located at the horizon. This change
simply means that the special solution we shall work with is
\bea
\label{specialsolution}
{\eta_s}_{(\alpha j,\beta \ell')}^{(1,\ell)}(x)&=& -{\eta_1}_{\alpha j}^{\ell}(x)\int^{\infty}_xdx'\frac{ {\eta_2}_{\alpha j}^{\ell}
X_{(\alpha j,\beta \ell')}^{\ell}}{{x'}^4 W(x')}+{\eta_2}_{\alpha j}^{\ell}(x)\int^{\infty}_xdx'\frac{
{\eta_1}_{\alpha j}^{\ell} X_{(\alpha j,\beta \ell')}^{\ell}}{{x'}^4
W(x')}. % \nonumber\\
% &=& {\eta_1}_{\alpha j}^{\ell}(x)\int_{0}^xdx'\frac{ {\eta_2}_{\alpha j}^{\ell}
% X_{(\alpha j,\beta \ell')}^{\ell}}{{x'}^4 W(x')}-{\eta_2}_{\alpha
% j}^{\ell}(x)\int^{x}_0 dx'\frac{
% {\eta_1}_{\alpha j}^{\ell} X_{(\alpha j,\beta \ell')}^{\ell}}{{x'}^4
% W(x')} + c_1 {\eta_1}_{\alp j}^{\ell}(x)
\eea

Thus the most general solution with the infalling boundary condition in the near region is
\be
\label{freedom}
\Phi_{\alpha j}^{(1,\ell)}(x)= {\eta_s}_{\alpha
  j}^{(1,\ell)}(x)+ c_4 {\eta_1}_{\alpha j}^{(\ell)}(x),
\ee where $c_4$ is an arbitrary constant. The choice of $c_4$ just corresponds to changing the {\em normalization} of the infalling boundary conditions of the homogeneous solution at first order in $\eps$.  Since the final Green's functions (or spectral functions) do not depend on the normalization, we may freely set $c_4=0$.
% % that we mentioned in the last subsection 
% to $1+\epsilon c_3$ infalling
% solutions, where $c_3$ is some constant of order unity. We will keep the
% limits of integration fixed and account for this freedom to add the
% infalling homogeneous solutions to our first order special solutions
% by an explicit extra term.

Now we need to see how this infalling special solution behaves at the
boundary of the near region $x \ll 1$ in order to match it to the far
region. To do so, we split the integration into two parts
%\comment{We substitute value of Wronskian here, check. Also minus sign}
\bea\label{etas1alpha}
{\eta_s}_{(\alpha j,\beta \ell')}^{(1,\ell)}(x)&=&-{\eta_1}_{\alpha j}^{\ell}\int^{\infty}_\varepsilon dx'\frac{{\eta_2}_{\alpha j}^{\ell}
X_{(\alpha j,\beta \ell')}^{\ell}}{{x'}^4 W(x')}+{\eta_2}_{\alpha j}^{\ell}\int^{\infty}_\varepsilon dx'\frac{
{\eta_2}_{\alpha j}^{\ell} X_{(\alpha j,\beta \ell')}^{\ell}}{{x'}^4 W(x')}\nonumber\\
&&-{\eta_1}_{\alpha j}^{\ell}\int^\varepsilon_x dx'\frac{ {\eta_2}_{\alpha j}^{\ell}
X_{(\alpha j,\beta \ell')}^{\ell}}{{x'}^4 W(x')}+{\eta_2}_{\alpha j}^{\ell}\int^\varepsilon_x dx'\frac{
{\eta_1}_{\alpha j}^{\ell} X_{(\alpha j,\beta \ell')}^{\ell}}{{x'}^4 W(x')},
\eea
where $\varepsilon$ is a very small but nonzero constant parameter. Because $\varepsilon$ is a constant, $\int^{\infty}_\varepsilon dx\frac{{\eta_{i}}_{\alpha j}^{\ell}
X_{(\alpha j,\beta \ell')}^{\ell}}{x^4 W(x)},~i=1,2$ are also
constants. Thus the $x$-dependence of first two terms in (\ref{etas1alpha}) % ${\eta_{i}}_{\alpha j}^{\ell}(x)\int_{\infty}^\varepsilon dz\frac{{\eta_{1(2)}}_{\alpha j}^{\ell}
% X_{(\alpha j,\beta \ell')}^{\ell}}{z^2 h_{\alpha j}}$ is just
is direcly determined by the $x$-dependence of the homogeneous
solutions ${\eta_{i}}_{\alpha j}^{\ell}(x)$. 

In the second part, if both $x$ and $\varepsilon$ are very small, we
can replace the integrands by their asymptotic values. For $\omega \ll
r-1$ these asymptotic values are (see \eqref{matchingphi0})
\begin{eqnarray}
  \label{eq:1}
  \eta_{1\alpha j}\sim \ome^{\nu_{k_\ell}}
  \left(x^{-\nu_{k_\ell}} +  \tilde{\cG}_{IR} x^{\nu_{k_\ell}}\right), ~~~
\eta_{2\alpha j}\sim \ome^{\nu_{k_\ell}}
\left(x^{-\nu_{k_\ell}} +
    \tilde{\cG}_{IR}^{\dagger} x^{\nu_{k_\ell}}\right).
\nonumber~~~
% X^{j}_{\alp \ell, \beta \ell'} \sim x h_{\alp j} {\eta_{1}}_{\beta j}(x)
%, ~~~ \eta_{2\alpha j}\sim x^{\nu_{k_\ell}},~~\text{when}~x\to 0; ~~~~\text {so%}~~ 
%c_{\alpha j}^\ell\simeq \frac{2 \nu_{k_\ell}}{\frac{q\mu_0}{6}+(-1)^{\alpha+j+1}\frac{k_\ell}{\sqrt{6}}}.
\end{eqnarray}
%\comment{CHECK}
%\comment{CHECK ALSO DEFINITION OF $X$}\commently{How did you choose $\eta_2$?}\comment{$\eta_2$ is complex conjugate of $\eta_1$, right?}% \commently{To get the specific solution, we only need that $ \eta_{1\alpha j},   \eta_{2\alpha j}$ are two linear independant solutions of the HOMOGENEOUS eqn of eqn. 4.19. We can choose $ \eta_{1\alpha j}$ as infalling zeroth order solution and $ \eta_{2\alpha j}$ as another solution, e.g. outgoing solution. }
The convenient way to substitute this into \eqref{etas1alpha} is to 
note that our choice of the special solution \eqref{specialsolution}
has been made such that under a linear transformation to a different
basis of the homogeneous equations
\be \eta_{1\alpha j}=a_{1\alpha j}\tilde\eta_{1\alpha j}+a_{2\alpha j}\tilde\eta_{2\alpha j},~~~\eta_{2\alpha j}=b_{1\alpha j}\tilde\eta_{1\alpha j}+b_{2\alpha j}\tilde\eta_{2\alpha j},\ee
the special solution is invariant% \comment{IF $a^2-b^2=1$ RIGHT?}\commently{No. It is true for arbitrary constants a\&b. }\comment{You are right}
. % Thus once we fix the lower limit of the integration, we can freely choose $\eta$'s to be whatever we want. Here for simplicity of the computations near $x \to 0$, we choose $\eta_{1\alpha j}$ and $\eta_{2\alpha j}$ to be the solutions that
Choosing %${{\tilde{\eta}}_{i}}_{\alp j}$ 
$\tilde{\eta_{1}}_{\alpha j}$
to be the two independent scaling solutions
\be\label{etahere} 
\tilde{\eta_{1}}_{\alpha j}\sim x^{-\nu_{k_\ell}}, ~~~ \tilde{\eta_{2}}_{\alpha j}\sim x^{\nu_{k_\ell}},~~\text{when}~x\to 0,
\ee
% \comment{ I DO NOT UNDERSTAND THIS SECTION:
% Remember at the section about boundary conditions, we have said that we impose the boundary condition of ``1" infalling solution for the corresponding field. Here we can see that though the solutions (\ref{infalling}) do not depend on $\omega$, they are not the most convenient choices because at $x\to 0$, the solutions approach $ x^{-\nu}+\tilde G_{IR} x^\nu$ and $x^{-\nu}$ would lead to divergences because $\omega\to 0$. Thus here we define ``1" infalling solution to be the solution which goes to $\omega^\nu (x^{-\nu}+\tilde G_{IR} x^\nu)$, i.e. $\omega^\nu$ times the W function. In this way, the result on the boundary would be finite.
% }
it is now straightforward to extract the leading $\ome$-dependence of
$\Phi^{(1,\ell)}$. All dependence on the frequency $\ome$ is contained
in $x = \frac{\ome}{r-1}$ plus the explicit factors
$\omega^{2\nu_{k_\ell}}$ in $\tilde{\eta}_{i\alp j}^{(0,\ell)}$. % the solutions $\eta_1$ and $\eta_2$ do not depend on $\omega$ in the $x$ coordinate. Then we can see that the dependance of the term $\int_{\infty}^\varepsilon$ on $\omega$ is $\omega^{\nu_{k_{\ell'}}+\nu_{k_\ell}}$ and
% $\omega^{\nu_{k_{\ell'}}-\nu_{k_\ell}}$. Thus we mainly need to
% focus on the term $\int_\varepsilon^x$ with $x\to 0$. From
% (\ref{etahere}) we can obtain the result for this part in
% (\ref{etas1alpha}). After some tedious calculation, one can get
One readily obtains 
(when $x\to 0$)
\bea
\label{x0} {\eta_s}_{(\alpha j,\beta \ell')}^{(1,\ell)}\simeq &&x^{\nu_{k_\ell}}\omega^{\nu_{k_{\ell'}}}{n_1}_{\alpha j\beta \ell'}^\ell
+x^{-\nu_{k_\ell}}\omega^{\nu_{k_{\ell'}}}{n_2}_{\alpha j\beta \ell'}^\ell\nonumber\\
&&+x^{\nu_{k_{\ell'}}}\omega^{\nu_{k_{\ell'}}}{n_3}_{\alpha j\beta \ell'}^\ell
+x^{-\nu_{k_{\ell'}}}\omega^{\nu_{k_{\ell'}}}{n_4}_{\alpha j\beta
  \ell'}^\ell +\ldots
\eea
%\commently{Here $\cO(x)$ should mean $x^{-\nu_{k_{\ell'}}}(1+\cO(x))$ etc. Maybe it is not necessary to write it out just as zeroth order case.}
where the coefficients $n_{i} ~(i=1,...,4)$ do not depend on $\omega$ or $x$.
The explicit solutions to ${\eta^{(1,\ell)}_s}_{\alp j,\beta \ell'}$ are given
in the appendix (\ref{co1}).

The next step is more subtle. Note, firstly, that the normalization
with an explicit $\ome^{\nu_{k_\ell}}$ chosen in \eqref{eq:1} is
convenient as the function $\Phi^{(0,\ell)}(r)$ is then manifestly
analytic around $\ome=0$. The special solution \eqref{x0} scales to
leading order as $\omega^{\nu_{k_{\ell'}}+\nu_{k_\ell}}$, $\omega^{\nu_{k_{\ell'}}-\nu_{k_\ell}}$,
 $\omega^{2\nu_{k_{\ell'}}}$ or $\omega^{0}$ respectively. 
 %\comment{THE
   %PREVIOUS VERSION CONTAINED DIFFERENT POWERS: DOUBLE CHECK}. 
 This is not analytic if $\nu_{k_{\ell'}}-\nu_{k_\ell}<0$. However, this
 cannot be corrected by chosing a new normalization: for the special solution \eqref{x0} the normalization
is {\em not} free.  We can, however, eliminate this
 divergence using the freedom to add a homogeneous solution, i.e.
the actual solution 
\be
\Phi^{(1,\ell)}_{\alp j, \beta \ell'}(x) = {\eta_s}^{(1,\ell)}_{\alp j,
  \beta \ell'}(x)+c_4\eta^{(0,\ell)}_{\alp j}(x)
\ee 
with $c_4$ a function of $\omega$ chosen such that $\Phi^{(1,\ell)}_{\alp j, \beta \ell'}(x)$ is analytic in
$\omega$. 
Recall that $\eta^{(0,\ell)}_{\alp j}=(x^{-\nu_{k_\ell}} +
     \tilde{\cG}_{IR} x^{\nu_{k_\ell}})+\ldots$.
    This is therefore the case if 
  \be c_4=-\omega^{\nu_{k_{\ell'}}}{n_2}_{\alpha j\beta \ell'}^\ell.\ee 
%\commently{$c_4$ is given here. Eqn 3.29 is the same as eqn 3.24.}  
  
Though the insistence on analyticity in $\ome$ may appear
arbitrary, it is clear that this is the only sensible solution to be
able to use the matching method with the far region
$\frac{\ome}{r-1}\ll 1$.
Doing so, one finds that
the first-order solution near $x\to 0$ equals 
 \bea \label{constant}{\Phi}_{(\alpha j,\beta \ell')}^{(1,\ell)}&\simeq& x^{\nu_{k_\ell}}\omega^{\nu_{k_{\ell'}}}\tilde{n_1}_{\alpha j\beta \ell'}^\ell
+x^{\nu_{k_{\ell'}}}\omega^{\nu_{k_{\ell'}}}{n_3}_{\alpha j\beta \ell'}^\ell
+x^{-\nu_{k_{\ell'}}}\omega^{\nu_{k_{\ell'}}}{n_4}_{\alpha j\beta
  \ell'}^\ell
\non 
&\simeq& \big(\frac{1}{r-1}\big)^{\nu_{k_\ell}}\omega^{\nu_{k_\ell}+\nu_{k_{\ell'}}}\tilde{n_1}_{\alpha j\beta \ell'}^\ell+\big(\frac{1}{r-1}\big)^{\nu_{k_{\ell'}}}\omega^{2\nu_{k_{\ell'}}}{n_3}_{\alpha j\beta \ell'}^\ell
+\big(\frac{1}{r-1}\big)^{-\nu_{k_{\ell'}}}{n_4}_{\alpha j\beta \ell'}^\ell,\nonumber\\
\eea
where
\be \tilde{n_1}_{\alpha j\beta \ell'}^\ell=\bigg({n_1}_{\alpha j\beta \ell'}^\ell
-(-1)^j\tilde{\mathcal{G}}_{\text{IR}\alpha}^{(0,\ell)} {n_2}_{\alpha
  j\beta \ell'}^\ell\bigg)\ee 
and in the second line we have substituted $x=\frac{\ome}{r-1}$.

The solution above is for general momenta, but it is not valid in the
degenerate case when $\nu_{k_\ell}=\nu_{k_{\ell'}}$. This condition
% $\nu_{k_\ell}=\nu_{k_{\ell'}}$
precisely gives the two points on the
boundary of the first Brillouin zone (see Fig. \ref{cartoonnu}). Thus we encounter here the
similar situation to what happens in the standard band structure in
condensed matter physics. At the first Brillouin boundary,
$k_x=\pm \frac{K}{2}
% =\pm \frac{2\pi}{2a}
$,% \comment{CHECK}
the modes with $\ell=0$ are degenerate
with modes $\ell=\mp 1$. We will discuss the
physics at this special boundary in more detail in the subsection \ref{sec:near-regi-solut} after we construct the second order non-degenerate near horizon solution.

%%%%%%%%%%%%%%%%%%%%%%%%%%%%%%%%%%%%%%
\subsubsection{Near region solutions for $\Phi_{\alpha}^{(2,\ell)}$}\label{3213}
%%%%%%%%%%%%%%%%%%%%%%%%%%%%%%%%%%%%%%

Finally, as we argued above eq. \eqref{matrices} we also need the
second order solution $\Phi_{\alpha}^{(2,\ell)}$.
Its equation of motion in the near horizon region is
\bea
&&\partial_r\Phi_\alpha^{(2,\ell)}-\frac{i}{6(r-1)^2}(\omega+q\mu_0(r-1))\sigma^2\Phi_\alpha^{(2,\ell)}
-(-1)^\alpha\frac{1}{\sqrt{6}(r-1)}k_\ell\sigma^1\Phi_{\alpha}^{(2,\ell)}\nonumber\\
&&~~~~~~~-iq\sigma^2\frac{1}{6(r-1)}\sum_{\beta;\ell'=\ell\pm 1}M_{\alpha\beta,\ell\ell'}\Phi_\beta^{(1,\ell')}=0. 
\eea
Transforming to the coordinate $x=\frac{\ome}{r-1}$ again, we find
\be
\partial_x^2\Phi_{\alpha j}^{(2,\ell)}+(\frac{2}{x}-\frac{\partial_x h_{\alpha j}}{h_{\alpha j}})\partial_x
\Phi_{\alpha j}^{(2,\ell)}
+\frac{h_{\alpha 1}h_{\alpha 2}}{x^4}\Phi_{\alpha j}^{(2,\ell)}+ \frac{1}{x^4}
Y_{\alpha j}^{\ell}=0,
\ee
with $h_{\alpha j}$ as before in eq. \eqref{hnew}. 
%\be h_{\alpha j}=-\frac{x}{\omega}(\frac{x}{6}+\frac{q\mu_0}{6}+(-1)^{\alpha+j+1}\frac{k_\ell}{\sqrt{6}}),\ee
The inhomogenous term now equals
%\be Y_{\alpha j}^{\ell}=
%-\frac{qh_{\alpha j}x}{6}\sum_{\beta; \ell'=\ell\pm 1}M_{\alpha\beta,\ell\ell'}\Phi_{\beta j}^{(1,\ell')}
 %-(-1)^j x^2\sum_{\beta;\ell'=\ell\pm 1}\partial_x\bigg[\frac{qx}{6 h_{\alpha j}}
% M_{\alpha\beta,\ell\ell'}\Phi_{\beta j}^{(1,\ell')}\bigg] h_{\alpha j}\ee
% \comment{SHOULDN'T THIS BE (Subscript changed in last $\Phi$)}
\be Y_{\alpha j}^{\ell}=
-\frac{qh_{\alpha j}x}{6}\sum_{\beta; \ell'=\ell\pm 1}M_{\alpha\beta,\ell\ell'}\Phi_{\beta j}^{(1,\ell')}
 -(-1)^j x^2\sum_{\beta;\ell'=\ell\pm 1}\partial_x\bigg[\frac{qx}{6 h_{\alpha j}}
 M_{\alpha\beta,\ell\ell'}\Phi_{\beta l}^{(1,\ell')}\bigg] h_{\alpha j}\ee
with $l=3-j$. In contrast to the first order case, both components in the $\ell\pm
1$ sectors will now contribute.% \comment{WHAT DOES THIS MEAN}\commently{It is realted to the statement in the boundary condition sector: We pick up only one zeroth order field $\Phi_{\alpha}^{(0,\ell)}$; at the first order $\Phi_{\beta}^{(1,\ell\pm 1)}$ is nonzero. When we calculate the specific solution for  $\Phi_{\beta}^{(1,\ell\pm 1)}$, we only need to take into account the contribution of  $\Phi_{\alpha}^{(0,\ell)}$ although there are two source terms in eqn. 3.19. But When we calculate the specific solution for  2nd order solution $\Phi_{\beta}^{(2,\ell)}$, we should take into account the contribution of  $\Phi_{\alpha}^{(1,\ell\pm 1)}$ to the source terms in eqn. 3.34.}

Through the Green's function (\ref{gfinhom}), the special solution can be written as:
\bea
\label{specialsolution2}
{\eta_s}_{(\alpha j)}^{(2,\ell)}(x)&=& -{\eta_1}_{\alpha j}^{\ell}(x)\int^{\infty}_xdx'\frac{ {\eta_2}_{\alpha j}^{\ell}
Y_{\alpha j}^{\ell}}{{x'}^4 W(x')}+{\eta_2}_{\alpha j}^{\ell}(x)\int^{\infty}_xdx'\frac{
{\eta_1}_{\alpha j}^{\ell} Y_{\alpha j}^{\ell}}{{x'}^4
W(x')}. 
\eea
%\be
%{\eta_s}_{(\alpha j)}^{(2,\ell)}=-\tilde \eta_{2\alpha j}^{\ell}\int^{\infty}_xdz
%\frac{ \tilde \eta_{1\alpha j}^{\ell} Y_{\alpha j}^{\ell}}{c_{\alpha j}^\ell  z^2 h_{\alpha j}}
%+\tilde \eta_{1\alpha j}^{\ell}\int^{\infty}_xdz
%\frac{ \tilde \eta_{2\alpha j}^{\ell} Y_{\alpha j}^{\ell}}{c_{\alpha j}^\ell z^2 h_{\alpha j}},
%\ee
%\comment{COMPARED TO THE PREVIOUS VERSION I SWITCHED THE LIMITS ON THE INTEGRAL: DOES THIS GIVE A MINUS SIGN %PROBLEM?}\commently{The definition of Wronskian $W$ is different from the previous version. It should take the same form as 3.22.}
%where $\Phi_{\beta j}^{(1,\ell)}={\eta_s}^{(1,\ell')}_{(\beta j,\alpha\ell)}$.
It is an infalling solution
as long as the upper bound of the integration in the second term is $\infty$.
Using the method similar to last subsection, one can obtain the
behavior of the special solution near $x=0$:
\bea\label{eta2special} {\eta_s}_{(\alpha j)}^{(2,\ell)}\simeq &&x^{\nu_{k_\ell}}\omega^{\nu_{k_\ell}}{s_1}_{\alpha j}^\ell
+x^{-\nu_{k_\ell}}\omega^{\nu_{k_\ell}}{s_2}_{\alpha j}^\ell+\sum_{\ell'=\ell\pm 1}\big(
x^{\nu_{k_{\ell'}}}\omega^{\nu_{k_\ell}}{s_3}_{\alpha j\ell'}^\ell\big)
\nonumber\\
&&
+x^{\nu_{k_\ell}}\omega^{\nu_{k_\ell}}\ln(\frac{1}{r-1}) {s_4}_{\alpha j}^\ell+x^{-\nu_{k_\ell}}\omega^{\nu_{k_\ell}}\ln(\frac{1}{r-1}) {s_5}_{\alpha j}^\ell
+x^{\nu_{k_\ell}}\omega^{\nu_{k_\ell}}(\ln\omega) {s_4}_{\alpha
  j}^\ell\nonumber\\
&&+x^{-\nu_{k_\ell}}\omega^{\nu_{k_\ell}}(\ln\omega)
{s_5}_{\alpha j}^\ell+\ldots
\eea
%\commently{$\cO(x)$: $x^\#\to x^{\#}(1+\cO(x))$}
where the coefficients ${s_i}_{\alp j}^{\ell}$ do not depend on $x$ or
$\omega$ and the concrete expressions can be found in
(\ref{secondordercoeff}).
Note again the degeneracy at
$\nu_{k_\ell}=\nu_{k_{\ell'}}$.

Similar to the first order case, the special solution is not regular in the $\omega\to 0$ limit. Making use of the freedom of adding arbitrary infalling solutions% of order $\epsilon^2$
, one can obtain the
regular second order solution:
\bea {\Phi}_{(\alpha j)}^{(2,\ell)}\simeq &&x^{\nu_{k_\ell}}\omega^{\nu_{k_\ell}}\tilde{s_1}_{\alpha j}^\ell
+\sum_{\ell'=\ell\pm 1}\big(
x^{\nu_{k_{\ell'}}}\omega^{\nu_{k_\ell}}{s_3}_{\alpha j\ell'}^\ell\big)
+x^{\nu_{k_\ell}}\omega^{\nu_{k_\ell}}\ln(\frac{1}{r-1}) {s_4}_{\alpha j}^\ell
\nonumber\\
&&+x^{-\nu_{k_\ell}}\omega^{\nu_{k_\ell}}\ln(\frac{1}{r-1}) {s_5}_{\alpha j}^\ell
+x^{\nu_{k_\ell}}\omega^{\nu_{k_\ell}}(\ln\omega) {\tilde {s_4}}_{\alpha
  j}^\ell +\ldots.
\eea
Substituting $x=\frac{\ome}{r-1}$ to display the explicit leading dependence
on $\ome$ we find 
\bea
\label{matchingphi2b} 
{\Phi}_{(\alpha j)}^{(2,\ell)}
\simeq &&\bigg(\frac{1}{r-1}\bigg)^{\nu_{k_\ell}}\omega^{2\nu_{k_\ell}}\tilde{s_1}_{\alpha j}^\ell+\sum_{\ell'=\ell\pm 1}\bigg[
\bigg(\frac{1}{r-1}\bigg)^{\nu_{k_{\ell'}}}\omega^{\nu_{k_{\ell'}}+\nu_{k_\ell}}{s_3}_{\alpha j\ell'}^\ell\bigg]
\nonumber\\
&&
+\bigg(\frac{1}{r-1}\bigg)^{\nu_{k_\ell}}\omega^{2\nu_{k_\ell}}\ln(\frac{1}{r-1})
{s_4}_{\alpha
  j}^\ell+\bigg(\frac{1}{r-1}\bigg)^{-\nu_{k_\ell}}\ln(\frac{1}{r-1})
{s_5}_{\alpha j}^\ell \non
&&
+\bigg(\frac{1}{r-1}\bigg)^{\nu_{k_\ell}}\omega^{2\nu_{k_\ell}}(\ln\omega) {\tilde {s_4}}_{\alpha j}^\ell +\ldots
\eea
where
\be
{\tilde{s_1}}_{\alpha j}^\ell={s_1}_{\alpha j}^\ell-(-1)^j{s_2}_{\alpha j}^\ell
\tilde{\mathcal{G}}_{\alpha}^{\ell},~~~
{\tilde{s_4}}_{\alpha j}^\ell={s_4}_{\alpha j}^\ell-(-1)^j{s_5}_{\alpha j}^\ell
\tilde{\mathcal{G}}_{\alpha}^{\ell}.
\ee
%
% The above formula means that at the matching point we have
% \bea\label{matchingphi2} {\eta_s}_{(\alpha j)}^{(2,\ell)}.\nonumber\\
% \eea

%Finally, we can get the behavior of the near region solution near the matching region:
%\be\Phi_{11}^{(1,\ell)}\sim \#_1 x^{\nu_{k_\ell}}+ \#_2 x^{-\nu_{k_\ell}}+\sum_{\ell'=\ell\pm 1}\bigg(
 %\#_3x^{-\nu_{k_{\ell'}}}+ \#_4 x^{\nu_{k_{\ell'}}}+ \#_5 x^{2\nu_{k_\ell}-\nu_{k_{\ell'}}}+ \#_6 x^{2\nu_{k_\ell}
 %+\nu_{k_{\ell'}}}+ \#_7x^{-2\nu_{k_\ell}-\nu_{k_{\ell'}}}+ \#_8
 %x^{-2\nu_{k_\ell}+\nu_{k_{\ell'}}}\bigg)\ee

\subsubsection{Near region solutions for the Degenerate case $\nu_{k_\ell}=\nu_{k_{\ell'}}$}
\label{sec:near-regi-solut}
As pointed out in the previous subsections, a degeneracy occurs
whenever $\nu_{k_\ell}=\nu_{k_{\ell'}}$ where $\ell'=\ell\pm 1$ and
this case must be treated in more detail. This degeneracy happens at
the edges of the Brillioun zones when $k_x=\frac{1}{2a}$, $\ell=0, -1$
and $k_x=-\frac{1}{2a}$,% \comment{CHECK} 
~$\ell=0, 1$ for any fixed $k_y$ (see Fig. \ref{cartoonnu}). % In the following, we will focus on the case
% $k_x=\frac{1}{2a}$, $\ell=0, 1$, i.e. $\nu_{k_0}=\nu_{k_{-1}}$. It
% will be clear that the solution immediately generalizes to other zone edges.

Specifically what goes wrong when $\nu_{k_\ell}=\nu_{k_{\ell'}}$ is
that the coefficients ${n_i}^{\ell}_{\alp j}$ in \eqref{constant} (explicitly given in (\ref{co1})) and
${s_i}^{\ell}_{\alp j}$ in eq. \eqref{matchingphi2b} (explicitly given
in eq.\eqref{secondordercoeff}) become divergent. The solutions we
obtain in this section can be seen as the continuous $\nu\to \nu'$
limit of the near-horizon solutions in previous two
subsections. However, in the next subsection where we address far
region solutions, we will see that at the boundary the degenerate case
is qualitatively quite different.

Repeating the procedure in the previous subsections, for this special
case $\nu_{k_{\ell'}}=\nu_{k_\ell}$, we can obtain the leading $\ome$-behavior of the infalling first order degenerate special solution near the matching point as % (when $\nu_{k_\ell}=\nu_{k_{\ell'}}$)
\bea\label{degetas1} {\eta_s}_{(\alpha j,\beta \ell')}^{(1,\ell)}\simeq &&x^{\nu_{k_\ell}}\omega^{\nu_{k_\ell}}{d_1}_{\alpha j\beta \ell'}^\ell
+x^{-\nu_{k_\ell}}\omega^{\nu_{k_\ell}}{d_2}_{\alpha j\beta \ell'}^\ell
+x^{\nu_{k_\ell}}\omega^{\nu_{k_\ell}}(\ln{\omega}){d_3}_{\alpha j\beta \ell'}^\ell
+x^{-\nu_{k_\ell}}\omega^{\nu_{k_\ell}}(\ln{\omega}){d_4}_{\alpha j\beta \ell'}^\ell
\nonumber\\
&&
+x^{\nu_{k_\ell}}\omega^{\nu_{k_\ell}}{d_3}_{\alpha j\beta \ell'}^\ell\ln{\frac{1}{r-1}}
+x^{-\nu_{k_\ell}}\omega^{\nu_{k_\ell}}{d_4}_{\alpha j\beta \ell'}^\ell\ln{\frac{1}{r-1}}+\ldots
\eea
where the coefficients ${d_i}_{\alpha j\beta \ell'}^\ell$ do not
depend on $\omega$ or $x$ and their expressions can be found in the
appendix, eq. (\ref{degco1}).
Removing the potential divergence in $\ome$ as $\ome \rar 0$, related
to the term $x^{-\nu_{k_\ell}}\omega^{\nu_{k_\ell}}\ln{\omega}$ by
adding a zeroth order infalling solution, we find:% \comment{SHOULDN'T
  % $\ell'=\ell$ BELOW? guess:no}\commently{No. There is $M_{\ell\ell'}$ in the coefficients which is not equal to $M_{\ell\ell}$}
\bea\label{degbnd1} {\Phi_{\text{degen}}}_{(\alpha j,\beta \ell')}^{(1,\ell)}\simeq &&\bigg(\frac{1}{r-1}\bigg)^{\nu_{k_\ell}}\omega^{2\nu_{k_\ell}}\tilde{d_1}_{\alpha j\beta \ell'}^\ell
+2\bigg(\frac{1}{r-1}\bigg)^{\nu_{k_\ell}}\omega^{2\nu_{k_\ell}}(\ln{\omega}) d_{3\alpha j\beta \ell'}^\ell\\
\nonumber
&&
+\bigg(\frac{1}{r-1}\bigg)^{\nu_{k_\ell}}\omega^{2\nu_{k_\ell}}d_{3\alpha j\beta \ell'}^\ell\ln{\frac{1}{r-1}}
+\bigg(\frac{1}{r-1}\bigg)^{-\nu_{k_\ell}}d_{4\alpha j\beta \ell'}^\ell\ln{\frac{1}{r-1}}+ \ldots
\eea
where
\be
{\tilde{d_1}}_{\alpha j\beta \ell'}^\ell={d_1}_{\alpha j\beta \ell'}^\ell-(-1)^j{d_2}_{\alpha j\beta \ell'}
\tilde{\mathcal{G}}_{\alpha}^{\ell}.
\ee
% \comment{HOW IS THIS CONTINUOUS TO non-degen}\commently{1)
The statement that this is the continous limit from the non-degenerate case is that the novel logarithmic term originates in from the combination $\frac{\omega^{\nu+\nu'}-\omega^{2\nu'}}{\nu-\nu'}$. It is precisely this combination that arises when we subsitute in the exact values of the coefficients ${n_i}^{\ell}_{\alpha j \beta \ell'}$.

We similarly obtain the infalling second order special solution near the matching point.
As pointed out in Sec. \ref{3213} , both $\Phi^{(1,\ell\pm
  1)}$ source terms contribute to
second order special solution. Only one of them can at any one time be degenerate in the sense that
$\nu_{k_\ell}=\nu_{k_{\ell'}}$ with $\ell'=\ell+1$ or $\ell'=\ell-1$. %\comment{CHECK}. %  We will focus
% on the first sector (when $\nu_{k_\ell}=\nu_{k_{\ell'}}$):
% \comment{I
%   DO NOT REALLY FOLLOW THE PREVIOUS PARAGRAPH. OR THE STEP BELOW
% Splitting 
% \be{\eta_s}_{(\alpha j,\beta \ell)}^{(1,\ell')}=\omega^{\nu_{k_\ell}}
% F(x)+\omega^{\nu_{k_\ell}} \ln{\omega}G(x)
% \ee
% }\commently{same question as the comment at the end of page 21.}
% \comment{WHICH COMMENT?}
% % the special solutin 
% It
Following the same steps as before, the leading $\ome$-behavior of the infalling second order degenerate solution 
behaves near the matching point as:
\newcommand{\degens}{\mathtt{s}}
\bea\label{degetas2} {\eta_s}_{(\alpha j,\ell')}^{(2,\ell)}\simeq &&x^{\nu_{k_\ell}}\omega^{\nu_{k_\ell}}{\degens_1}_{\alpha j \ell'}^\ell
+x^{-\nu_{k_\ell}}\omega^{\nu_{k_\ell}}{\degens_2}_{\alpha j\ell'}^\ell
+x^{\nu_{k_\ell}}\omega^{\nu_{k_\ell}}{\degens_3}_{\alpha j \ell'}^\ell\ln{\omega}
+x^{-\nu_{k_\ell}}\omega^{\nu_{k_\ell}}{\degens_4}_{\alpha j \ell'}^\ell\ln{\omega}
\nonumber\\
&&
+x^{\nu_{k_\ell}}\omega^{\nu_{k_\ell}}{\degens_5}_{\alpha j\ell'}^\ell\ln{\frac{1}{r-1}}
+x^{-\nu_{k_\ell}}\omega^{\nu_{k_\ell}}{\degens_6}_{\alpha j \ell'}^\ell\ln{\frac{1}{r-1}}
+x^{\nu_{k_\ell}}\omega^{\nu_{k_\ell}}{\degens_7}_{\alpha j \ell'}^\ell\ln\omega\ln{\frac{1}{r-1}}
\nonumber\\
&&
+x^{\nu_{k_\ell}}\omega^{\nu_{k_\ell}}{\degens_8}_{\alpha j\ell'}^\ell(\ln\omega)^2
+x^{-\nu_{k_\ell}}\omega^{\nu_{k_\ell}}{\degens_{9}}_{\alpha j \ell'}^\ell(\ln\omega)^2
\nonumber\\
&&
+x^{\nu_{k_\ell}}\omega^{\nu_{k_\ell}}{\degens_{10}}_{\alpha j \ell'}^\ell(\ln{\frac{1}{r-1}})^2
+x^{-\nu_{k_\ell}}\omega^{\nu_{k_\ell}}{\degens_{11}}_{\alpha j\ell'}^\ell(\ln{\frac{1}{r-1}})^2+\ldots.
\eea
The explicit expressions for the $\omega,x$-independent coefficients
${\degens_{i}}_{\alpha j\ell'}^\ell$  are given in 
(\ref{degco2}).
 The $\omega \rar 0$-regular second order degenerate solution is:
\bea\label{degbnd2} {\Phi_{\text{degen}}}_{(\alpha j,\ell')}^{(2,\ell)}\simeq &&\bigg(\frac{1}{r-1}\bigg)^{\nu_{k_\ell}}\omega^{2\nu_{k_\ell}}\tilde{\degens_1}_{\alpha j \ell'}^\ell
+\bigg(\frac{1}{r-1}\bigg)^{\nu_{k_\ell}}\omega^{2\nu_{k_\ell}}\ln{\omega}{\tilde{\degens_3}}_{\alpha j \ell'}^\ell
+\bigg(\frac{1}{r-1}\bigg)^{\nu_{k_\ell}}\omega^{2\nu_{k_\ell}}{\degens_5}_{\alpha j\ell'}^\ell\ln{\frac{1}{r-1}}
\nonumber\\
&&
+\bigg(\frac{1}{r-1}\bigg)^{-\nu_{k_\ell}}{\degens_6}_{\alpha j \ell'}^\ell\ln{\frac{1}{r-1}}
+\bigg(\frac{1}{r-1}\bigg)^{\nu_{k_\ell}}\omega^{2\nu_{k_\ell}}\ln\omega{\degens_7}_{\alpha j \ell'}^\ell\ln{\frac{1}{r-1}}
\nonumber\\
&&
+\bigg(\frac{1}{r-1}\bigg)^{\nu_{k_\ell}}\omega^{2\nu_{k_\ell}}(\ln\omega)^2{\tilde{\degens_8}}_{\alpha j\ell'}^\ell
+\bigg(\frac{1}{r-1}\bigg)^{\nu_{k_\ell}}\omega^{2\nu_{k_\ell}}{\degens_{10}}_{\alpha j \ell'}^\ell(\ln{\frac{1}{r-1}})^2
\nonumber\\
&&
+\bigg(\frac{1}{r-1}\bigg)^{-\nu_{k_\ell}}{\degens_{11}}_{\alpha j\ell'}^\ell(\ln{\frac{1}{r-1}})^2
\eea
where
\be
{\tilde{\degens_1}}_{\alpha j}^\ell={\degens_1}_{\alpha j}^\ell-(-1)^j{\degens_2}_{\alpha j}^\ell
\tilde{\mathcal{G}}_{\alpha}^{\ell},~~~
{\tilde{\degens_3}}_{\alpha j}^\ell={\degens_3}_{\alpha j}^\ell-(-1)^j{\degens_4}_{\alpha j}^\ell
\tilde{\mathcal{G}}_{\alpha}^{\ell},~~~
{\tilde{\degens_8}}_{\alpha j}^\ell={\degens_8}_{\alpha j}^\ell-(-1)^j{\degens_9}_{\alpha j}^\ell
\tilde{\mathcal{G}}_{\alpha}^{\ell}.
\ee
% The final second order solution is a sum of the degenerate sector and the non-degenerate sector with considering only the $\ell''$ sector source term:
% \be{\eta_s}_{(\alpha j)}^{(2,\ell)}={\eta_s}_{(\alpha j,\ell')}^{(2,\ell)}+{\eta_s}_{(\alpha j,\ell'')}^{(2,\ell)}.
% \ee
% \comment{I DO NOT UNDERSTAND THIS LAST SENTENCE}

%%%%%%%%%%%%%%%%%%%%%%%%%%%%%%%%%%%%%%
\subsubsection{The far region}
%%%%%%%%%%%%%%%%%%%%%%%%%%%%%%%%%%%%%%

From the original equation of motion \eqref{eom2} it is immediately
clear that in the far region, $r-1 \gg \omega$, the terms in the equations of motion
containing $\omega$ are automatically subleading in
$\frac{1}{r-1}$. The $1/(r-1)$ expansion is therefore correlated with
an expansion in the frequency $\omega$. The leading, $\omega$-independent, order equation of motion for $\Phi_\alpha^{(0,\ell)}$ in the far region equals
\be
(\partial_r+m\sqrt{g_{rr}}\sigma^3)\Phi_\alpha^{(0,\ell)}-i\sqrt{\frac{g_{rr}}{g_{tt}}}(q\mu_0(1-\frac{1}{r}))\sigma^2\Phi_\alpha^{(0,\ell)}
-(-1)^\alpha\sqrt{\frac{g_{rr}}{g_{xx}}}k_\ell\sigma^1\Phi_\alpha^{(0,\ell)} =0,
\ee
the leading order far region equation of motion for $\Phi_\alpha^{(1,\ell)}$ equals
\bea\label{farfirst}
&&(\partial_r+m\sqrt{g_{rr}}\sigma^3)\Phi_\alpha^{(1,\ell)}-i\sqrt{\frac{g_{rr}}{g_{tt}}}(q\mu_0(1-\frac{1}{r}))\sigma^2\Phi_\alpha^{(1,\ell)}-(-1)^\alpha\sqrt{\frac{g_{rr}}{g_{xx}}}k_\ell \sigma^1\Phi_\alpha^{(1,\ell)} \nonumber\\
&&~~~~~~~-iq\sqrt{\frac{g_{rr}}{g_{tt}}}\sigma^2(1-\frac{1}{r})\sum_{\beta; \ell'=\ell\pm1}M_{\alpha\beta,\ell\ell'}\Phi_\beta^{(0,\ell')}=0,
\eea
and the leading order far region equation of motion for
$\Phi_\alpha^{(2,\ell)}$ is similar. Recalling that
$\sqrt{g_{rr}}\sim 1/\sqrt{g_{tt}}\sim 1/\sqrt{g_{xx}}\sim 1/r$ we see
that at the boundary of the AdS spacetime $r\to\infty$, the equations
for all orders in the modulation $\eps$ linearize and all
fields $\Phi^{(s)}_{\alp}$ for any order behave as 
% We can solve the EOM order by order perturbatively in $\omega$ \cite{Faulkner:2009wj}. EOM for $\Phi_\alpha^{(0,\ell)}$ in the far region is
%% and near the boundary $r\to\infty$ we have
\be
\Phi_{\alpha}^{(s,\ell)}= \cA_\alpha^{(s,\ell)} r^m  \begin{pmatrix}
0 \\
1
\end{pmatrix}+B_\alpha^{(s,\ell)} r^{-m}
\begin{pmatrix}
1\\
0
\end{pmatrix}+\ldots,~~~~s=0,1,2....
\ee
%\be
%\Phi_{\alpha}^{(1,\ell)}\sim \cA_\alpha^{(1,\ell)} r^m  \begin{pmatrix}
%0 \\
%1
%\end{pmatrix}+b_\alpha^{(1,\ell)} r^{-m}
%\begin{pmatrix}
%1\\
%0
%\end{pmatrix},
%\ee
%\be
%\Phi_{\alpha}^{(2,\ell)}\sim \cA_\alpha^{(2,\ell)} r^m  \begin{pmatrix}
%0 \\
%1
%\end{pmatrix}+b_\alpha^{(2,\ell)} r^{-m}
%\begin{pmatrix}
%1\\
%0
%\end{pmatrix},
%\ee
The coefficients $\cA_{\alpha}^{(s,\ell)}$ 
and $B_{\alpha}^{(s,\ell)}$ are determined by matching these solutions
at the inner boundary of the far region
with the $x\rar 0$ value of the near
horizon infalling solutions obtained in the previous subsections: Eqs
(\ref{matchingphi0}), (\ref{constant}) and (\ref{matchingphi2b}) or (\ref{degbnd1})  and (\ref{degbnd2}). The
equation of the zeroth order field $\Phi^{(0,\ell)}_\alp(r)$ is homogenous and linear, so under the near horizon $\alpha\ell$-th boundary condition the boundary value of the zeroth order solution can be obtained using the matching method to be
\be\label{farzero}
\cA_{\alpha}^{(0,\ell)}=\cA_{\alpha\ell,\alpha\ell}^{(0)-}v_{-\alpha2}^{(0,\ell)}+\mathcal{G}_{\alpha\text{IR}}^{(0,\ell)}
\cA_{\alpha\ell,\alpha\ell}^{(0)+} v_{+\alpha2}^{(0,\ell)},~~~
\cB_{\alpha}^{(0,\ell)}=\cB^{(0)-}_{\alpha\ell,\alpha\ell}v_{-\alpha1}^{(0,\ell)}+\mathcal{G}_{\alpha\text{IR}}^{(0,\ell)}\cB^{(0)+}_{\alpha\ell,\alpha\ell}v_{+\alpha1}^{(0,\ell)}
\ee
where $\cA_{\alp\ell,\alp\ell}^{(0)\pm}$,
$\cB_{\alp\ell,\alp\ell}^{(0)\pm}$ are matrices (at zeroth order proportional
to the identity) determined by the value of the coefficients of $\big(\frac{1}{r-1}\big)^{\pm\nu}$
of the zeroth order near-horizon $\alp\ell$-th solution with near
horizon initial condition at the matching point.
It is in principle straightforward to include the corrections of first order of $\omega$ in the coefficients, by using the next order expansion in the far region $r-1$. This yields $\cA^{(0)\pm}(\ome)=\cA^{(0)\pm}(0)+\omega \pa_{\ome}\cA^{(0)\pm}(0)+...$. This matching procedure makes the dependence of these coefficients on $\omega$ quite clear.

Similar to the zeroth order solution, we can also obtain the
coefficients $\cA^{(i)}_{\alp\ell,\beta\ell'}$ and
$\cB^{(i)}_{\alp\ell,\beta\ell'}$ from matching the first and second
order far region equations of motion with the near-horizon solutions. The difference is two-fold. (1)  Due to the nonlinear behavior of the near-horizon solutions, the
matrices are no-longer proportional to the identity. At first and
higher order in $\eps$ the $\alp\ell$-th
infalling boundary condition can ``source'' a
$\beta\neq\alpha,\ell\neq\ell'$-th far region solution.
%\comment{I DO NOT UNDERSTAND THIS: the first or second order equations are linear but inhomogenous, so we have to be very careful about the source terms for those initial functions at the matching point, and once the source function is found at the matching point, the solution of the zeroth order equation with this source function at the matching point as the initial condition would be the source term in the whole far region. Because the far region equations do not depend on $\omega$ and are still linear, we can still get similar expressions for those boundary coefficients of the first order or second order solutions.}  
(2) The first or second order equations are inhomogenous, so we have to also be careful about matching the inhomogenous  terms at the matching point. The far region equations, however, do not depend on $\omega$ and are still linear. If we are only interested in the $\omega$-behavior, we can therefore still get expressions for the boundary coefficients of the first order or second order solutions that are as concise as the zeroth order one.
%\comment{CHECK} 

%Near the matching region, one can obtain basis of solutions $\phi_\alpha^{(0,\ell)}\sim c_{\pm\alpha}^{(0,\ell)}\big(\frac{1}{r-1}\big)^{\pm\nu_{k_\ell}}$.

This constructs order by order in the modulation $\epsilon$ the
matrices $\cA=\cA^{(0)}+\epsilon\cA^{(1)}+\ldots$ and $\cB=\cB^{(0)}+\epsilon\cB^{(1)}+\ldots$ whose ratio determines the
boundary Green's function. To obtain the Green's function up to the second order, we can truncate the
series for the diagonal elements of the matrices at second order
$\cA_{\alpha\ell, \alpha \ell}=\cA_{\alpha\ell, \alpha
  \ell}^{(0)}+\epsilon^2 {\cA}^{(2)}_{\alpha\ell, \alpha \ell}$,
$\cB_{\alpha\ell,\alpha\ell}=\cB_{\alpha\ell,\alpha\ell}^{(0)}+\eps^2{\cB}^{(2)}_{\alpha\ell,\alpha\ell}$
(we use here the fact that first order contribution to the diagonal terms vanishes.)
and off-diagonal terms to first order. The latter means that only
terms with $\ell'=\ell\pm1$ are nonvanishing and equal
$\cA_{\alp\ell,\alp'\ell\pm 1} = \epsilon \cA^{(1)}_{\alpha\ell, \alpha' \ell\pm 1}$ or $\cB_{\alp\ell,\alp'\ell\pm 1}=\epsilon \cB^{(1)}_{\alpha\ell, \alpha' \ell\pm 1}$.
%Later we will show that the $\tilde{a}_{\alpha\ell, \alpha \ell}$ terms are very important.
%Similar notation is taken in the matrix $\mathcal{B}$. We have infinity copies of such structure where `$\bullet$'  is in the diagonal line of the matrix.
%Thus $\mathcal{B}=G_{R}\mathcal{A}$, or $G_{R}=\mathcal{B}\mathcal{A}^{-1}.$
% Using the fact that the zeroth order term is diagonal and we are only
% interested in the diagonal part of the second order term, plus the
% fact that the first order term is always off-diagonal, we can
% introduce the shorthand $a_{\alpha\ell,\alpha\ell}\equiv
% a^{(0)}_{\alpha\ell,\alpha\ell},
% \tilde{a}_{\alpha\ell,\alpha\ell}=a^{(2)}_{\alpha\ell,\alpha\ell},
% a_{\alpha\ell,\alpha'\ell'}=a^{(1)}_{\alpha\ell,\alpha\ell}$.
After
matching to the near region we can then write out the relevant elements of these matrices as (here $\ell'=\ell\pm 1$)
\bea
\label{eq:k11}
\cA_{\alpha\ell,\alpha\ell}^{(0)}&=&\cA_{\alpha\ell,\alpha\ell}^{(0)-}v_{-\alpha2}^{(0,\ell)}+\tilde{\mathcal{G}}_{\alpha\text{IR}}^{\ell}
\cA_{\alpha\ell,\alpha\ell}^{(0)+}v_{+\alpha2}^{(0,\ell)}\ome^{2\nu_{k_\ell}},\\
\cB_{\alpha\ell,\alpha\ell}^{(0)}&=&\cB_{\alpha\ell,\alpha\ell}^{(0)-}v_{-\alpha1}^{(0,\ell)}+\tilde{\mathcal{G}}_{\alpha\text{IR}}^{\ell}\cB_{\alpha\ell,\alpha\ell}^{(0)+}v_{+\alpha1}^{(0,\ell)}\ome^{2\nu_{k_\ell}},\nonumber\\
\cA_{\alpha\ell,\beta\ell'}^{(1)}&=&\cA_{\alpha\ell,\beta\ell'}^{(1,1)}\tilde{n_1}^{\ell}_{\alpha2\beta\ell'}\omega^{\nu_{k_\ell}+\nu_{k_{\ell'}}}+\cA_{\alpha\ell,\beta\ell'}^{(1,3)}{n_3}^{\ell}_{\alpha2\beta\ell'}\omega^{2\nu_{k_{\ell'}}}+\cA_{\alpha\ell,\beta\ell'}^{(1,4)}{n_4}^{\ell}_{\alpha2\beta\ell'},\nonumber\\
\cB_{\alpha\ell,\beta\ell'}^{(1)}&=&\cB_{\alpha\ell,\beta\ell'}^{(1,1)}\tilde{n_1}^{\ell}_{\alpha1\beta\ell'}\omega^{\nu_{k_\ell}+\nu_{k_{\ell'}}}+\cB_{\alpha\ell,\beta\ell'}^{(1,3)}{n_3}^{\ell}_{\alpha1\beta\ell'}\omega^{2\nu_{k_{\ell'}}}+\cB_{\alpha\ell,\beta\ell'}^{(1,4)}{n_4}^{\ell}_{\alpha1\beta\ell'},\nonumber\\
{\cA}_{\alpha\ell,\alpha\ell}^{(2)}&=&\big(\cA_{\alpha\ell}^{(2,1)}\tilde{s_1}^{\ell}_{\alpha2}+\cA_{\alpha\ell}^{(2,4)}{s_4}^{\ell}_{\alpha2}\big)\omega^{2\nu_{k_\ell}}+\sum_{\ell'}\cA_{\alpha\ell,\ell'}^{(2,3)}{s_3}^{\ell}_{\alpha2\ell'}\omega^{\nu_{k_\ell}+\nu_{k_{\ell'}}}+\tilde{\cA}_{\alpha\ell}^{(2,4)}\tilde{s_4}^{\ell}_{\alpha2}\omega^{2\nu_{k_\ell}}\ln{\omega}+\cA_{\alpha\ell}^{(2,5)}{s_5}^{\ell}_{\alpha2},\nonumber\\
\cB_{\alpha\ell,\alpha\ell}^{(2)}&=&\big(\cB_{\alpha\ell}^{(2,1)}\tilde{s_1}^{\ell}_{\alpha1}+\cB_{\alpha\ell}^{(2,4)}{s_4}^{\ell}_{\alpha1}\big)\omega^{2\nu_{k_\ell}}+\sum_{\ell'}\cB_{\alpha\ell,\ell'}^{(2,3)}{s_3}^{\ell}_{\alpha1\ell'}\omega^{\nu_{k_\ell}+\nu_{k_{\ell'}}}+\tilde{\cB}_{\alpha\ell}^{(2,4)}\tilde{s_4}^{\ell}_{\alpha1}\omega^{2\nu_{k_\ell}}\ln{\omega}+\cB_{\alpha\ell}^{(2,5)}{s_5}^{\ell}_{\alpha1}\nonumber
\eea
for the nondegenerate case% ,\comment{SHOULDN'T THE UNBARRED $a^{(4)}$
  % BE BARRED $\tilde{a}^{(4)}$?}\commently{$a^{(4)}$ is correct.}
. The parenthesized superscripts now refer to the different
building blocks ${n_i}_{\alp j \beta \ell'}^{\ell}$ and ${s_i}_{\alp j
  \beta \ell'}^{\ell}$. For the degenerate case when
$\nu_{k_\ell}=\nu_{k_{\ell'}}$
we find
\bea
\label{eq:k12}
\cA_{\alpha\ell,\alpha\ell}^{(0)}&=&\cA_{\alpha\ell,\alpha\ell}^{(0)-}v_{-\alpha2}^{(0,\ell)}+\tilde{\mathcal{G}}_{\alpha\text{IR}}^{\ell}
\cA_{\alpha\ell,\alpha\ell}^{(0)+}v_{+\alpha2}^{(0,\ell)}\ome^{2\nu_{k_\ell}}, \\
\cB_{\alpha\ell,\alpha\ell}^{(0)}&=&\cB_{\alpha\ell,\alpha\ell}^{(0)-}v_{-\alpha1}^{(0,\ell)}+\tilde{\mathcal{G}}_{\alpha\text{IR}}^{\ell}\cB_{\alpha\ell,\alpha\ell}^{(0)+}v_{+\alpha1}^{(0,\ell)}\ome^{2\nu_{k_\ell}},\nonumber\\
\cA_{\alpha\ell,\beta\ell'}^{(1)}&=&\big(\cA_{\alpha\ell,\beta\ell'}^{(1,1)}\tilde{d_1}^{\ell}_{\alpha2\beta\ell'}\omega^{2\nu_{k_\ell}}+\cA_{\alpha\ell,\beta\ell'}^{(1,4)}{d_4}^{\ell}_{\alpha2\beta\ell'}\omega^{2\nu_{k_\ell}}\big)+\cA_{\alpha\ell,\beta\ell'}^{(1,2)}{d_3}^{\ell}_{\alpha2\beta\ell'}+2\cA_{\alpha\ell,\beta\ell'}^{(1,2)}{d_3}^{\ell}_{\alpha2\beta\ell'}\omega^{2\nu_{k_\ell}}\ln\omega,\nonumber\\
\cB_{\alpha\ell,\beta\ell'}^{(1)}&=&\big(\cB_{\alpha\ell,\beta\ell'}^{(1,1)}\tilde{d_1}^{\ell}_{\alpha1\beta\ell'}\omega^{2\nu_{k_\ell}}+\cB_{\alpha\ell,\beta\ell'}^{(1,4)}{d_4}^{\ell}_{\alpha1\beta\ell'}\omega^{2\nu_{k_\ell}}\big)+\cB_{\alpha\ell,\beta\ell'}^{(1,2)}{d_3}^{\ell}_{\alpha1\beta\ell'}+2\cB_{\alpha\ell,\beta\ell'}^{(1,2)}{d_3}^{\ell}_{\alpha1\beta\ell'}\omega^{2\nu_{k_\ell}}\ln\omega,\nonumber\\
\cA_{\alpha\ell,\alpha\ell}^{(2)}&=&\big(\cA_{\alpha\ell}^{(2,1)}\tilde{\degens_1}^{\ell}_{\alpha2}+\cA_{\alpha\ell}^{(2,5)}{\degens_5}^{\ell}_{\alpha2}+\cA_{\alpha\ell}^{(2,10)}{\degens_{10}}^{\ell}_{\alpha2}\big)\omega^{2\nu_{k_\ell}}+\big(\cA_{\alpha\ell}^{(2,3)}\tilde{\degens_3}^{\ell}_{\alpha2\ell'}+\cA_{\alpha\ell}^{(2,7)}\tilde{\degens_7}^{\ell}_{\alpha2\ell'}\big)\omega^{2\nu_{k_\ell}}\ln{\omega}
\nonumber\\&&~~+\big(\cA_{\alpha\ell}^{(2,6)}{\degens_{6}}^{\ell}_{\alpha2}+\cA_{\alpha\ell}^{(2,11)}{\degens_{11}}^{\ell}_{\alpha2}\big)+\cA_{\alpha\ell}^{(2,8)}\tilde{\degens_8}^{\ell}_{\alpha2\ell'}\omega^{2\nu_{k_\ell}}(\ln{\omega})^2+(\text{terms from non-degenerate $\ell''$}),\nonumber\\
\cB_{\alpha\ell,\alpha\ell}^{(2)}&=&\big(\cB_{\alpha\ell}^{(2,1)}\tilde{\degens_1}^{\ell}_{\alpha1}+\cB_{\alpha\ell}^{(2,5)}{\degens_5}^{\ell}_{\alpha1}+\cB_{\alpha\ell}^{(2,10)}{\degens_{10}}^{\ell}_{\alpha1}\big)\omega^{2\nu_{k_\ell}}+\big(\cB_{\alpha\ell}^{(2,3)}\tilde{\degens_3}^{\ell}_{\alpha1\ell'}+\cB_{\alpha\ell}^{(2,7)}\tilde{\degens_7}^{\ell}_{\alpha1\ell'}\big)\omega^{2\nu_{k_\ell}}\ln{\omega}\nonumber\\&&~~+\big(\cB_{\alpha\ell}^{(2,6)}{\degens_{6}}^{\ell}_{\alpha1}+\cB_{\alpha\ell}^{(2,11)}{\degens_{11}}^{\ell}_{\alpha1}\big)+\cB_{\alpha\ell}^{(2,8)}\tilde{\degens_8}^{\ell}_{\alpha1\ell'}\omega^{2\nu_{k_\ell}}(\ln{\omega})^2+(\text{terms from non-degenerate $\ell''$})\nonumber
\eea
%for the degenerate case where $\ell'=\ell\pm 1$.

All the coefficients $\cA^{(n,j)},~\cB^{(n,j)}$ are real numbers, which depend on the
momentum $k$, but are only the leading term in an analytic expansion in
$\ome$. Including the corrections due to sub-leading terms in order
$1/(r-1)$ in the equations obtains the subleading powers of $\omega$ for
each of these coefficients $\cA^{(n,j)}=\sum_{l=0}^{\infty} \cA^{(n,j,l)}\omega^l$ etc.

%The zeroth order coefficients can be obtained in the above way stated below (\ref{farzero}), and the first and second order coefficients can be determined as the following way:  For the first order far region EOM (\ref{farfirst}), one need to impose the boundary condition for $\Phi^{(1,\ell)}$ and figure out the corresponding source term $\Phi^{(0,\ell')}$ in the whole far region which can generate such boundary condition, thus one can solve the inhomogeneous linear equation for $\Phi^{(1,\ell)}$. Similarly one can determine the UV datas which are related to the inhomogeneous equation for $\Phi^{(2,\ell)}$.  The lesson is that these coefficients can be determined numerically. We leave numerically obtaining these coefficients for future  work.  Note here we take the normalization that at the matching point we only choose the initial boundary condition for each components of the Dirac fields are related to the function of $r$, such as $(\frac{1}{r-1})^{\pm\nu}$ etc.
%In this paper, we will extract the most interesting physics based on our knowledge of
%$\omega$ dependance.

%%%%%%%%%%%%%%%%%%%%%%%%%%%%%%%%%%%%%%
\subsection{The full retarded Green's function}
%%%%%%%%%%%%%%%%%%%%%%%%%%%%%%%%%%%%%%

The final non-trivial step to obtain the full retarded Green's functions, is to
invert the matrix $\cA$. 
For non-degenerate momenta $\nu_{k_\ell}\neq \nu_{k_\ell'}$, up to the second
order in $\epsilon$ the determinant of $\cA$ equals
%\comment{I AM MISSING THE SECOND ORDER DIAGONAL CONTRIBUTION?}
\bea\label{deta}
\det{{\cA}}&=&\prod_{\alpha\ell}(A_{\alpha\ell,\alpha\ell}^{(0)}+\eps^2A_{\alpha\ell,\alpha\ell}^{(2)})\bigg(1-\epsilon^2\big[\sum_{\ell'\alpha'}
\frac{\cA_{\alpha'\ell',\alpha'\ell'+1}^{(1)}\cA_{\alpha'\ell'+1,\alpha'\ell'}^{(1)}}{\cA_{\alpha'\ell',\alpha'\ell'}^{(0)}\cA_{\alpha'\ell'+1,\alpha'\ell'+1}^{(0)}}\nonumber\\&&+\sum_{\ell'}\frac{\cA_{1\ell',2\ell'+1}^{(1)}\cA_{2\ell'+1,1\ell'}^{(1)}}{\cA_{1\ell',1\ell'}^{(0)}\cA_{2\ell'+1,2\ell'+1}^{(0)}}+\sum_{\ell'}\frac{\cA_{2\ell',1\ell'+1}^{(1)}\cA_{1\ell'+1,2\ell'}^{(1)}}{\cA_{2\ell',2\ell'}^{(0)}\cA_{1\ell'+1,1\ell'+1}^{(0)}}\big]\bigg)\eea
% \bea
% \label{deta}
% \det{{\cA}}&=&\prod_{\alpha\ell}A_{\alpha\ell,\alpha\ell}^{(0)}\bigg(1-\epsilon^2\big[\sum_{\ell'\alpha'}
% \frac{\cA_{\alpha'\ell',\alpha'\ell'+1}^{(1)}\cA_{\alpha'\ell'+1,\alpha'\ell'}^{(1)}}{\cA_{\alpha'\ell',\alpha'\ell'}^{(0)}\cA_{\alpha'\ell'+1,\alpha'\ell'+1}^{(0)}}\nonumber\\&&+\sum_{\ell'}\frac{\cA_{1\ell',2\ell'+1}^{(1)}\cA_{2\ell'+1,1\ell'}^{(1)}}{\cA_{1\ell',1\ell'}^{(0)}\cA_{2\ell'+1,2\ell'+1}^{(0)}}+\sum_{\ell'}\frac{\cA_{2\ell',1\ell'+1}^{(1)}\cA_{1\ell'+1,2\ell'}^{(1)}}{\cA_{2\ell',2\ell'}^{(0)}\cA_{1\ell'+1,1\ell'+1}^{(0)}}\big] +\eps^2 \sum_{\ell'}\frac{\cA^{(2)}_{\alp\ell',\alp\ell'}}{\cA^{(0)}_{\alp\ell',\alp\ell'}}
% \bigg)\eea
%\bea\label{deta}
%\det{\cA}=\det\cA^{(0)}(1-\eps^2\text{Tr})
%\eea
and up to same order $\epsilon^2$ the algebraic cofactors of $\cA$ are
%\Delta_{\alpha\ell-1,\alpha\ell}&=&-\epsilon\frac{a_{\alpha\ell,\alpha\ell-1}}{a_{\alpha\ell-1,\alpha\ell-1}a_{\alpha\ell,\alpha\ell}}\prod_{\alpha'\ell'}a_{\alpha'\ell',\alpha'\ell'};\nonumber\\
%\Delta_{\alpha\ell+1,\alpha\ell}&=&-\epsilon\frac{a_{\alpha\ell,\alpha\ell+1}}{a_{\alpha\ell,\alpha\ell}a_{\alpha\ell+1,\alpha\ell+1}}\prod_{\alpha'\ell'}a_{\alpha'\ell',\alpha'\ell'};\nonumber\\
%\Delta_{1\ell,2\ell-1}&=&-\epsilon\frac{a_{2\ell-1,1\ell}}{a_{1\ell,1\ell}a_{2\ell-1,2\ell-1}}\prod_{\alpha'\ell'}a_{\alpha'\ell',\alpha'\ell'};\nonumber\\
%\Delta_{2\ell,1\ell+1}&=&-\epsilon\frac{a_{1\ell+1,2\ell}}{a_{1\ell+1,1\ell+1}a_{2\ell,2\ell}}\prod_{\alpha'\ell'}a_{\alpha'\ell',\alpha'\ell'};\nonumber\\
%\Delta_{1\ell-1,2\ell}&=&-\epsilon\frac{a_{2\ell,1\ell-1}}{a_{1\ell-1,1\ell-1}a_{2\ell,2\ell}}\prod_{\alpha'\ell'}a_{\alpha'\ell',\alpha'\ell'};\nonumber\\
%\Delta_{2\ell+1,1\ell}&=&-\epsilon\frac{a_{1\ell,2\ell+1}}{a_{2\ell+1,2\ell+1}a_{1\ell,1\ell}}\prod_{\alpha'\ell'}a_{\alpha'\ell',\alpha'\ell'};\nonumber\\
\bea
\Delta_{\alpha\ell,\beta\ell\pm1}&=&-\epsilon\frac{\cA_{\beta\ell\pm1,\alpha\ell}^{(1)}}{\cA_{\alpha\ell,\alpha\ell}^{(0)}\cA_{\beta\ell\pm1,\beta\ell\pm1}^{(0)}}\prod_{\alpha'\ell'}\cA_{\alpha'\ell',\alpha'\ell'}^{(0)};\nonumber\\
\Delta_{\alpha\ell,\alpha\ell}&=&\frac{\prod_{\alpha'\ell'}(\cA_{\alpha'\ell',\alpha'\ell'}^{(0)}+\eps^2\cA_{\alpha\ell,\alpha\ell}^{(2)})}{(\cA_{\alpha\ell,\alpha\ell}^{(0)}+\eps^2\cA_{\alpha\ell,\alpha\ell}^{(2)})}\bigg(1-\epsilon^2\big[\sum_{\ell'\alpha'}\frac{\cA_{\alpha'\ell',\alpha'\ell'+1}^{(1)}\cA_{\alpha'\ell'+1,\alpha'\ell'}^{(1)}}{\cA_{\alpha'\ell',\alpha'\ell'}^{(0)}\cA_{\alpha'\ell'+1,\alpha'\ell'+1}^{(0)}}+\sum_{\ell'}\frac{\cA_{1\ell',2\ell'+1}^{(1)}\cA_{2\ell'+1,1\ell'}^{(1)}}{\cA_{1\ell',1\ell'}^{(0)}\cA_{2\ell'+1,2\ell'+1}^{(0)}}\nonumber\\&&+\sum_{\ell'}\frac{\cA_{2\ell',1\ell'+1}^{(1)}\cA_{1\ell'+1,2\ell'}^{(1)}}{\cA_{2\ell',2\ell'}^{(0)}\cA_{1\ell'+1,1\ell'+1}^{(0)}}\big]+\epsilon^2\sum_{\beta}\big(\frac{\cA_{\beta\ell-1,\alpha\ell}^{(1)}\cA_{\alpha\ell,\beta\ell-1}^{(1)}}{\cA_{\beta\ell-1,\beta\ell-1}^{(0)}\cA_{\alpha\ell,\alpha\ell}^{(0)}}+\frac{\cA_{\beta\ell+1,\alpha\ell}^{(1)}\cA_{\alpha\ell,\beta\ell+1}^{(1)}}{\cA_{\beta\ell+1,\beta\ell+1}^{(0)}\cA_{\alpha\ell,\alpha\ell}^{(0)}}\big)\bigg);\nonumber\\
\eea
%\commently{Need to change $\cA_{\alpha\ell,\alpha'\ell'}\to  \cA_{\alpha\ell,\alpha'\ell'}^{(1)}$}
where $\beta=\alpha, 3-\alpha.$ %We have the similar expression for matrix ${\cB}$. 
The full retarded Green's function is then
\bea
\label{GR}
G_{\text{R}\alpha\ell,\alpha\ell}&=&\frac{1}{\det{{\cA}}}\bigg(\sum_{\beta=1,2}({\cB}_{\alpha\ell,\beta\ell-1}\Delta_{\alpha\ell,\beta\ell-1}+{\cB}_{\alpha\ell,\beta\ell+1}\Delta_{\alpha\ell,\beta\ell+1})+{\cB}_{\alpha\ell,\alpha\ell}\Delta_{\alpha\ell,\alpha\ell}\bigg).
\eea

In all expressions above, we have assumed that the diagonal elements
in the ${\cA}$ matrix are composed of both order 1 and order
$\epsilon^2$ parts and the off diagonal elements are of order
$\epsilon$ so that the diagonal elements are always more important
than the off diagonal ones. But we know that near the Fermi momentum
of the zeroth order solution, the zeroth order part in the
corresponding diagonal element becomes neglible --- it vanishes at $k_F$ --- and the above expansion used
to construct the Green's function does not hold. % and (one  diagonal elements
                                % (for the correct level) would become
                                % of order $\epsilon^2$. 
For the degenerate case, two of the diagonal elements are identical
and both become subleading of order $\epsilon$ near the Fermi
momentum. Degenerate eigenvalues is the familiar case from solid state
textbooks and we will now show, that the intuition that a correct
re-diagonalization will cause the opening of a band gap is correct. 
In the non-degenerate case, no band gap opens up --- there is no
eigenvalue mixing. However, correcting the violation of the
naive expansion in $\eps$ will show that there are important second order
corrections to the self-energy at the quasi-particle pole in the non-degenerate case.

%%%%%%%%%%%%%%%%%%%%%%%%%%%%%%%%%%%%%%
%%%%%%%%%%%%%%%%%%%%%%%%%%%%%%%%%%%%%%
 \section{Band structure and hybridized dissipation}\label{sec4}
%%%%%%%%%%%%%%%%%%%%%%%%%%%%%%%%%%%%%%
%%%%%%%%%%%%%%%%%%%%%%%%%%%%%%%%%%%%%%

We shall first discuss the physically most straightforward case: with
degenerate eigenvalues any interaction will cause eigenvalue repulsion
and the opening up of a band gap.

%%%%%%%%%%%%%%%%%%%%%%%%%%%%%%%%%%%%%%
\subsection{Band gap for the degenerate case}
%%%%%%%%%%%%%%%%%%%%%%%%%%%%%%%%%%%%%%

%\subsubsection{At the degenerate point}

For the degenerate case, $\nu_{k_\ell}=\nu_{k_\ell'}$ with
$\nu_{k_\ell}=\frac{1}{\sqrt{6}}\sqrt{k_{\ell}^2-q^2\frac{{\mu_0}^2}{6}}$
and $k_{\ell}=\sqrt{(k_x+\ell K)^2+k_y^2}$ the $k_y=0$ and $k_y\neq 0$
cases are quite different. We first consider the special case $k_F=\frac{K}{2}$ ($i.e.$ $k_x=\pm \frac{K}{2}, k_y=0$).
In this case, different spinor
components do not interact (see eq. \ref{eom1}), so we can deal with the different spinor
components independently.  By definition the Fermi momentum is the
value $k=k_F$ where the Green's function has a pole, i.e. where the numerator
$A_{\alpha\ell,\alpha\ell}(\ome=0)$ vanishes at zero frequency, % and 
% % Near the Fermi momentum $k_F$
i.e.
 $\cA_{\alpha\ell,\alpha\ell}(k=k_F)\sim \ome a^{(0,1)}_{\alpha\ell,\alpha\ell}+
 \mathcal{O}(\ome^2,\epsilon^2)$.
% \comment{THE FOLLOWING WHY $\beta=3-\alpha.$
%   IS NOT EXPLAINED ANYWHERE}
% .\commently{Just from the eqn. 3.15 for $k_x=\pm K/2$}
Recalling eq. (\ref{eom1}) for this special case $k_y=0$, we note that the equations for the two spinor components $\alpha$ decouple and are identical for each. Then it is straightforward to see that at zeroth order the momentum coefficient in the second Brillioun-zone ($\ell=1$ with $k_x=-K/2$ or $\ell=-1$ with $k_x=K/2$)  of the component $\alpha$ is equal to the momentum component of the other spin component $\beta=3-\alpha$ in the first Brillioun-zone ($\ell=0$ and $k_x=-K/2$ or $k_x=K/2$ respectively). The
degenerate term with $A_{\alp\ell\alp\ell}$ is therefore % \comment{REALLY??? THIS STILL DOES NOT EXPLAIN WHY $\beta=3-\alpha.$ IS THE DEGENERATE. IT SEEMS TO BE THE OPPOSITE SPIN COMPONENT IN THE SAME BRILLIOUN ZONE. THE ARGUMENT/STEPS IN THE DERIVATION/REASON WHY HAS TO BE EXPLAINED, PLEASE PUT IT IN THE TEXT CAREFULLY}
the {\em other}
spinor 
component $\beta=3-\alpha$ in the neighboring Brillioun zone
$\ell'=\ell-1$ % \commently{here to make these two modes degenerate, we have to fix $k_x,\ell$.}
and hence $A_{\beta\ell-1;\beta\ell-1}=A_{\alpha\ell,\alpha\ell}\sim \mathcal{O}(\epsilon^2)$.
%$\cA_{\beta,\ell-1;\beta,\ell-1}(k=k_F)\sim \ome a^{(0,1)}_{\beta,\ell-1;\beta,\ell-1}+
%\mathcal{O}(\ome^2,\epsilon^2)$.
Let us assume that it is both $\cA_{1\ell,1\ell}$ and
$\cA_{2\ell-1,2\ell-1}$ which are of order $\ome a^{(1)}+\cO(\eps^2)$
and much smaller than any other diagonal component of $\cA_{\alpha\ell,\beta\ell'}$. Substituting
this in the expression for the retarded Green's function \eqref{GR},
we find
\bea\label{GR2}
G_{\text{R}1\ell,1\ell}&=&\frac{\cB_{1\ell,1\ell}^{(0)}+\mathcal{O}(\epsilon^2)}{
  A^{(0)}_{1\ell,1\ell}+\epsilon^2{A}_{1\ell,1\ell}^{(2)}-\epsilon^2\big(\frac{A_{1\ell,1\ell-1}^{(1)}A_{1\ell-1,1\ell}^{(1)}}{A_{1\ell-1,1\ell-1}^{(0)}}+\frac{A_{1\ell,1\ell+1}^{(1)}A_{1\ell+1,1\ell}^{(1)}}{A_{1\ell+1,1\ell+1}^{(0)}}\big)+\mathcal{O}(\epsilon^4)},\nonumber\\
G_{\text{R}2\ell,2\ell}&=&\frac{1-\epsilon^2\frac{A_{2\ell-1,2\ell-2}^{(1)}A_{2\ell-2,2\ell-1}^{(1)}}{A_{2\ell-2,2\ell-2}^{(0)}A_{2\ell-1,2\ell-1}^{(0)}}}{1-\epsilon^2\big(\frac{A_{2\ell-1,2\ell}^{(1)}A_{2\ell,2\ell-1}^{(1)}}{A_{2\ell,2\ell}^{(0)}A_{2\ell-1,2\ell-1}^{(0)}}+\frac{A_{2\ell-1,2\ell-2}^{(1)}A_{2\ell-2,2\ell-1}^{(1)}}{A_{2\ell-2,2\ell-2}^{(0)}A_{2\ell-1,2\ell-1}^{(0)}}\big)}\bigg[\frac{B_{2\ell,2\ell}^{(0)}}{A_{2\ell,2\ell}^{(0)}}+\mathcal{O}(\epsilon^2) \bigg].
\eea
From the above formulae we see that near $k_F$, $G_{\text{R}2\ell,2\ell}$ is regular while 
\be\label{gfdegnogap}
G_{\text{R}1\ell,1\ell}\simeq\frac{\cB_{1\ell,1\ell}^{(0)}}{\omega -v_F (k_\ell-k_F)+\Sigma(\omega)},
\ee
where $\Sigma(\omega)$ denotes the
self-energy. So there is no gap in this case. This may seem surprising. It is however a direct
consequence of the fact that the holographic fermion spectral function
has a chiral decomposition for $k_y=0$. In the basis of Gamma-matrices \eqref{gammat} spin up
particles ($\alpha=1$) only have right-moving dispersion $\ome=k_x$,
whereas spin down ($\alpha=2$) particles only have left moving dispersion
$\ome=-k_x$.\footnote{The spin structure for holographic fermions has been studied in \cite{alexandrov, {Herzog:2012kx}}.} 
As noted in \cite{Faulkner:2009am}, in order to have a gap
this chiral structure must be broken and interactions must be present
that mix the two.

In the absence of a lattice any nonzero $k_y$ solution can be again rotated to a chiral basis. In the presence of a lattice this rotation freedom is broken and 
these interactions occur for any $k_y\neq 0$ as we can directly see in (\ref{eom2}).
Now the degeneracy $\nu_{{k}_{\ell}}=\nu_{k_{\ell'}}$ leads to
$A_{\alpha\ell,\alpha\ell}=
A_{\alpha\ell-1,\alpha\ell-1}$. (For any $k_y\neq0$, $k_{\ell}$ is always positive and \eqref{eom2}, than immediately shows that now the same spinor component is degenerate between the first $\ell=0$ and second $\ell=\pm 1$ Brillioun zones). 
As a result, near a zeroth order Fermi
surface, both $\cA^{(0)}_{\alpha\ell,\alpha\ell}$ and
$\cA^{(0)}_{\alpha\ell-1,\alpha\ell-1}$ become very small and comparable to the modulation order $\eps$, 
and assuming that all other diagonal components of $\cA_{\alpha
  \ell,\beta\ell'}$ are much larger, the retarded Green's function equals
\bea
G_{\text{R}\alpha\ell,\alpha\ell}&=&\frac{\cA_{\alpha\ell,\alpha\ell}^{(0)}\cB_{\alpha\ell,\alpha\ell}^{(0)}+\mathcal{O}(\epsilon^2)}{(\cA_{\alpha\ell,\alpha\ell}^{(0)}+\eps^2\cA^{(2)}_{\alpha\ell,\alpha\ell})^2-\epsilon^2 \cA_{\alpha\ell,\alpha\ell-1}^{(1)}\cA_{\alpha\ell-1,\alpha\ell}^{(1)}+\mathcal{O}(\epsilon^3)}.
\eea
Near $k_\ell=k_F, \omega=0$, we have $\cA_{\alpha\ell,\alpha\ell}^{(0)}+\eps^2\cA^{(2)}_{\alpha\ell,\alpha\ell}=\omega -v_F
 (k_\ell-k_F)+i(c_1-ic_2)\omega^{2\nu_{k_\ell}}$ with $v_F, c_1,c_2$ are real and
 dependent on UV data. 
Assuming $2\nu_{k_\ell}>1$, i.e. the original theory possessed a
regular quasiparticle with linear dispersion liquid, we can ignore
$c_2$ and find that
Umklapp for degenerate momenta modifies the Green's function to
\bea\label{bandgap1}
G_{\text{R}\alpha\ell,\alpha\ell}&\simeq&\frac{(\omega -v_F (k_\ell-k_F))\cB_{\alpha\ell,\alpha\ell}}{(\omega -v_F (k_\ell-k_F))^2-\Delta^2+ic_1(\omega -v_F (k_\ell-k_F))\omega^{2\nu_{k_\ell}}}
\eea
where \be\label{gapp}\Delta^2=\epsilon^2 \cA_{\alpha\ell,\alpha\ell-1}^{(1)}\cA_{\alpha\ell-1,\alpha\ell}^{(1)}=\epsilon^2\big(1-\frac{1}{\sqrt{1+(2ak_y)^2}}\big)\mathfrak{a}_{k_\ell}^2\ee
with $\mathfrak{a}_{k_\ell}$ a function of $k_\ell$.
% Note here we work in the extended Brillouin zone scheme.\comment{HOW
%   DO I SEE THIS}.\commently{When we calcluated the spectral function, we did not sum over the different Brillouin zone. So it is not the simple Brillouin zone scheme.}
The denominator shows a classic band gap\footnote{To be explicit, all points on
these two bands with finite $\omega$ are not strict poles because the
imaginary part of the self-energy doesn't vanish but it is small. The
existence of bands only depends on the real part of the denominator
and the self energy can be ignored as long as it stays small throughout.} with two peaks at $\omega=v_F (k_\ell-k_F)\pm \Delta.$ 
Note that the bandgap already appears at first order in $\eps$, the
second order corrections will only make a small quantitative
change. The width of gap is therefore proportional to $\epsilon$. The
pole has shifted to $\ome \sim \Delta \sim \eps$ and its width is
therefore of order $\epsilon^{2\nu_{k_\ell}}$.  Recall that here
$2\nu_{k_\ell}> 1$, so the width of the peaks is negligible comparing
to the width of the gap.    Thus the peaks are sharp.

Assuming the original quasiparticle was non-Fermi liquid-like,
i.e. $2\nu_{k_\ell}<1$, and near $k_\ell=k_F, \omega=0$,
$\cA_{\alpha\ell,\alpha\ell}=c_2\omega^{2\nu_{k_\ell}} -v_F
(k_\ell-k_F)+ic_1\omega^{2\nu_{k_\ell}}$ 
we find the Umklapp modified Green's function
\bea\label{bandgap2}
G_{\text{R}\alpha\ell,\alpha\ell}&\simeq&\frac{(\omega^{2\nu_{k_\ell}} -v_F (k_\ell-k_F))\cB_{\alpha\ell,\alpha\ell}}{(\omega^{2\nu_{k_\ell}} -v_F (k_\ell-k_F))^2-\Delta^2+ic_1(\omega^{2\nu_{k_\ell}} -v_F (k_\ell-k_F))\omega^{2\nu_{k_\ell}}}.
\eea 
This is a gapped system for non-linear dispersion with
two peaks at $\omega^{2\nu_{k_\ell}}=v_F (k_\ell-k_F)\pm \Delta$. The
instability of the original non-Fermi liquid quasiparticle now
reflects itself in that the width of peak and the gap are of the same
order $\epsilon^{1/2\nu_{k_\ell}}$. For the marginal case
$2\nu_{k_\ell}=1$, where near $k_\ell=k_F, \omega=0$ we have
$A_{\alpha\ell,\alpha\ell}=\omega+c_2\omega\ln\omega -v_F
(k_\ell-k_F)+ic_1\ome\ln \ome$ one similarly concludes the presence of a bandgap.  

We emphasize that the mechanism which leads to a gap is identical to
the standard band structure in condensed matter theories: the
occurence of degenerate eigenvalues which repel each other upon the
inclusion of interactions between the different levels.  
In that sense the mechanism is similar to 
the holographic BCS gap in
\cite{Faulkner:2009am,Gubser:2009dt} and
\cite{Vegh:2010fc,Benini:2010pr}. %  as well as the Mott gap
% cite{Edalati:2010ge,{Guarrera:2011my}}.
In essence,
the fact that the holographic spectral function can be reproduced from
a Dyson resummation of a free fermion coupled to the local quantum
critical AdS${}_2$, a.k.a. semi-holography \cite{arXiv:1001.5049},
guarantees that all these situations should simply recover the
textbook gap. As we will see in the next subsection, however, the dissipative properties also change due to the lattice. And, correlated to this, we can then also show that the Green's function in the latticized AdS${}_2$ metal do not possess a gap.

%In the formula above, we can see that because $-\nu_{k_\ell}+\nu_{k_{\ell-1}}<0$ and $-\nu_{k_{\ell+1}}+\nu_{k_\ell}<0$ %while other exponents are positive, these two terms are the leading terms in the $\epsilon ^2$ terms. We %also have assumed that $\epsilon^2 \omega^{\text{Min}[-\nu_{k_\ell}+\nu_{k_{\ell-1}},-\nu_{k_{\ell+1}}+\nu_{k_\ell}]}\ll 1$, %so that we can get another pole at $\omega_0$ which satisfies that the imaginary part of the self-energy is %0 and $\omega_0^{2 \nu_l}$ should be of the same order as $\epsilon^2 \omega_0^{\text{Min}[-\nu_{k_\ell}+%\nu_{k_{\ell-1}},-\nu_{k_{\ell+1}}+\nu_{k_\ell}]} $, and this is consistent with the assumption that $\epsilon^2 \omega^
%{\text{Min}[-\nu_{k_\ell}+\nu_{k_{\ell-1}},-\nu_{k_{\ell+1}}+\nu_{k_\ell}]}\ll 1$. Thus this $\omega_0$ gives a gap scale. %But the above result is not valid for $\omega=0$ because there is no meaning of near region. When
%$\omega=0$, there are only $\omega^0$ and $\omega^{\nu_{k_\ell}+\nu_{k_{\ell'}}}$ terms in the $\epsilon^2$ %terms.  So $\omega=0$ should still be a pole.

%%%%%%%%%%%%%%%%%%%%%%%%%%%%%%%%%%%%%%
\subsection{Hybridized dissipation around the gap and the  AdS$_2$-metal Greens function on the lattice}\label{sec-self}
%%%%%%%%%%%%%%%%%%%%%%%%%%%%%%%%%%%%%%

The band gap in the presence of a Fermi surface is a first order in $\epsilon$ effect. 
Our analysis of the perturbative form of the retarded Green's function in section \ref{sec3} already showed that the change in the dispersive properties occurs at second order in the modulation $\epsilon$. In the case of a Fermi surface the dispersive properties show up as a modification of the self-energy, whereas for the AdS$_2$ metal they directly affect the defining characteristics of the spectral function. We shall discuss the case of a Fermi-surface (with a band gap) first.

\subsubsection{Near the Fermi surface}

Near the pole, the leading component $\cA_{\alpha\ell,\alpha\ell}^{(0)}$ becomes of the same order as the formally subleading $\cO(\epsilon^2)$ corrections, so we should be very careful  with small terms in the bracket in the formula (\ref{deta}) of $\det{\cA}$. We define
\be\label{cdef}
C=\sum_{\beta=1,2}\bigg(\frac{A_{\beta\ell-1,\alpha\ell}^{(1)}A_{\alpha\ell,\beta\ell-1}^{(1)}}{\cA_{\beta\ell-1,\beta\ell-1}\cA_{\alpha\ell,\alpha\ell}}+\frac{\cA_{\beta\ell+1,\alpha\ell}^{(1)}\cA_{\alpha\ell,\beta\ell+1}^{(1)}}{\cA_{\beta\ell+1,\beta\ell+1}\cA_{\alpha\ell,\alpha\ell}}\bigg).
\ee
Then for both the non-degenerate case as well as the degenerate case with $k_y=0$ the retarded Green's function becomes
\bea\label{GRagain}
G_{\text{R}\alpha\ell,\alpha\ell}&\simeq&\frac{1}{(\cA^{(0)}_{\alpha\ell,\alpha\ell}+\eps^2\cA^{(2)}_{\alp\ell,\alp\ell})(1-\epsilon^2 C)}\bigg[\cB_{\alpha\ell,\alpha\ell}-\epsilon^2\sum_{\alpha'=1,2}\bigg(\frac{\cA_{\alpha'\ell-1,\alpha\ell}^{(1)}\cB_{\alpha\ell,\alpha'\ell-1}^{(1)}}{\cA_{\alpha'\ell-1,\alpha'\ell-1}^{(0)}}+\frac{\cA_{\alpha'\ell+1,\alpha\ell}^{(1)}\cB_{\alpha\ell,\alpha'\ell+1}^{(1)}}{\cA_{\alpha'\ell+1,\alpha'\ell+1}^{(0)}}\bigg)\bigg]\nonumber\\
&=&\frac{\cB_{\alpha\ell,\alpha\ell}^{(0)} +{{\cO}}(\epsilon^2)}{\cA_{\alpha\ell,\alpha\ell}^{(0)}+\epsilon^2({\cA}_{\alpha\ell,\alpha\ell}^{(2)}-\cA_{\alpha\ell,\alpha\ell}^{(0)} C)+{\mathcal{O}}(\epsilon^4)}.
\eea
A consistency check for the expression (\ref{GRagain}) is that it does not depend on the freedom to add arbitrary infalling homogenous solution at  first or second order which we used to cancel the possible divergence related to $\omega\to 0$. The operation of adding them is equal to the transformation ${\cA}\to {\cA}U$ and ${\cB}\to {\cB}U$ with $U$ a matrix of two-component spinors of which the non-zero elements are: ${U_1}_{\alpha'\ell',\alpha'\ell'}=1, (\forall \alpha',\ell'), {U_1}_{\beta\ell+1,\alpha\ell}=\epsilon \zeta^{(1)}$ or ${U_1}_{\beta\ell-1,\alpha\ell}=\epsilon \zeta^{(1)}$ and  ${U_2}_{\alpha'\ell',\alpha'\ell'}=1, (\forall \alpha',\ell'\neq \alpha\ell), {U_2}_{\alpha\ell,\alpha\ell}=1+\epsilon^2 \zeta^{(2)}$ with $\zeta^{(i)}$  a constant. % $AU$ will include arbitrary zeroth order solution automatically.}\comment{I DO NOT UNDERSTAND THIS AT ALL. I AM HAPPY TO LEAVE IT IN. JUST MAKE SURE IT IS CORRECT.}
It is easy to show that formula (\ref{GRagain}) is invariant under these two operations up to order $\epsilon^2$.  

Here we are focussing on the diagonal elements of the Green's
function. The off-diagonal elements of the Green's function will also
contribute to the spectral weight in the coordinate space. 
% , but because the zeroth order off-diagonal elements are zero an
However, following the characteristics of the building blocks $\cA_{\alp\ell,\beta\ell'}$ and $\cB_{\alp\ell,\bet\ell'}$ 
the leading order contribution of the off-diagonal Green's function elements is of order $\epsilon$. We will not study them explicitly in this paper and focus here only on the diagonal part.

%We will keep the most relevant terms in the formula (\ref{GR})
%\bea\label{final}
%G_{\text{R}\alpha\ell,\alpha\ell}&=&\frac{1}{\#_1+\#_2\omega^{2\nu_{k_\ell}}+\#_3\epsilon^2 g(\omega)}\bigg[
%\#_4+\#_5\omega^{2\nu_{k_\ell}}+\#_6\epsilon^2 g(\omega)+\epsilon^2\frac{\#_9+\#_{10}\omega^{\nu_{k_\ell}+\nu_{k_{\ell-1}}}}{\#_7+%\#_8\omega^{2\nu_{k_\ell}}}\bigg]
%\nonumber\\
%&\simeq&\frac{1}{\#+[\#\omega^{2\nu_{k_\ell}}-\epsilon^2(\#\omega^{-\nu_{k_\ell}+\nu_{k_{\ell-1}}}+\#\omega^{-%\nu_{k_{\ell+1}}+
%\nu_{k_\ell}})]}
%\eea
%The numbers $\#_n$ are order 1 and they can be determined directly: e.g. $\#_1=
%a_{-\alpha\ell,\alpha\ell}v_{-\alpha2}^{(0,\ell)}$ etc.
%It is important to note that $\#_1$ is real while $\#_{2,3}$ are complex, and $\#_n=\#_n^{(0)}+\omega\#_n^{(1)}+...$.
%The precise formula for $g(\omega)$ depends on $k$ and $\ell$ and does not depend on $\epsilon$. We will consider it in %the next subsection.

For the non-degenerate case, we have (ungapped) quasi-particle poles when the denominator in (\ref{GRagain}) vanishes. Substituting the expression for $C$, this happens whenever
\be \label{59}  \cA_{\alpha\ell,\alpha\ell}^{(0)}+\epsilon^2\cA_{\alpha\ell,\alpha\ell}^{(2)}
-\epsilon^2\sum_{\beta=1,2}\bigg(\frac{\cA_{\beta\ell-1,\alpha\ell}^{(1)}\cA_{\alpha\ell,\beta\ell-1}^{(1)}}{\cA_{\beta\ell-1,\beta\ell-1}^{(0)}}+\frac{\cA_{\beta\ell+1,\alpha\ell}^{(1)}\cA_{\alpha\ell,\beta\ell+1}^{(1)}}{\cA_{\beta\ell+1,\beta\ell+1}^{(0)}}\bigg)=0. \ee
The essence is now to extract the $\ome$-dependence from this expression using the perturbative answers (\ref{eq:k11}) we obtained for $\cA_{\alp\ell,\alp\ell}$, etc. 
Denoting the diagonal component of the retarded Green's function in its standard form
\begin{eqnarray}
  \label{spectralvb1}
  G_{\alp\ell,\alp\ell} = \frac{Z}{\omega-v_F(k-k_F)+\Sigma}+\ldots
\end{eqnarray}
one finds after some laborious algebra that the self-energy
\begin{eqnarray}
  \label{selfen1}
\Sigma & =& % \cG \equiv  
\alpha_{\vec{k}} \omega^{2 \nu_{\vec{k}}}  +\beta^{(1)}_{\vec{k}} \ome^{2\nu_{\vec{k}-\vec{K}}}+
\beta^{(2)}_{\vec{k}} \ome^{{\nu_{\vec{k}}}+{\nu_{\vec{k}-\vec{K}}}}+ \beta^{(3)}_{\vec{k}} \ome^{2\nu_{\vec{k}}}\ln\ome
\nonumber\\&&+\beta^{(4)}_{\vec{k}} \ome^{{\nu_{\vec{k}}}+\nu_{\vec{k}+\vec{K}}}+\beta^{(5)}_{\vec{k}} \ome^{2\nu_{\vec{k}+\vec{K}}}+\dots
\end{eqnarray}
%\comment{GUYS, THIS IS ****NOT**** EQUAL TO EQ. (2.5)}
%is equal to the infrared Green's function $\cG$ \commently{We have no idea what the IR Green's function would be. The %coefficients in $\Sigma$ depend on UV data.}
where $ \alpha_{\vec{k}}\sim{\mathcal{O}}(1)$ and $  \beta^{(i)}_{\vec{k}} \sim{\mathcal{O}}(\epsilon^2)$ with $i=1,...5$. These coefficients are complex functions of $k$.  The concrete formula of the parameters can be found in the appendix (\ref{appb8}-\ref{appb12}).

This is our main result whose physics we discussed in the beginning. For this, note that the Brillioun zone mixing terms $\ome^{\nu_{\vec{k}}+\nu_{\vec{k}\pm \vec{K}}}$ are never the leading terms in the IR. For the purpose of the IR physics, we may therefore truncate the self-energy to
\begin{equation}
\label{eq:2}
\Sigma  = % \cG \equiv  
\alpha_{\vec{k}} \omega^{2 \nu_{\vec{k}}}  +\beta^{(1)}_{\vec{k}} \ome^{2\nu_{\vec{k}-\vec{K}}}% +
% \beta^{(2)}_{\vec{k}} \ome^{{\nu_{\vec{k}}}  +{\nu_{\vec{k}-\vec{K}}}}
+ \beta^{(3)}_{\vec{k}} \ome^{2\nu_{\vec{k}}}\ln\ome
% \nonumber\\&&+\beta^{(4)}_{\vec{k}} \ome^{{\nu_{\vec{k}}}+\nu_{\vec{k}+\vec{K}}}
+\beta^{(5)}_{\vec{k}} \ome^{2\nu_{\vec{k}+\vec{K}}}+\dots
\end{equation}
This is eq. \eqref{CFT1general}.

%with \comment{NEED TO FILL IN EXACT ANSWERS}.
%is equal to the infrared Green's function $\cG$
%with \comment{NEED TO FILL IN EXACT ANSWERS}
%\begin{eqnarray}
 % Z&=& \\
%v_F&=& \\
%\beta^{(-)}_{k} &=&\\
%\beta^{(0)}_{k} &=&\\
%\beta^{(+)}_{k} &=&
%\end{eqnarray}

Performing the same steps for the degenerate case with $k_y=0$, one finds the diagonal Green's function (see eq. (\ref{gfdegnogap}))
\begin{eqnarray}
  \label{spectralvb2}
  G_{\alp\ell,\alp\ell} = \frac{Z}{\omega-v_F(k-k_F)+\Sigma}+\ldots
\end{eqnarray}
with % \comment{NEED TO FILL IN EXACT ANSWERS}
\begin{eqnarray}
  \label{selfen2}
\Sigma % & =& \cG 
&=&\alpha_{\vec{k}} \omega^{2 \nu_{\vec{k}}}  +\beta^{(1)}_{\vec{k}} \ome^{2\nu_{\vec{k}-\vec{K}}}+
\beta^{(2)}_{\vec{k}} \ome^{{\nu_{\vec{k}}}+{\nu_{\vec{k}-\vec{K}}}}+ \beta^{(3)}_{\vec{k}} \ome^{2\nu_{\vec{k}}}\ln\ome
+{\tilde{\beta}}_{\vec{k}} \ome^{2\nu_{\vec{k}}}(\ln\ome)^2
\nonumber\\&&+\beta^{(4)}_{\vec{k}} \ome^{{\nu_{\vec{k}}}+\nu_{\vec{k}+\vec{K}}}+\beta^{(5)}_{\vec{k}} \ome^{2\nu_{\vec{k}+\vec{K}}}+\dots
% \end{eqnarray}
% with
% \begin{eqnarray}
\end{eqnarray}
where $ \alpha_{\vec{k}}\sim{\mathcal{O}}(1)$ and ${\tilde{\beta}}$, $  \beta^{(i)}_{\vec{k}} \sim{\mathcal{O}}(\epsilon^2)$ with $i=1,...5$. The concrete formula of the parameters can be obtained similarly as the non-degenerate case. 
The distinction with the non-degenerate case is that the logarithmic term correction to the naive dispersion is now of the form $(\ln\ome)^2$.  Here we only list the result for $\tilde{\beta}_{\vec{k}}$
\begin{eqnarray} \label{appb13}
\tilde{\beta}_{\vec{k}}\sim\frac{\cA_{\alpha\ell}^{(2,8)}\tilde{\degens_8}^{\ell}_{\alpha2\ell'}}{A_{\alpha\ell,\alpha\ell}^{(0+)}v_{+\alpha 2}^{(0,\ell)}\tilde{\mathcal{G}}_{\alpha}^{\ell}}.
\end{eqnarray}
For the degenerate case with $k_y\neq 0$ where the poles repel to form a gap, the dispersion part changes due to the formation of the gap, see  (\ref{bandgap1}) and (\ref{bandgap2}). In this case, the self energy will also get corrections with the 
formulae (\ref{selfen2}). 

We emphasize here that the form of the self-energy means that it cannot be simply interpreted as some local quantum critical (i.e. pure AdS$_2$) Green's function. The solution is no longer a pure power of $\omega$, but a polynomial. At the same time the near-far matching construction does reflect that the self-energy is governed by the IR-dynamics. This, however, is a latticized version of the near-horizon AdS$_2$ region to which we now turn.

\subsubsection{AdS${}_2$ metal}

In the AdS${}_2$ metal regime of parameters the full spectral function is related to the IR-spectral function. In essence this is due to the fact that there is no pole in the spectral function, i.e. $\cA_{\alp\ell,\alp\ell}\neq 0$ always. This allows us to immediately extract the spectral function in a perturbation  expansion in $\epsilon$. 
The Green's function itself equals
\bea\label{GR2again}
G_{\text{R}\alpha\ell,\alpha\ell}
&=&\frac{\cB_{\alpha\ell,\alpha\ell}^{(0)}}{\cA_{\alpha\ell,\alpha\ell}^{(0)}}+\epsilon^2\bigg(\frac{\cB^{(0)}_{\alpha\ell,\alpha\ell}}{\cA^{(0)}_{\alpha\ell,\alpha\ell}}C
-\sum_{\beta}\frac{\cB_{\alpha\ell,\beta\ell\pm 1}^{(1)}\cA_{\beta\ell\pm 1, \alpha\ell}^{(1)}}{\cA_{\alpha\ell,\alpha\ell}^{(0)}\cA_{\beta\ell\pm 1,\beta\ell\pm 1}^{(1)}}\bigg)
+{\mathcal{O}}(\epsilon^4)
\eea
where $C$ is defined as in (\ref{cdef}), 
and the spectral function at low $\omega$ is thus
\begin{eqnarray}
  \label{eq:4}
  \mathrm{Im}G_{\text{R}\alpha\ell,\alpha\ell} \simeq  \mathrm{Im}\Sigma
\end{eqnarray}
with $\Sigma$ the same formula of the self-energy %infrared Green's function 
as above in \eqref{selfen1} and \eqref{selfen2} with different values of coefficients. As in \cite{Faulkner:2009wj}, this directly follows from the fact that the analytic terms in $\ome$ in $\cA^{(0)}_{\alp\ell,\alp\ell}$ are real (for real $\nu_k$) and that all non-analytic terms in $\cB^{(0)}_{\alp\ell,\alp\ell}$ are of higher order in $\omega$.
%\commently{CHECK}

% Just as expected, here the dependance of the imaginary part of the
% full Green's function on $\omega$ is similar to the imaginary part
% of self-energy up to the relative constant ratio coefficient:
% Both of them are $w^{2\nu}+\#\epsilon^2 g(\omega)$, but the numbers $\#$
% are different and much more complicated now. Here $g(\omega)$ takes the same formula as the previous subsection.
%\comment{It is strange: the 0th source generate 0th vev, we can get 0th's GF.  The 0th %source also can generate 2nd order source as well as 2nd order vev. We can analyze %this procedure in detail, but it is still difficult to define $G_{IR}^{(2)}$}.

%%%%%%%%%%%%%%%%%%%%%%%%%%%%%%%%%%%%%%
%\begin{figure}[h!]
%\begin{center}
%\begin{tabular}{cc}
%\includegraphics[width=0.7\textwidth]{figure.pdf}
%\end{tabular}
%\caption{A cartoon of our results:}
%\label{cartoon}
%\end{center}
%\end{figure}
%%%%%%%%%%%%%%%%%%%%%%%%%%%%%%%%%%%%%%

%%%%%%%%%%%%%%%%%%%%%%%%%%%%%%%%%%%%%%
%%%%%%%%%%%%%%%%%%%%%%%%%%%%%%%%%%%%%%
\section{Technical Conclusions}\label{sec5}
%%%%%%%%%%%%%%%%%%%%%%%%%%%%%%%%%%%%%%
%%%%%%%%%%%%%%%%%%%%%%%%%%%%%%%%%%%%%%

Eqns. \eqref{bandgap1}, \eqref{selfen1}, \eqref{selfen2} and \eqref{eq:4} are our main conclusions of lattice effects on holographic single fermion spectral functions. Here the lattice is encoded by a periodic correction to the gauge potential in RN AdS black hole background  dual to a strongly coupled $2+1$ dimensional field theory at finite density.  
We have already discussed the main physics of these results in the beginning, section~\ref{sec2sum}. In the quasi-Fermi-surface regime ($q\gg m$) one finds a gap and
the mechanism for the presence of the gap is similar to the standard condensed matter physics: the degeneracy from the interaction between different levels gives the gap. The gap disappears when such kinds of interaction turned off. The more interesting physics is that related to the AdS$_{2}$ sector. In the quasi-Fermi surface regime this is responsible for the dispersive properties of the Fermi-surface excitations. In the opposite ($q\ll m$) AdS${}_2$ metal regime it itself controls the IR physics. The effect of the lattice is to allow Umklapp contributions $\ome^{2\nu_{\vec{k}\pm \vec{K}}}$ to the characteristic local quantum critical scaling $\ome^{2\nu_{\vec{k}}}$. These essentially shift more spectral weight to lower energies. % These are the most interesting physics related to the dispertion relation of the spectral function and they come from the first order corrections. The second order correction has important  effect on the self-energy. 
% Our result indicates the possible pseudogap at arbitrary  momentum. This might be something new in this holographic set-up.
% For $2\nu_{k_\ell}<1$ case, the possible bandgap at general momentum is also new.

At the technical level we made a number of assumptions to obtain our results. 
For convenience we have chosen the dual spacetime geometry as RN black hole with only $A_t$ modified to encode the lattice effect, i.e. we neglected the backreaction of the lattice effect to the metric field. One immediate question is what would happen when the first order gravity corrections appears. We have argued in section~\ref{sec2} that the understanding of level repulsion and lattice symmetries should not change the qualitative features of our result. Nevertheless it would be good to confirm this. Such kinds of geometry where the lattice effect to the metric field was also studied has been recently considered in \cite{Maeda:2011pk}  (see also \cite{Nakamura:2009tf,Donos:2011bh,Bergman:2011rf,Horowitz:2012ky}). It would be very interesting to study the spectral functions of the probe fermion in these backgrounds. 

In addition there are several ways to generalize our methods. 
\begin{itemize}
\item The most obvious next step is the compute the spatially local
  Green's function $G(\ome;x,x)$ relevant for scanning tunneling
  spectroscopy rather then ARPES. In momentum space this implies that we shall also need
  to evaluate the $G(\ome,k, k+K)$ Green's function.

\item
One can consider a more general potential, such as 
the gauge potentials with higher order Fourier components
$\mu_1(x+2\pi a)=\mu_1(x)=\sum_n\epsilon_n \cos\frac{nx}{a}$ or other
lattice structure, e.g. the canonical two-dimensional square or rectangular lattice $\mu_1(x+2\pi a,y+2\pi b)=\mu_1(x,y)$. 

\item In the approach here we worked perturbatively in the weak potential limit. In condensed matter physics, another interesting approach to the band structure is the tight binding model. % It would be interesting to realize the tight-binding case where tightly bound fermions in the potential. \commently{??}
  Holographically these are presumably closer related to brane-models of the type considered in \cite{Kachru:2009xf,arXiv:1009.3268}. Perhaps there might be a way to interpolate between them.

% \item  Finally, the most direct generalization is to consider three dimensional case and perform the  corresponding numerical calculations. 
\end{itemize}
We hope to return to these issues in the future.

%\commently{IMPORTANT NOTE TO ADD: (1) periodicity of $G(x,x',\omega)$; (2) Diagnalization and resummation; (3) regularity of the UV coeffcient: smoothly arguments}

\section*{Acknowledgments}
We thank A. Beekman, B. van Rees, A. Parnachev, S. Sachdev, J-H She, A. O. Starinets, K. Wu for useful discussions. 
This research is supported in part by a Spinoza Award (J. Zaanen)
from the Netherlands Organisation for Scientific Research (NWO) and by
the Dutch Foundation for Fundamental Research on Matter (FOM).

\appendix

%%%%%%%%%%%%%%%%%%%%%%%%%%%%%%%%%%%%%%

\section{Coefficients}
%%%%%%%%%%%%%%%%%%%%%%%%%%%%%%%%%%%%%%
%%%%%%%%%%%%%%%%%%%%%%%%%%%%%%%%%%%%%%

In this appendix, we collect the coefficients appeared in eqns. (\ref{x0},
\ref{eta2special}, \ref{degetas1}, \ref{degetas2}, \ref{selfen1}). These coefficients
do not depend on $\omega$ and $\varepsilon$ and are finite.

\begin{itemize}
\item The coefficients in (\ref{x0})
\bea\label{co1}
{n_1}_{\alpha j\beta \ell'}^\ell&=&\omega^{-\nu_{k_{\ell'}}}\int_{\infty}^\varepsilon dz\frac{\eta_{1\alpha j}^\ell X_{(\alpha j,\beta \ell')}^{\ell}}{c^\ell_{\alpha j}h_{\alpha j}z^2}
-\frac{qM_{\alpha\beta,\ell\ell'}}{12\nu_{k_\ell}}\bigg(\frac{q\mu_0}{3}
+(-1)^{\alpha+j+1}\frac{k_\ell}{\sqrt{6}}\nonumber\\
&&~~~~~-(-1)^{\beta+j+1}\frac{k_{\ell'}}{\sqrt{6}}
\bigg)\bigg[\frac{\varepsilon^{-\nu_{k_\ell}-\nu_{k_{\ell'}}}}{-\nu_{k_\ell}-\nu_{k_{\ell'}}}v^{(0,\ell')}_{-\beta j}+\frac{\varepsilon^{-\nu_{k_\ell}+\nu_{k_{\ell'}}}}{-\nu_{k_\ell}+\nu_{k_{\ell'}}}\tilde{\mathcal{G}}_{\beta}^{\ell'}v^{(0,\ell')}_{+\beta j}\bigg];\nonumber\\
{n_2}_{\alpha j\beta \ell'}^\ell&=&-\omega^{-\nu_{k_{\ell'}}}\int_{\infty}^\varepsilon dz\frac{\eta_{2\alpha j}^\ell X_{(\alpha j,\beta \ell')}^{\ell}}{c^\ell_{\alpha j}h_{\alpha j}z^2}
+\frac{qM_{\alpha\beta,\ell\ell'}}{12\nu_{k_\ell}}\bigg(\frac{q\mu_0}{3}
+(-1)^{\alpha+j+1}\frac{k_\ell}{\sqrt{6}}\nonumber\\
&&~~~~~-(-1)^{\beta+j+1}\frac{k_{\ell'}}{\sqrt{6}}
\bigg)\bigg[\frac{\varepsilon^{\nu_{k_\ell}-\nu_{k_{\ell'}}}}{\nu_{k_\ell}-\nu_{k_{\ell'}}}v^{(0,\ell')}_{-\beta j}+\frac{\varepsilon^{\nu_{k_\ell}+\nu_{k_{\ell'}}}}{\nu_{k_\ell}+\nu_{k_{\ell'}}}\tilde{\mathcal{G}}_{\beta}^{\ell'}v^{(0,\ell')}_{+\beta j}\bigg];\nonumber\\
{n_3}_{\alpha j\beta \ell'}^\ell&=&-\frac{qM_{\alpha\beta,\ell\ell'}}{6(\nu_{k_\ell}^2-\nu_{k_{\ell'}}^2)}
\bigg(\frac{q\mu_0}{3}+(-1)^{\alpha+j+1}\frac{k_\ell}{\sqrt{6}}
-(-1)^{\beta+j+1}\frac{k_{\ell'}}{\sqrt{6}}\bigg)\tilde{\mathcal{G}}_{\beta}^{\ell'}v^{(0,\ell')}_{+\beta j};\nonumber\\
{n_4}_{\alpha j\beta \ell'}^\ell&=&-\frac{qM_{\alpha\beta,\ell\ell'}}{6(\nu_{k_\ell}^2-\nu_{k_{\ell'}}^2)}\bigg(\frac{q\mu_0}{3}
+(-1)^{\alpha+j+1}\frac{k_\ell}{\sqrt{6}}-(-1)^{\beta+j+1}\frac{k_{\ell'}}{\sqrt{6}}\bigg)v^{(0,\ell')}_{-\beta j}
\eea

where %\comment{IS THIS EXACT OR NOT; WHAT IS THE APPROXIMATION?}
\be c_{\alpha j}^\ell= \frac{2 \nu_{k_\ell}}{\frac{q\mu_0}{6}+(-1)^{\alpha+j+1}\frac{k_\ell}{\sqrt{6}}}.\ee

\item  The coefficients in (\ref{eta2special})

\bea\label{secondordercoeff}
{s_1}_{\alpha j}^\ell&=&\omega^{-\nu_{k_\ell}}\int_\infty^\varepsilon dz\frac{
\tilde{\eta}_{1\alpha j}^\ell Y_{\alpha j}^{\ell}}{c_{\alpha j}^\ell z^2 h_{\alpha j}}
+\frac{q}{6c_{\alpha j}^\ell}\sum_{\beta;\ell'=\ell\pm 1}M_{\alpha\beta,\ell\ell'}\bigg[
\frac{\varepsilon^{\nu_{k_{\ell'}}-\nu_{k_\ell}}}{\nu_{k_{\ell'}}-\nu_{k_\ell}}{N_1}_{\beta j,\alpha\ell}^{\ell'}\nonumber\\&&~~~~~+ {N_3}_{\beta j,\alpha\ell}^{\ell'}(\ln\varepsilon+\frac{1}{2\nu_{k_\ell}})
-\frac{\varepsilon^{-2\nu_{k_\ell}}}{2\nu_{k_\ell}}{N_4}_{\beta j,\alpha\ell}^{\ell'}
\bigg];
\nonumber\\
{s_2}_{\alpha j}^\ell&=&-\omega^{-\nu_{k_\ell}}\int_\infty^\varepsilon dz\frac{
\tilde{\eta}_{2\alpha j}^\ell Y_{\alpha j}^{\ell}}{c_{\alpha j}^\ell z^2 h_{\alpha j}}
-\frac{q}{6c_{\alpha j}^\ell}\sum_{\beta,\ell'=\ell\pm 1}M_{\alpha\beta,\ell\ell'}\bigg[
\frac{\varepsilon^{\nu_{k_{\ell'}}+\nu_{k_\ell}}}{\nu_{k_{\ell'}}+\nu_{k_\ell}}{N_1}_{\beta j,\alpha\ell}^{\ell'}
\nonumber\\&&~~~~~~+ {N_3}_{\beta j,\alpha\ell}^{\ell'}\frac{\varepsilon^{2\nu_{k_\ell}}}{2\nu_{k_\ell}}
+{N_4}_{\beta j,\alpha\ell}^{\ell'}\big(\ln\varepsilon-\frac{1}{2\nu_{k_\ell}}\big)\bigg];\nonumber\\
{s_3}_{\alpha j\ell'}^\ell&=&\frac{q}{6c_{\alpha j}^\ell}\sum_{\beta}M_{\alpha\beta,\ell\ell'}
\frac{2\nu_{k_\ell}}{\nu_{k_\ell}^2-\nu_{k_{\ell'}}^2}{N_1}_{\beta j,\alpha\ell}^{\ell'};\nonumber\\
{s_4}_{\alpha j}^\ell&=&-\frac{q}{6c_{\alpha j}^\ell}\sum_{\beta,\ell'=\ell\pm 1}M_{\alpha\beta,\ell\ell'}{N_3}_{\beta j,\alpha\ell}^{\ell'};\nonumber\\
{s_5}_{\alpha j}^\ell&=&\frac{q}{6c_{\alpha j}^\ell}\sum_{\beta,\ell'=\ell\pm 1}M_{\alpha\beta,\ell\ell'}{N_4}_{\beta j,\alpha\ell}^{\ell'}\eea

with
\bea
{N_1}^{\ell'}_{\beta j,\alpha\ell}&=&\tilde{n_1}_{\beta j,\alpha\ell}^{\ell'}-\frac{(-1)^j\nu_{k_{\ell'}}}{\frac{q\mu_0}{6}+(-1)^{\alpha+j+1}\frac{k_\ell}{\sqrt{6}}}\tilde{n_1}_{\beta l,\alpha\ell}^{\ell'};\nonumber\\
{N_3}^{\ell'}_{\beta j,\alpha\ell}&=&{n_3}_{\beta j,\alpha\ell}^{\ell'}-\frac{(-1)^j\nu_{k_\ell}}{\frac{q\mu_0}{6}+(-1)^{\alpha+j+1}\frac{k_\ell}{\sqrt{6}}}{n_3}_{\beta l,\alpha\ell}^{\ell'};\nonumber\\
{N_4}^{\ell'}_{\beta j,\alpha\ell}&=&{n_4}_{\beta j,\alpha\ell}^{\ell'}+\frac{(-1)^j\nu_{k_\ell}}{\frac{q\mu_0}{6}+(-1)^{\alpha+j+1}\frac{k_\ell}{\sqrt{6}}}{n_4}_{\beta l,\alpha\ell}^{\ell'}.
\eea

\item The coefficients in (\ref{degetas1})

\bea\label{degco1}
{d_1}_{\alpha j\beta \ell'}^\ell&=&\omega^{-\nu_{k_{\ell'}}}\int_{\infty}^\varepsilon dz\frac{\eta_{1\alpha j}^\ell X_{(\alpha j,\beta \ell')}^{\ell}}{c^\ell_{\alpha j}h_{\alpha j}z^2}
-\frac{qM_{\alpha\beta,\ell\ell'}}{12\nu_{k_\ell}}\bigg(\frac{q\mu_0}{3}
+(-1)^{\alpha+j+1}\frac{k_\ell}{\sqrt{6}}\nonumber\\
&&~~~~~-(-1)^{\beta+j+1}\frac{k_{\ell'}}{\sqrt{6}}
\bigg)\bigg[\frac{\varepsilon^{-2\nu_{k_\ell}}}{-2\nu_{k_\ell}}v^{(0,\ell')}_{-\beta j}+\tilde{\mathcal{G}}_{\beta}^{\ell'}v^{(0,\ell')}_{+\beta j}\ln{\varepsilon}\bigg]
\nonumber\\
&&~~~~~-\frac{qM_{\alpha\beta,\ell\ell'}}{24\nu_{k_\ell}^2}\bigg(\frac{q\mu_0}{3}
+(-1)^{\alpha+j+1}\frac{k_\ell}{\sqrt{6}}-(-1)^{\beta+j+1}\frac{k_{\ell'}}{\sqrt{6}}\bigg)\tilde{\mathcal{G}}_{\beta}^{\ell'}v^{(0,\ell')}_{+\beta j};\nonumber\\
{d_2}_{\alpha j\beta \ell'}^\ell&=&-\omega^{-\nu_{k_{\ell'}}}\int_{\infty}^\varepsilon dz\frac{\eta_{2\alpha j}^\ell X_{(\alpha j,\beta \ell')}^{\ell}}{c^\ell_{\alpha j}h_{\alpha j}z^2}
+\frac{qM_{\alpha\beta,\ell\ell'}}{12\nu_{k_\ell}}\bigg(\frac{q\mu_0}{3}
+(-1)^{\alpha+j+1}\frac{k_\ell}{\sqrt{6}}\nonumber\\
&&~~~~~-(-1)^{\beta+j+1}\frac{k_{\ell'}}{\sqrt{6}}
\bigg)\bigg[v^{(0,\ell')}_{-\beta j}\ln{\varepsilon}+\frac{\varepsilon^{\nu_{k_\ell}+\nu_{k_{\ell'}}}}{\nu_{k_\ell}+\nu_{k_{\ell'}}}\tilde{\mathcal{G}}_{\beta}^{\ell'}v^{(0,\ell')}_{+\beta j}\bigg]\nonumber\\
&&~~~~~-\frac{qM_{\alpha\beta,\ell\ell'}}{24\nu_{k_\ell}^2}
\bigg(\frac{q\mu_0}{3}+(-1)^{\alpha+j+1}\frac{k_\ell}{\sqrt{6}}
-(-1)^{\beta+j+1}\frac{k_{\ell'}}{\sqrt{6}}\bigg)v^{(0,\ell')}_{-\beta j};\nonumber\\
{d_3}_{\alpha j\beta \ell'}^\ell&=&\frac{qM_{\alpha\beta,\ell\ell'}}{12\nu_{k_\ell}}
\bigg(\frac{q\mu_0}{3}+(-1)^{\alpha+j+1}\frac{k_\ell}{\sqrt{6}}
-(-1)^{\beta+j+1}\frac{k_{\ell'}}{\sqrt{6}}\bigg)\tilde{\mathcal{G}}_{\beta}^{\ell'}v^{(0,\ell')}_{+\beta j};\nonumber\\
{d_4}_{\alpha j\beta \ell'}^\ell&=&-\frac{qM_{\alpha\beta,\ell\ell'}}{12\nu_{k_\ell}}\bigg(\frac{q\mu_0}{3}
+(-1)^{\alpha+j+1}\frac{k_\ell}{\sqrt{6}}-(-1)^{\beta+j+1}\frac{k_{\ell'}}{\sqrt{6}}\bigg)v^{(0,\ell')}_{-\beta j}.\nonumber\\
\eea

\item The coefficients in (\ref{degetas2})

\bea
{\degens_1}_{\alpha j \ell'}^\ell&=&\int_\infty^\varepsilon dz\frac{\eta_1F(z)}{c_{\alpha j}^\ell z^2h_{\alpha j}}+
\frac{q}{6c_{\alpha j}^\ell}\sum_\beta M_{\alpha\beta,\ell\ell'}\bigg[\ln{\varepsilon}D_{1\beta}+\varepsilon^{-2\nu_{k_\ell}}(-\frac{D_{2\beta}}{2\nu_{k_\ell}}-\frac{D_{4\beta}}{4\nu_{k_\ell}^2})\nonumber\\&&~~~~+\frac{(\ln\varepsilon)^2}{2}D_{3\beta}-\frac{1}{2\nu_{k_\ell}}\epsilon^{-2\nu_{k_\ell}}\ln{\varepsilon}D_{4\beta}+(\frac{D_{1\beta}}{2\nu_{k_\ell}}-\frac{D_{3\beta}}{4\nu_{k_\ell}^2})
\bigg];
\nonumber\\
{\degens_2}_{\alpha j \ell'}^\ell&=&-\int_\infty^\varepsilon dz\frac{\eta_2F(z)}{c_{\alpha j}^\ell z^2h_{\alpha j}}-
\frac{q}{6c_{\alpha j}^\ell}\sum_\beta M_{\alpha\beta,\ell\ell'}\bigg[\ln{\varepsilon}D_{2\beta}+\varepsilon^{2\nu_{k_\ell}}(\frac{D_{1\beta}}{2\nu_{k_\ell}}-\frac{D_{3\beta}}{4\nu_{k_\ell}^2})\nonumber\\&&~~~~+\frac{(\ln\varepsilon)^2}{2}D_{4\beta}+\frac{1}{2\nu_{k_\ell}}\epsilon^{-2\nu_{k_\ell}}\ln{\varepsilon}D_{3\beta}+(-\frac{D_{2\beta}}{2\nu_{k_\ell}}-\frac{D_{4\beta}}{4\nu_{k_\ell}^2})
\bigg];
\nonumber
\eea
\bea
{\degens_3}_{\alpha j \ell'}^\ell&=&\int_\infty^\varepsilon dz\frac{\eta_1G(z)}{c_{\alpha j}^\ell z^2h_{\alpha j}}+
\frac{q}{6c_{\alpha j}^\ell}\sum_\beta M_{\alpha\beta,\ell\ell'}\bigg[(\ln{\varepsilon}+\frac{1}{\nu_{k_\ell}})D_{3\beta}+\varepsilon^{-2\nu_{k_\ell}}\frac{D_{4\beta}}{2\nu_{k_\ell}}-D_{1\beta}
\bigg];
\nonumber\\
{\degens_4}_{\alpha j \ell'}^\ell&=&-\int_\infty^\varepsilon dz\frac{\eta_2G(z)}{c_{\alpha j}^\ell z^2h_{\alpha j}}-
\frac{q}{6c_{\alpha j}^\ell}\sum_\beta M_{\alpha\beta,\ell\ell'}\bigg[-\ln{\varepsilon}D_{4\beta}+\varepsilon^{2\nu_{k_\ell}}\frac{D_{3\beta}}{2\nu_{k_\ell}}-D_{2\beta}
\bigg];\nonumber
\eea
\bea
{\degens_5}_{\alpha j \ell'}^\ell&=&-
\frac{q}{6c_{\alpha j}^\ell}\sum_\beta M_{\alpha\beta,\ell\ell'}\bigg[D_{1\beta}-\frac{D_{3\beta}}{2\nu_{k_\ell}}\bigg];
\nonumber\\
{\degens_6}_{\alpha j \ell'}^\ell&=&-\frac{q}{6c_{\alpha j}^\ell}\sum_\beta M_{\alpha\beta,\ell\ell'}\bigg[-D_{2\beta}-\frac{D_{4\beta}}{2\nu_{k_\ell}}\bigg];
\nonumber
\eea
\bea
{\degens_7}_{\alpha j \ell'}^\ell&=&-
\frac{q}{6c_{\alpha j}^\ell}\sum_\beta M_{\alpha\beta,\ell\ell'}2D_{3\beta};
\nonumber
\eea
\bea
{\degens_8}_{\alpha j \ell'}^\ell&=&-
\frac{q}{6c_{\alpha j}^\ell}\sum_\beta M_{\alpha\beta,\ell\ell'}\bigg[\frac{3}{2}D_{3\beta}\bigg];
\nonumber\\
{\degens_{9}}_{\alpha j \ell'}^\ell&=&-\frac{q}{6c_{\alpha j}^\ell}\sum_\beta M_{\alpha\beta,\ell\ell'}\bigg[\frac{1}{2}D_{4\beta}\bigg];
\nonumber
\eea
\bea
{\degens_{10}}_{\alpha j \ell'}^\ell&=&-
\frac{q}{6c_{\alpha j}^\ell}\sum_\beta M_{\alpha\beta,\ell\ell'}\bigg[\frac{1}{2}D_{3\beta}\bigg];
\nonumber\\
{\degens_{11}}_{\alpha j \ell'}^\ell&=&\frac{q}{6c_{\alpha j}^\ell}\sum_\beta M_{\alpha\beta,\ell\ell'}\bigg[\frac{1}{2}D_{4\beta}\bigg]
\nonumber
\eea

with
\bea\label{degco2}
{D_1}_\beta&=&\tilde{d_1}_{\beta j,\alpha\ell}^{\ell'}-\frac{(-1)^j\nu_{k_{\ell'}}}{\frac{q\mu_0}{6}+(-1)^{\alpha+j+1}\frac{k_\ell}{\sqrt{6}}}\tilde{d_1}_{\beta l,\alpha\ell}^{\ell'}+{d_3}_{\beta l,\alpha\ell}^{\ell'};\nonumber\\
{D_2}_\beta&=&{d_4}_{\beta l,\alpha\ell}^{\ell'};\nonumber\\
{D_3}_\beta&=&{d_3}_{\beta j,\alpha\ell}^{\ell'}-\frac{(-1)^j\nu_{k_{\ell'}}}{\frac{q\mu_0}{6}+(-1)^{\alpha+j+1}\frac{k_\ell}{\sqrt{6}}}{d_3}_{\beta l,\alpha\ell}^{\ell'};\nonumber\\
{D_4}_\beta&=&{d_4}_{\beta j,\alpha\ell}^{\ell'}-\frac{(-1)^j\nu_{k_{\ell'}}}{\frac{q\mu_0}{6}+(-1)^{\alpha+j+1}\frac{k_\ell}{\sqrt{6}}}{d_4}_{\beta l,\alpha\ell}^{\ell'}.\nonumber\\
\eea

\item The coefficients in (\ref{selfen1})

Note that near $\omega=0, k=k_F$, we expand 
\bea
&& \cA_{\alpha\ell,\alpha\ell}^{(0)-}v_{-\alpha2}^{(0,\ell)}+\epsilon^2\cA_{\alpha\ell}^{(2,5)}\tilde{s_5}_{\alpha2}^\ell-\epsilon^2\sum_{\beta,\ell'=\ell\pm 1}\frac{\cA_{\beta\ell',\alpha\ell}^{(1,4)}{n_4}_{\beta 2\alpha\ell}^{\ell'}\cA_{\alpha\ell,\beta\ell'}^{(1,4)}{n_4}_{\alpha2\beta\ell'}^{\ell}}{\cA_{\beta\ell',\beta\ell'}^{(0)-}v_{-\beta2}^{(0,\ell')}}\nonumber\\
&=&e_1 \omega- e_2 (k_\ell-k_F)+...
\eea
where in the above formulae, we have substituted $A^\text{UV}=A^{(0)}+\omega A^{(1)}+...$.
\begin{eqnarray} \label{appb8}
  Z e_1&=&B_{\alpha\ell,\alpha\ell}^{(0)-}v_{-\alpha1}^{(0,\ell)} ;\\
v_F e_1&=& e_2 ;\\
 \alpha_{\vec{k}}e_1&=&A_{\alpha\ell,\alpha\ell}^{(0)+}v_{+\alpha 2}^{(0,\ell)}\tilde{\mathcal{G}}_{\alpha}^{\ell} 
  +\epsilon^2\bigg(\cA_{\alpha\ell}^{(2,1)}\tilde{s_1}^{\ell}_{\alpha2}+\cA_{\alpha\ell}^{(2,4)}{s_4}^{\ell}_{\alpha2}\nonumber\\
  &&-\sum_{\beta\ell'=\ell\pm1}\frac{\cA_{\alpha\ell,\beta\ell'}^{(1,4)}{n_4}^{\ell}_{\alpha2\beta\ell'}
  \cA_{\alpha\ell',\beta\ell}^{(1,3)}{n_3}^{\ell'}_{\alpha2\beta\ell}}{\cA_{\alpha\ell',\alpha\ell'}^{(0)-}v_{-\alpha2}^{(0,\ell')}}\bigg)\\
\beta^{(1)}_{\vec{k}} e_1&=&-\epsilon^2\sum_{\beta\ell'=\ell-1}\frac{\cA_{\alpha\ell,\beta\ell'}^{(1,3)}{n_3}^{\ell}_{\alpha2\beta\ell'}
\cA_{\alpha\ell',\beta\ell}^{(1,4)}{n_4}^{\ell'}_{\alpha2\beta\ell}}{\cA_{\alpha\ell',\alpha\ell'}^{(0)-}v_{-\alpha2}^{(0,\ell')}} ;\\
\beta^{(2)}_{\vec{k}}e_1&=&\epsilon^2\cA_{\alpha\ell,\ell-1}^{(2,3)}{s_3}^{\ell}_{\alpha2\ell-1}
-\epsilon^2\sum_{\beta\ell'=\ell-1}\frac{\cA_{\alpha\ell,\beta\ell'}^{(1,1)}\tilde{n_1}^{\ell}_{\alpha2\beta\ell'}
\cA_{\alpha\ell',\beta\ell}^{(1,4)}{n_4}^{\ell'}_{\alpha2\beta\ell}
}{\cA_{\alpha\ell',\alpha\ell'}^{(0)-}v_{-\alpha2}^{(0,\ell')}}\nonumber\\&&
-\epsilon^2\sum_{\beta\ell'=\ell-1}\frac{\cA_{\alpha\ell,\beta\ell'}^{(1,4)}{n_4}^{\ell}_{\alpha2\beta\ell'}
\cA_{\alpha\ell',\beta\ell}^{(1,1)}\tilde{n_1}^{\ell'}_{\alpha2\beta\ell}
}{\cA_{\alpha\ell',\alpha\ell'}^{(0)-}v_{-\alpha2}^{(0,\ell')}};
\\
\beta^{(3)}_{\vec{k}} e_1&=&\epsilon^2\tilde{\cA}_{\alpha\ell}^{(2,4)}\tilde{s_4}^{\ell}_{\alpha2};\\
\beta^{(4)}_{\vec{k}} e_1&=&\epsilon^2\cA_{\alpha\ell,\ell+1}^{(2,3)}{s_3}^{\ell}_{\alpha2\ell+1}
-\epsilon^2\sum_{\beta\ell'=\ell+1}\frac{\cA_{\alpha\ell,\beta\ell'}^{(1,1)}\tilde{n_1}^{\ell}_{\alpha2\beta\ell'}
\cA_{\alpha\ell',\beta\ell}^{(1,4)}{n_4}^{\ell'}_{\alpha2\beta\ell}
}{\cA_{\alpha\ell',\alpha\ell'}^{(0)-}v_{-\alpha2}^{(0,\ell')}}\nonumber\\&&
-\epsilon^2\sum_{\beta\ell'=\ell+1}\frac{\cA_{\alpha\ell,\beta\ell'}^{(1,4)}{n_4}^{\ell}_{\alpha2\beta\ell'}
\cA_{\alpha\ell',\beta\ell}^{(1,1)}\tilde{n_1}^{\ell'}_{\alpha2\beta\ell}
}{\cA_{\alpha\ell',\alpha\ell'}^{(0)-}v_{-\alpha2}^{(0,\ell')}};
\\ \label{appb12}
\beta^{(5)}_{\vec{k}} e_1&=&-\epsilon^2\sum_{\beta\ell'=\ell+1}\frac{\cA_{\alpha\ell,\beta\ell'}^{(1,3)}{n_3}^{\ell}_{\alpha2\beta\ell'}
\cA_{\alpha\ell',\beta\ell}^{(1,4)}{n_4}^{\ell'}_{\alpha2\beta\ell}}{\cA_{\alpha\ell',\alpha\ell'}^{(0)-}v_{-\alpha2}^{(0,\ell')}}.
\end{eqnarray}
%for the degenerate case where $\ell'=\ell\pm 1$.

% Now we consider the degenerate case $k_x=\frac{1}{2a}, k_y=0$. The retarded Green's function is in (\ref{GR2}).   When $\omega\to 0 $ and $2\nu_{k_\ell}>1$, the real part of $G_{\alpha\ell,\alpha\ell}^{-1}$ is 
% \bea
% \text{Re}\big[G_{\alpha\ell,\alpha\ell}^{-1}\big]&\sim&
% a_{-\alpha\ell,\alpha\ell}v_{-\alpha2}^{(0,\ell)}+\epsilon^2\tilde{a}_{\alpha\ell}^{(5)}\tilde{s_5}_{\alpha2}^\ell-\epsilon^2\sum_{\ell'=\ell\pm 1}\frac{a_{\alpha\ell',\alpha\ell}^{(4)}{n_4}_{\alpha 2\alpha\ell}^{\ell'}a_{\alpha\ell,\alpha\ell'}^{(4)}{n_4}_{\alpha2\alpha\ell'}^{\ell}}{a_{-\alpha\ell',\alpha\ell'}v_{-\alpha2}^{(0,\ell')}}\nonumber\\
% &\sim&\omega-v_F(k_\ell-k_F).
% \eea 
% The imaginary part of $G_{\alpha\ell,\alpha\ell}^{-1}$ or the self-energy is 
% \be \text{Im}\big[G_{\alpha\ell,\alpha\ell}^{-1}\big]\sim \omega^{2\nu_{k_\ell}}\bigg(1+\epsilon^2 \#'' (\ln\omega)^2\bigg)\ee
% where 
% \be \#=\frac{\tilde{a}_{\alpha\ell}^{(8)}\tilde{s_8}_{\alpha 2\ell-1}}{a_{+\alpha\ell,\alpha\ell}v_{+\alpha 2}^{(0,\ell)}\tilde{\mathcal{G}}_{\alpha}^{\ell}}.\ee
% Poles are at  $ \omega^{2\nu_{k_\ell}}(1+\epsilon^2\#''(\ln\omega)^2)=0,$
% i.e. $\omega=0$ and $\omega\sim e^{-1/\epsilon\sqrt{-\#''}}$.
% As to the degenerate case $k_x=\frac{1}{2a}$ with $k_y\neq 0$, the second order correction is negligible compared to the first order correction which gives bandgap. 

\end{itemize}

\end{document}